\documentclass[12pt,preprint]{aastex}

\usepackage{graphicx}
\usepackage{epsfig}
\usepackage{epstopdf}
\usepackage{amsmath}

\def\gs{\mathrel{\raise0.35ex\hbox{$\scriptstyle >$}\kern-0.6em
\lower0.40ex\hbox{{$\scriptstyle \sim$}}}}
\def\ls{\mathrel{\raise0.35ex\hbox{$\scriptstyle <$}\kern-0.6em
\lower0.40ex\hbox{{$\scriptstyle \sim$}}}}

\def\spose#1{\hbox to 0pt{#1\hss}}
\def\simlt{\mathrel{\spose{\lower 3pt\hbox{$\mathchar"218$}}
     \raise 2.0pt\hbox{$\mathchar"13C$}}}
\def\simgt{\mathrel{\spose{\lower 3pt\hbox{$\mathchar"218$}}
     \raise 2.0pt\hbox{$\mathchar"13E$}}}

\newcommand{\asec}{$^{\prime\prime}$}

\shorttitle{5\,--\,90\,GHz SEDs of radio galaxies}
\shortauthors{}

\begin{document}

\title{High frequency radio SEDs and polarization fractions of sources in an ACT survey field}

\author{Anna Sajina\altaffilmark{1,2}, Bruce Partridge\altaffilmark{2}, Tyler Evans\altaffilmark{2,3}, Shannon Stefl\altaffilmark{4}, Nicholas~Vechik\altaffilmark{2},  Steve~Myers\altaffilmark{5}, Simon~Dicker\altaffilmark{6},Phillip~Korngut\altaffilmark{6}}

\altaffiltext{1}{Tufts University, Medford, MA, 02155, USA}
\altaffiltext{2}{Haverford College, Haverford, PA, 19041, USA}
\altaffiltext{3}{Swinbourne University, Melbourne, Australia}
\altaffiltext{4}{Kent State University}
\altaffiltext{5}{National Radio Astronomy Observatory, Soccorro, NM, 87801, USA}
\altaffiltext{6}{University of Pennsylvania, Philadelphia, PA, 19104, USA}

\begin{abstract}

We present flux densities and polarization percentages of 159 radio galaxies based on nearly simultaneous Very Large Array (VLA) observations at four frequencies, 4.86, 8.46, 22.46 and 43.34\,GHz.  This sample is selected from the high-frequency Australia Telescope 20\,GHz (AT20G) survey and consists of all sources with flux density $S_{20\,GHz}$\,$>$\,40\,mJy in an equatorial field of the ongoing Atacama Cosmology Telescope (ACT) survey.  For a subset of 25 of these sources we used the Green Bank Telescope (GBT) to obtain 90\,GHz data. The goals of this program are: 1) a characterization of the spectra, polarization and variability of high frequency-selected radio sources; 2) extrapolating from the few\,GHz regime to the $\sim$\,150\,GHz regime of the ACT survey, allowing for more accurate removal of the radio source signal in our particular field; and 3)  providing a data set that will allow more accurate modeling of the high-frequency radio source contamination in current and future SZ and CMB experiments. We find that, as expected, this sample consists of flatter spectrum and more compact or point-like sources than low frequency-selected samples.  In the K-band, variability is typically $\ls$\,20\%, although there are exceptions. The higher frequency data is well suited to the detection of extreme Giga-Hertz Peak spectrum Sources (GPS). The inclusion of the 43\,GHz data causes the relative fraction of inverted spectrum sources to go down and of peaked spectrum sources to go up when compared with the AT20G survey results. The trend largely continues with the inclusion of the 90\,GHz data, although $\sim$\,10\% of the sources with GBT data show a spectral upturn from 43\,GHz to 90\,GHz. The measured polarization fractions are typically $<$\,5\%, although in some cases they are measured to be up to $\sim$\,20\%. For sources with detected polarized flux in all four bands, about 40\% of the sample, the polarization fractions typically increase with frequency.  This trend is stronger for steeper spectrum sources as well as for the lower flux density sources.
\end{abstract}


\section{Introduction}

The high frequency spectral energy distributions (SEDs) of radio galaxies are still poorly understood, in part because millimeter wave observations are more difficult than those at centimeter wavelengths \citep[for a recent review see][]{dezotti10}. And yet, better characterization of these extragalactic sources, including their polarization properties, across a wider frequency range, will be invaluable for testing models of the physics of accreting black holes and radio jet formation. In addition to increasing our understanding of the physics of radio sources, knowledge of the high frequency SEDs of foreground radio sources is becoming increasingly important as measurements of small-scale fluctuations in the cosmic microwave background achieve higher sensitivity and higher resolution. The same is true for increasingly sensitive searches for the Sunyaev-ZelÕdovich (SZ) effect in clusters of galaxies. Radio sources both in clusters and in the field are a principal source of confusion in such experiments \citep[see][]{lueker09,reich09,sharp09,sievers09,vieira09,fowler10,marriage10}. 

The 10-meter South Pole Telescope \citep[SPT;][]{carlstrom09} is conducting a survey for the SZ effect produced by clusters of galaxies \citep{stan09,lueker09}, and is also measuring small  fluctuations in the microwave background. This group detects nearly 200 foreground sources, three quarters of which are dominated by apparent synchrotron radiation, with a small fraction, mostly faint sources, dominated by thermal dust emission \citep{vieira09}. The Atacama Cosmology Telescope (ACT) is carrying out a similar survey over a larger survey area. Preliminary results reported by \citet{marriage10} at 148\,GHz reveal more than 150, again mostly synchrotron, sources. In both cases, virtually all the detected sources are identified with radio sources contained in catalogs constructed at lower radio frequencies. And of course, the ongoing {\sl ESA Planck} mission is conducting an all-sky CMB experiment over an unprecedented frequency range from 30 to 857\,GHz \citep[see e.g.][]{tauber10,mandolesi10}. 

A better understanding of the high frequency SEDs of foreground radio sources would permit a better statistical analysis of their effect on both the high-$l$ power spectrum of CMB anisotropies and on increasingly sensitive surveys for the SZ effect. Data already available, for instance, suggest that the models of \citet{toffolatti98,toffolatti05} based on extrapolations from lower frequency radio observations overestimate the effect of radio sources at millimeter wavelengths \citep{marriage10}. 

High frequency observations are particularly relevant to CMB observations made at frequencies of 90\,GHz or above, often to avoid signals from foreground sources. The polarization properties of these same sources at high frequencies are also of considerable interest given the current emphasis on polarization measurements in CMB research \citep[see e.g.][]{battye10}. The characterization of polarized fluctuations in the CMB can help break some of the degeneracies between cosmological parameters, and hence can add to the accuracy of results obtained from the CMB temperature fluctuations.  An example of this is the slope of the primordial perturbation spectrum. To constrain this quantity, polarization measurements at small scales are important, and it is at small scales that foreground sources have a maximum impact. The CMB polarization percentage at high $l$ can reach $\sim$\,15-20\%; a rough measure of the effect of polarized foreground fluctuations is the comparison of this number to the typical polarization percentage of high frequency emission by extragalactic sources \citep{battye10}. Polarization of a source, and its dependence on the frequency of observation, also provides information about the physics of the emission process. For all these reasons, both SEDs and polarization measurements of foreground radio sources at high frequencies, equal to or approaching those used in current CMB studies, are of value. This paper addresses such observations.

This paper is organized as follows. In Section\,2, we present the sample. In Sections 3 and 4 we discuss, respectively, the VLA and GBT observations and data reduction. In Section 5 we deal with the issue of the variability of sources. Section 6 presents our findings, including both polarization percentages and SEDs. We discuss the implications of our work to mm-wave cosmology surveys, in particular the ACT survey, in Section 7. We summarize our conclusions in the final section. Throughout this paper we adopt the $\Lambda$CDM cosmology with $\Omega_{\rm{M}}$\,=\,0.3, $\Omega_{\Lambda}$\,=\,0.7, and $H_0$\,=\,70\,$\rm{km}\rm{s}^{-1}\rm{Mpc}^{-1}$ \citep{spergel03}.

\section{The sample \label{sec_sample}}

The ACT survey includes two equatorial fields (both avoiding the Galactic plane) as well as a field centered on -53$^{\circ}$ declination, which is too far south for the VLA. In part as a service in support of ACT and the {\sl Planck} mission, the Australia Telescope team \citep{murphy10} extended the Australia Telescope  20\,GHz survey (AT20G) of the southern sky up to the celestial equator. The AT20G survey thus covers one half of ACTÕs equatorial strip, or about 30\% of the total area of the ACT surveys. The AT20G survey limit is $\sim$\,40 mJy. While the majority of sources in the AT20G catalog \citep{murphy10} have follow-up observations at 5 and 8\,GHz, that is not true of the sources at declinations north of -15$^{\circ}$. Our observations were proposed to provide such low frequency follow-up, but with the added benefit of extending the SEDs to higher frequencies as well. We selected all sources in the ACT equatorial regions that appeared in the preliminary version of the AT20G catalog. Slightly fewer than 200 sources fit the description, of which we have obtained observations for 159.  Note that a couple of our targets which were in in the preliminary version of the AT20G catalog were dropped from the final published version \citep{murphy10}. We did not observe every source in the survey area because telescope time was limited; however, the selection of those to omit from our survey was not based on flux or other properties of the source, but only on location. Thus there should be  no bias in the observed properties of our sample. 

\begin{center}
\begin{deluxetable}{ccc ccc c}
\tablecolumns{7}
\tablewidth{7in}
\tabletypesize{\scriptsize}
\tablecaption{\label{table_obs} Observation summary}
\tablehead{\colhead{Band} & \colhead{Central freq.} & \colhead{Dates} &  \colhead{Int time} & \colhead{Typical sensitivity} & \colhead{Configuration}  & \colhead{Approx. beam size}  \\
\colhead{} & \colhead{GHz} & \colhead{} & \colhead{[s]} & \colhead{[mJy/beam]} & \colhead{} & \colhead{[arcsec]} }
\startdata
C & 4.86 &  2008,June19-21 & 60 & 0.7 & DnC & 9\,$\times$\,16\\
X & 8.46 &  " & 45 & 0.3 & DnC & 4.5\,$\times$\,7.7 \\
K & 22.46 & 2008,July25-Sept.6 & 60 & 0.9 &  D & 3.1\,$\times$\,4.4 \\
Q & 43.34 & 2008, July17-Aug.8 & 330 & 1.2 & D & 1.7\,$\times$\,2.1\\
K & 22.46 & 2009, Nov.3-4 & 60 & 0.5 & D & 3.1\,$\times$\,4.6 \\
Q & 43.34 & " & 300 & 0.6 & D & 1.6\,$\times$\,2.4 \\
W & 90.0 & 2009 December & $\gs$\,120 & 20.0 & --- & 9\ \\
    &  &   2010 January & $\gs$\,120 & & --- & 9 \\
    &  &   2010 March & $\gs$\,120 & & --- & 9  
\enddata
\end{deluxetable}
\end{center}

\section{VLA observations  and data reduction \label{sec_obs}}

Measurements in all four spectral bands were made at default VLA frequencies, centered at 43.34, 22.46, 8.46 and 4.86\,GHz. These are the Q, K, X and C bands respectively. The observations for the bulk of the sample were scattered over an interval from June 19 to September 6, 2008. For about a third of the sample, the Q-band observations were made later on November 3 and 4, 2009. We also re-observed these sources in the K-band, in order to be able to estimate the level of variability between these two observing epochs (separated by 14\,--\,15 months). 

For any given source, observations at the lowest two frequencies (C and X) were separated by no more than 3 days, and the observations for the highest two frequencies (K and Q) were made no more than 13 days apart.  The observations discussed in this paper are summarized in Table\,\ref{table_obs}.

The VLA was in a hybrid DnC configuration for the 2008 June 19-22 observations at the two lower frequencies, which were made first. Since the north-south baselines were on average longer than east-west baselines in this configuration, an elliptical synthesized beam resulted (see Table 1). For the later July-September observations at 22 and 43\,GHz, the array was in its most compact D configuration. Thus the synthesized beam size at 22\,GHz was very roughly equal in solid angle to the beam at 8\,GHz. Because of work related to the transition to the EVLA, we typically had  only 22-24 antennas available rather than 27, resulting in a 10-20\% reduction in sensitivity. The weather for these summer observations was mixed. Even though the observations were made in compact configurations of the VLA, the scatter induced by atmospheric turbulence was clearly visible in the data. Most of the effects of turbulence were mitigated by self-calibration (see below).
 
In the case of observations at the two highest frequencies, we employed fast switching between the source of interest and a nearby, bright, phase calibrator source. The integration times on source and calibrator were set to be approximately equal to or less than the atmospheric phase coherence time at 43\,GHz. For the K-band, the integration times were 60 sec. on source and 30 sec. on the calibrator. For the Q-band observations, we used  110 sec. on source and 40 sec. on a calibrator. Given the relatively low sensitivity of the Q-band receivers, we repeated that cycle three times for each source at Q-band. The integration times on source listed in column 4 of Table\,\ref{table_obs} give approximate values for each band.
	
Roughly 70 of our sources did not get observed at 43\,GHz as planned in the summer and early fall of 2008 (that is, during the first round of observations). Both weather and scheduling problems got in the way. The sources omitted were those in the right ascension range 04h08m to 13h22m. In a second application to NRAO, we received additional time for follow-up observations on November 3-4 2009 to complete our catalog. Again, time limitations prevented us from observing all 70 sources, but all 58 sources we did observe at  43\,GHz we also reobserved at 22\,GHz to allow us to assess the level of variability of these sources between the two observing epochs (September 2008 and November 2009). The integration times for these follow-up runs, conducted in the most compact D configuration, amounted to roughly 1 min. in the K band and 5 min. in the Q band for each source (see Table\,\ref{table_obs}). 

\subsection{Data reduction and calibration \label{sec_datared}}
The data reduction was done using the standard VLA package AIPS (Automatic Image Processing Software\footnote{http://www.aips.nrao.edu/cook.html}).  In each band, the raw amplitude and phase data were flagged for shadowing of one antenna by another, interference, noisy correlators, malfunctioning antennas, and so on. In general, this flagging process removed only a few percent of the raw data. Corrections for baseline errors in the array were applied. Since all of our sources are quite bright, we were able to use self-calibration to improve the phases of our UV data, with the exception of  6 faint sources at 43\,GHz. All our images, and consequently all of the flux densities, are determined from self-calibrated data, except for the 6 weakest Q band sources as mentioned above. We corrected for elevation dependent atmospheric absorption using standard VLA procedures and values in each band. In the K band, the typical value of zenith extinction was 0.07-0.11 for the summer 2008 observations and 0.07 for the November 2009 observations.
	
The flux density scale was fixed by observing one or both of two standard NRAO flux calibrators, 0137+331 (3C48) or 1331+305 (3C286), during each observing run.  For 0137+331, we employ the standard NRAO flux densities of 5.4320\,Jy, 3.1543\,Jy, 1.1188\,Jy  and 0.5297\,Jy for the C, X, K and Q bands, respectively.  Our measured flux densities, taken from our images of these sources, are 5.415$\pm$\,0.010, 3.154$\pm$\,0.013, 1.122$\pm$\,0.004 and 0.528$\pm$\,0.005\,Jy for the same four bands.  For 1331+305, the standard NRAO 
flux densities are 7.486, 5.2053, 2.5192 and 1.4555 Jy; we measured 7.474\,$\pm$\,0.030, 5.215\,$\pm$\,0.020, 2.507\,$\pm$\,0.017 and 1.470\,$\pm$\,0.014\,Jy.  A variety of secondary, phase calibrators near our equatorial regions were observed frequently (and were used for fast switching for the K and Q-band observations). 
	
Each source in each frequency band was imaged using standard AIPS procedures. In this process, we selected a pixel size for each frequency band to fully sample the synthesized beam size ($\sim$ angular resolution) at that frequency. These were 2\asec, 1\asec, 0.3\asec, 0.15\asec\ for the C, X, K and Q-band images, respectively. In all cases, we constructed 1024\,$\times$\,1024 images. In most cases, the raw images of sources were lightly cleaned of interferometric side lobes using standard NRAO procedures in AIPS. The exceptions to this procedure were the 26 clearly resolved or multiple sources; these images required heavier cleaning (in some cases, up to 2,000 clean components were removed). We experimented with different levels of cleaning, and found no significant change in the flux densities of the unresolved sources, or unresolved components of more complex sources.

\subsection{Flux Densities}

For the 133 sources that are not resolved, or barely resolved, we use the standard AIPS program IMFIT to fit a two-dimensional Gaussian to the image, and to derive the integrated flux density and its associated uncertainty (which includes the rms of the image and the fit uncertainty).  In addition,we estimate  calibration uncertainties from the scatter in values from different measurements of VLA calibrators made during our runs. These are: 1.0\% for the C and Q-band measurements, 0.8\% for the K band, and 0.7\% for the X band measurements.   The error associated with each measurement of the total flux density is the quadrature sum of the IMFIT error and the uncertainty in the overall flux scale in each frequency band. Given the high signal-to-noise ratios, the calibration uncertainty term dominates. The integrated flux densities and these total errors are the values given in Table\,\ref{table_fluxes}.
	
\begin{figure}[!ht]
\begin{center}
\includegraphics[width=8cm]{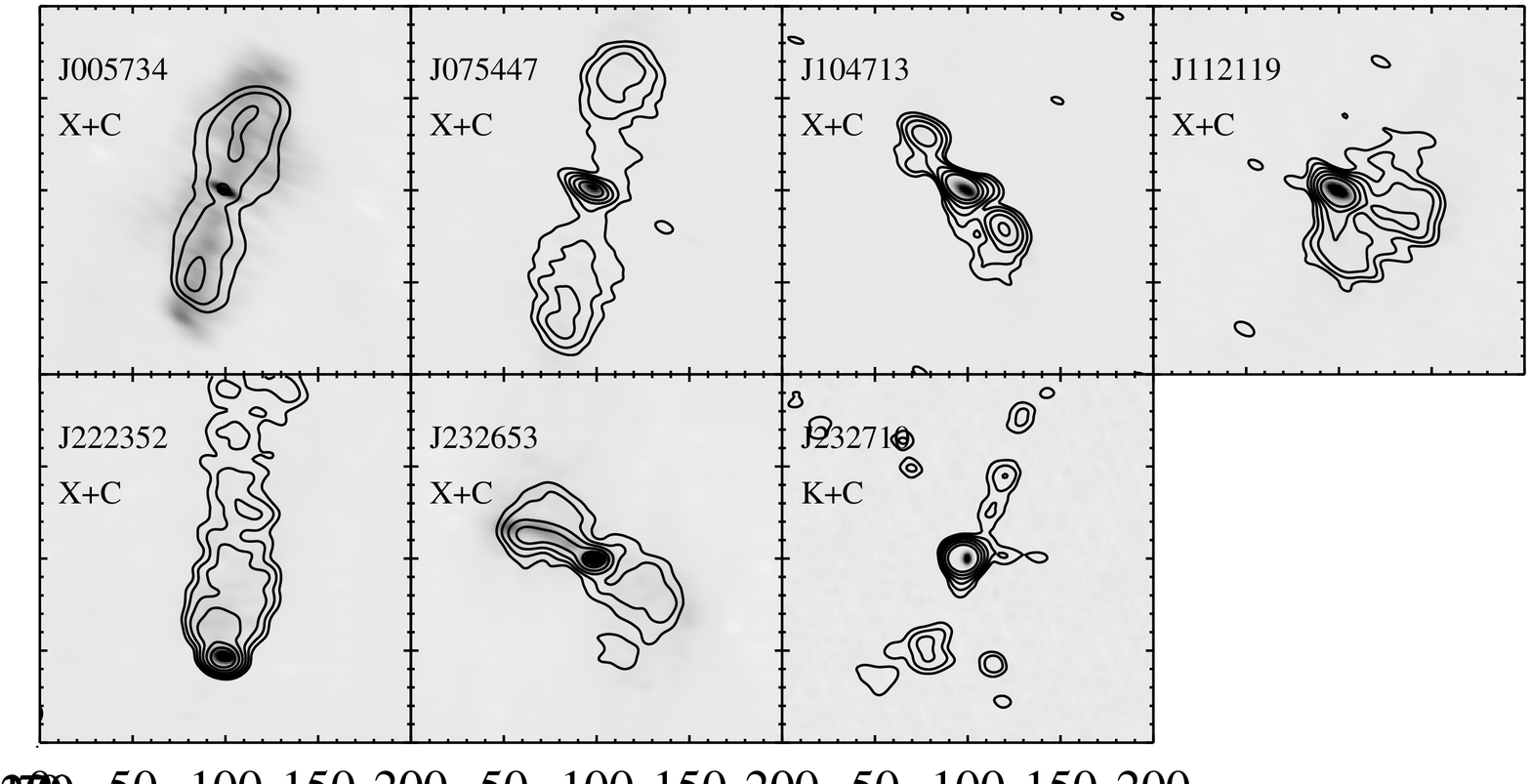} \\
\includegraphics[width=8cm]{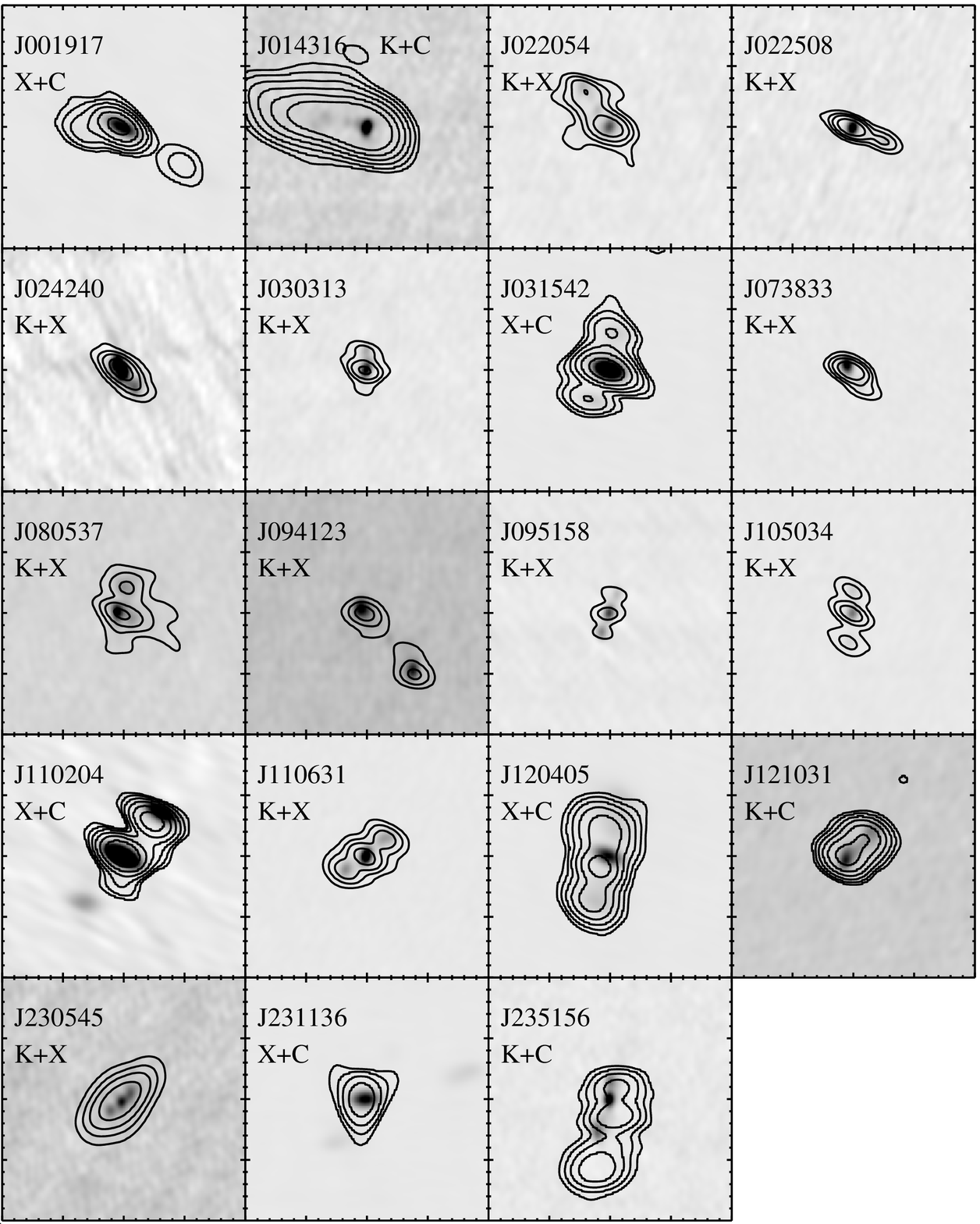}
\end{center}
\caption{ Images of the 26 sources resolved at one or more of our frequencies. The images are not all to the same scale. In each case the grey-scale image is taken from the higher resolution (higher frequency) image, and the contours from the lower resolution image of the same source. Which two frequency bands are imaged is indicated in each panel. For source J005734, for instance, the grey-scale image is the X-band (8\,GHz) image; the contours are shown from the C-band (5\,GHz) image.\label{fig_extended}}
\end{figure}

The situation for the 26 resolved sources shown in Figure\,\ref{fig_extended} is more complicated, and often frequency dependent.  For many sources, at least at some frequencies, a bright unresolved or barely resolved core is clearly visible (e.g. J075447-024734). Flux densities of these cores are denoted by ÒCÓ in column 2 of Table 2. These flux densities were obtained using IMFIT and a tight fitting box around the core. To obtain the total fluxes at each frequency, we marked a large box enclosing all visible emission and then used the AIPS program IMSTAT to compute the total flux. These total flux densities are denoted by "T" in column 2 of Table\,\ref{table_fluxes}.  Both the values and the errors of total flux for such sources should be used with care, since flux was almost certainly being resolved out.  

\subsection{Polarization Calibration and Flux Densities \label{sec_polcal}}

Our observations were typically made in short time intervals (a couple of hours). Most sources (including calibration sources) were observed near the meridian. For both reasons, the change in parallactic angle of potential polarization calibrators, such as 1331+305 and 0137+331, was small. Consequently, we use the known polarization fraction and polarization angle of these sources to make an approximate determination of the leakage and R-L phase difference from single scans, which we then apply to the rest of the data.  Normally, use of polarized sources for polarization calibration requires 2 or 3 scans (if source polarization is unknown) at different parallactic angles in order to break the degeneracies between source and instrumental polarization.  Our procedure uses a step in the calibration where the first-order antenna-based leakage terms average to zero over the array \footnote{Actually, we assume the average of $d_{iR}+d^\ast_{jL}$ is zero over the array. See the CASA Manual for details on polarization calibration.}.  This procedure was carried out using the CASA software package (http://casa.nrao.edu). In order to break these degeneracies and get a viable solution.  However, this procedure will leave a small residual of the calibrator polarization in the leakage solutions, which will lead to an overall offset to the derived source polarizations (e.g., a small bias in polarization) that should be factored in to interpretation of these results.

Because of the residual polarization biases described above, our procedure is not as accurate or robust as using observations of the same source that span a range of parallactic angles. Therefore, we carry out a number of tests. First, we compare our values of total polarized flux for a few calibrators to those tabulated in polarization calibration tables maintained at the NRAO\footnote{http://www.vla.nrao.edu/astro/calib/polar/calsources.shtml}. This is shown in the upper half of Table\,\ref{table_polcal}. For 0137+331 and 1331+305, our primary flux calibrators, where we have multiple observations we show both the average value of our observations and the average of the NRAO tabulated values closest in time to our observations.  For J042315-012033, we used the January 13th 2008 tabulated measurement for the C and X-band and the December 14th, 2009, measurement for the K and Q-band. Note however that this source is strongly variable, as we discuss in Section\,\ref{sec_strongvar}. The only other calibrator source which we observed and is monitored by the NRAO is 2136+006. However, this source was only observed in the K-band and showed messy/extended structure. However, for comparison the NRAO observation of July 28th 2008 gives its K-band polarization as 2.5\,$\pm$\,0.1, while we find 4.2\,$\pm$\,1.3 from our July 25th 2008 observations. With the exception of our Q-band measurement for 1331+305, the agreement for the two primary calibrators is acceptable. However, there is an evident trend for our polarized flux densities to run a few percent high, as expected for our polarization calibration procedure, but we do not attempt to correct our values for this trend. In Table\,\ref{table_polcal}, we also compare polarization percentages for a few of the sources with measured polarizations in the AT20G survey \citep{murphy10}.  Bearing in mind the possibility of variability, we again see reasonable agreement for these few sources. In light of the discussion in Section\,\ref{sec_compat20}, we note explicitly that we compare polarization percentages with the AT20G data, and not polarized flux. 
	
Here we discuss only the linear polarization component. While the study of circular polarization is beyond the scope of this paper, we did look at the Stokes V images for a small number of sources, including all those showing $>$10\% linear polarization in a given band.  In all cases, we found that the circular polarization percentage  is consistent with being $\ls$\,1\%.  

We use IMFIT to measure the integrated flux densities ($S_Q$, and $S_U$) and their uncertainties ($\sigma_Q$, and $\sigma_U$)  from the Stokes Q and U images.  This was done even if no polarized flux was visible (in Section\,\ref{sec_polest} we discuss the determination of the linear polarization fraction including upper limits). For extended sources, we measure the polarization of the core only. The IMFIT fitting box was determined from the Stokes-I images and then kept fixed for the much smaller signal-to-noise Stokes Q and U images. We skip sources at a given frequency where the core is not clearly defined. In a small number of cases, IMFIT failed, and we used the AIPS program IMSTAT to estimate the polarized flux within our fitting box.  We look at the spread in Stokes Q and Stokes U values in a few cases of multiple observations of the same calibration source. Combined with the results of Table\,\ref{table_polcal}, these suggest systematic uncertainties in the value of polarized flux densities of $\sim$\,8\%, 6\%, 10\% and 8\% in the C, X, K and Q-bands, respectively. As in the case of the Stokes I flux, these internal uncertainties were combined in quadrature with the IMFIT errors.  Unlike the case for Stokes I, this calibration uncertainty dominates the error budget only for the  brightest or most strongly polarized sources. 

\begin{center}
\begin{deluxetable}{c cc cc cc cc}
\tablecolumns{7}
\tablewidth{7in}
\tabletypesize{\scriptsize}
\tablecaption{\label{table_polcal} Calibrator Polarization Percentage}
\tablehead{\colhead{Source} & \multicolumn{2}{c}{C-band} & \multicolumn{2}{c}{X-band} &
\multicolumn{2}{c}{K-band} & \multicolumn{2}{c}{Q-band} \\
\colhead{} & \colhead{\% Prior} & \colhead{\% This paper} & \colhead{\% Prior} &
\colhead{\% This paper} & \colhead{\% Prior} & \colhead{\% This paper} & \colhead{\%
Prior} & \colhead{\% This paper} }

\startdata
{\bf NRAO} & & & & & & & & \\
\hline
0137+331   & 4.1 & 4.6 & 5.5 & 5.9 &  8.0  &  9.6 & 8.9 & 9.5 \\
1331+305 & 11.3 &  11.5 & 11.8 & 11.9 & 11.7 & 12.5 & 11.3 & 14.5 \\
J042315-012033\tablenotemark{b} & 1.37\,$\pm$\,0.03 & 1.1\,$\pm$\,0.1 & 1.23\,$\pm$\,0.05 & 2.6\,$\pm$\,0.2 & 1.44\,$\pm$\,0.05 & 3.0\,$\pm$\,0.3 & 2.31\,$\pm$\,0.21 & 2.0\,$\pm$\,0.2 \\
\hline
{\bf AT20G} & & & & & & & & \\
\hline
J000622-000423   & & & & & 2.6  & 2.8\,$\pm$\,0.5 & & \\
%0725-009 & & & & & 1.4  & 1.2$\pm$0.2  & & \\
0725-009\tablenotemark{a} & & & & & 1.4  & 0.6$\pm$0.1  & & \\
J074554-004418   & & & & & 3.5  & 4.0\,$\pm$\,0.4 & & \\
1150-003 & & & & & 2.6  & 1.3\,$\pm$\,0.2 & & \\
2134-018   & & & & & 7.4  & 7.6\,$\pm$\,0.8 & & \\
J220643-003103\tablenotemark{b}   & & & & & 6.5  & $<$\,4.4 & & \\
J230107-015804 & & & & & 5.2  & 7.6\,$\pm$\,0.8 & & 
\enddata
\tablenotetext{a}{The value given is for the November 2009 observations. For the September 6, 2008 observations (where the polarization calibration is highly uncertain), we find 2.7\,$\pm$\,0.5 for this source.}
\tablenotetext{b}{This is a strongly variable source.}
\end{deluxetable}
\end{center} 

\subsubsection{Problematic polarization calibration}

Two of the 12 batches of data (one K-band and one Q-band) presented problems with polarization calibration. These are addressed in turn below. 

The 17 July, 2008, Q-band observations (RA range 19-23:50h) were done in particularly bad atmospheric conditions, and hence required more aggressive flagging. As a consequence, the statistical errors in both total intensity and polarized flux are larger than for other data sets. In six cases, the data do not provide reliable polarization fluxes. The 17 July, 2008, Q-band data are also not baseline calibrated, unlike other data. We found that removing the baseline calibration for this dataset gave more reasonable polarization fractions for the calibrators although of course the total intensity values were then wrong (we use the baseline calibrated data only to get the Stokes I values).  Because of the poor quality of this batch of data, the Q-band polarization fractions for these sources are typically upper limits, with only the highest polarization fractions being detectable.

Using the polarization calibration procedure described above, our first attempt at polarization calibration of the 6 Sept., 2008, K-band data failed. While the total intensity values were reasonable, the polarization percentages for calibrators with known polarization were in error by a factor $>$\,2. These data include sources in the RA range 07\,--\,13hr, which set of sources were re-observed in November 2009. As the November 2009 data were taken in significantly better weather (and higher instrumental sensitivity), and moreover were near-simultaneous with the Q-band data for the same sources, we mostly use the later observations in our analysis.  We tried two different solutions to this. First, we used the raw Stokes Q and Stokes U images to obtain polarized flux densities.  With no allowance for instrumental polarization, the polarization angles derived in this way make no sense, but, given both small polarization fractions of our sources and small instrumental polarization of the VLA antennae, the quadrature sum of the Q and U images should give a reasonable representation of the total polarized flux.  We test this by comparing our September 2008 K-band polarization fractions with the November 2009 K-band polarization fractions for the same sources. We find a median ratio of 1.14 with significant scatter. On the other hand, the bright calibrators show a median ratio of  0.83. Excluding upper limits, $\sim$\,12\% of these sources have polarizations that are a factor of 2 higher in 2008 than in  2009.  All of these, also show strong variability in their Stokes I flux densities. The apparent one sidedness of the variability can be understood when we consider the significantly different sensitivities of the two datasets. A source whose polarization fraction grows from 2008 to 2009 would almost certainly appear as an upper limit in the 2008 data. The sources, whose polarization fraction changes significantly between the two epochs are: J073025-024125, J073245-022858, J091643-025910,J102834-023659, and J111439-024731. However, due to the problematic polarization calibration it is hard to judge whether these are scientifically interesting or merely reflect the poor quality of the September 6th, K-band dataset. 

We were later able to use the new NRAO data reduction package, CASA, to repeat the polarization calibration of this data set, but without the standard baseline calibration (as in the case of the 17 July 2008 Q-band data).  The comparison between the values obtained above and these new polarization fraction values showed a ratio between then two of $\sim$\,1 after three outliers with $>$\,50\% variability where excluded. These were all revealed to be extended sources, which we did not account for in the CASA reduction.  Apart from such outliers, the scatter is largely consistent with the quoted errors. Because of this good agreement and the more careful treatment of extended sources in our earlier reduction, we decided to stick with the values obtained from the raw Stokes Q and U images but note that performing polarization calibration on the data without baseline calibration yielded consistent answers. 

\subsection{Estimating the degree of linear polarization \label{sec_polest}}

Here we largely follow the prescription of \citet{ss85}, with the exception that we do not equate $\sigma_q$ and $\sigma_u$ \citep[see also][for an extended discussion]{topasna}. We begin by defining the normalized Stokes parameters: $q$\,=$S_Q$/$S_I$ and $u$\,=$S_U$/$S_I$, where $S_I$ is the Stokes I (i.e. total) flux density. The uncertainties on $q$ and $u$, $\sigma_q$ and $\sigma_u$ respectively, can be computed from the measured flux density errors by using the following expressions:
 
 \begin{equation}
 \small
 \sigma_q=q\sqrt{\Big(\frac{\sigma_Q}{S_Q}\Big)^2+\Big(\frac{\sigma_I}{S_I}\Big)^2},\space    \sigma_u=u\sqrt{\Big(\frac{\sigma_U}{S_U}\Big)^2+\Big(\frac{\sigma_I}{S_I}\Big)^2}
 \end{equation}

The measured degree of linear polarization, $p$, is then given by:
\begin{equation}
 p=(q^2+u^2)^{1/2}
 \label{eq_pmeas}
 \end{equation}

Assuming the errors on $q$ and $u$ are independent, the uncertainty on $p$ is given by: 

\begin{equation}
\sigma_p=\frac{1}{p}\sqrt{q^2\sigma_q^2+u^2\sigma_u^2}
\label{eq_sigp2}
\end{equation}
 
It is well known that this expression (Equation\,\ref{eq_pmeas}) is inherently biased as $p$\,$>$\,0 even if there is no true linear polarization, due to the presence of noise on $S_Q$ and $S_U$. Using $\sigma_p$ defined in Equation\,\ref{eq_sigp2}, the bias correction takes the form:
 
\begin{equation}
p_{corr}=\sqrt{p^2 - (K\sigma_p)^2},   
\label{eq_pcorr}
\end{equation}

where $K$ is the bias correction factor. Note that, $p_{corr}$ is only defined where $p$\,$>$\,$K\sigma_p$. \citet{ss85} compare several different approaches to estimate $K$.  For cases where $p$/$\sigma_p$\,$>$\,0.7), they advocate the Wardle $\&$ Kronberg estimator with  $K$\,=\,1.  For lower signal-to-noise ratios, the recommended value is $K$\,=\,1.41 based on the maximum likelihood estimator of the true value of $p$.  In practice, all our quoted polarization percentages are based on $K$\,=\,1.0. In Table\,\ref{table_fluxes} we quote the linear polarization percentages, i.e. 100($p_{corr}$\,$\pm$\,$\sigma_p$). Note also that our quoted $\sigma_p$ likely represents a confidence level close to but somewhat less than 68\%.  This difference is not crucial for our purposes and hence for simplicity it is ignored.  In order to assess whether or not a given source can be considered to have detectable polarization, we also look at the associated confidence intervals. \citet{ss85} provide a procedure for estimating these, but also quote conservative estimates based on the assumption that $p^2$ is $\chi^2$-distributed.  This assumption leads to a 68\%\ confidence level of [max(0,p-1.49$\sigma_p$),p+1.49$\sigma_p$]  and a 95\%\ confidence level of [max(0,p-2.45$\sigma_p$),p+2.45$\sigma_p$]. Note that \citet{ss85} argue that these are somewhat larger than the true confidence intervals (by roughly 20-40\%), and hence are a fairly conservative choice. We take any source with a 95\%\ lower confidence level of zero to be a non-detection. In such cases, Table\,\ref{table_fluxes} shows the 95\%\ upper limit (i.e. $p$+2.45\,$\sigma_p$). Sources detected in only one of Stokes Q or Stokes U still have an overall polarization percentage detection, while the upper limits on $p$ are typically associated with sources without a significant detection in either Q or U. 

\section{GBT 90GHz observations and data reduction}

For 25 of our sources, we also obtained flux measurements at 90~GHz using the MUSTANG receiver \citep{dicker08} on the 100~m diameter Green Bank Telescope (GBT).
MUSTANG is a 64 element array of TES bolometers with a bandpass of 81--99\,GHz. On the GBT it has a beam FWHM of 9$''$ and detectors are spaced by 0.7$f \lambda$\ to give a 
field of view of 45$''$ square.

Our observations were carried out in 3 night time sessions. Observing at night is necessary as during the day solar heating
distorts the shape of the GBT's primary mirror causing significant gain
changes on timescales faster than can be calibrated out (15 minutes or
less). At the start of each session, Out-Of-Focus holography (OOF) 
was used both to find the focus and to
measure the pointing offsets and residual large scale thermal
distortions of the mirror \citep{nikolic07}.  The thermal distortions were then
compensated for using the GBT's active surface.  This procedure increases
the forward gain of the telescope and hence the signal to noise by 10
to 30\%.

Each source was observed at least twice using 60 second ``daisy
 petal'' scans which modulate the source position on the array faster than drifts in
any detector. Most sources were easily detected in a single scan but
observing twice provided a consistency check and allowed better
 determination of our errors.  Quick look software was
 used to check for clear source detections and more scans taken if
 needed.  Observations of an absolute calibrator were carried out every
20 to 30 minutes.  In the first session our calibrator was Mars,  and in the second,
 Saturn, supplemented by the secondary calibrator 1058+0133 which was
 nearer to our sources. In the final session, we used Ceres and
 Neptune.  After observing each calibrator a short data file was taken
 while pulsing MUSTANG's internal blackbody calibrator; ``CAL''.  The
 amplitude of this response is used to calibrate the optical response
 of each bolometer.

Data analysis was carried out by making maps of each source and individual calibrator observations.  First the response of each detector was normalized by its response to ``CAL'' and dead or noisy detectors were flagged. This  timestream data is dominated by emission from the atmosphere, the telescope, and MUSTANG's internal optics which are all highly common mode between pixels.  Thus by fitting a common mode signal to all off-source pixels the majority of atmospheric and instrumental signal is removed. In addition we explicitly fit and remove a 1.4Hz sine wave (caused by internal temperature fluctuations in the receiver) and subtract a baseline from the timestream of each detector which has time scales longer than the modulation of the source (typically 0.5Hz). The data are then binned in elevation/cross-elevation.

 To convert from our arbitrary units of ``CALs'' to Janskys, the following steps were carried out:
\begin{itemize}
\item Gaussian fits to each map were used to obtain the source amplitudes
(in ``CALs'') and also to find the solid angle of the GBT's beam as a
function of time.
\item The opacity as a function of time was calculated using the
observing tools at Green 
Bank\footnote{For details see www.gb.nrao.edu/$\sim$rmaddale/Weather/}.
  These tools make use of
pressure, temperature, and humidity profiles of the atmosphere measured
by the National Weather Service.  
\item The expected brightness of the planets (in Jy) were calculated
assuming the blackbody temperatures in \citet{weiland10}, and
\citet{wright07}, angular diameters from the Horizons
ephemeris\footnote{available at http://ssd.jpl.nasa.gov/?horizons},the
measured beam sizes and the measured 18~GHz bandpass of MUSTANG.
\item For each observation of a calibrator, the conversion of ``CALs''
to Janskys was calculated and these values extrapolated to the
observation of each source.
\end{itemize}

The 90GHz flux densities with their associated 1\,$\sigma$ uncertainties of our 25 GBT targets are presented in Table\,\ref{table_gbt}. 

\section{Variability}

We had two means of detecting variability in the flux density of our sources:  comparison of our 22\,GHz measurements with the AT20G survey from which we drew our sample, and comparison of the 22\,GHz fluxes of the 58 sources observed in the summer of 2008 and then reobserved in early November, 2009. We refer to the latter as the internal comparison. 

\subsection{Variability detected internally in our VLA measurements}

\begin{figure}[!h]
\begin{center}
\includegraphics[width=8cm]{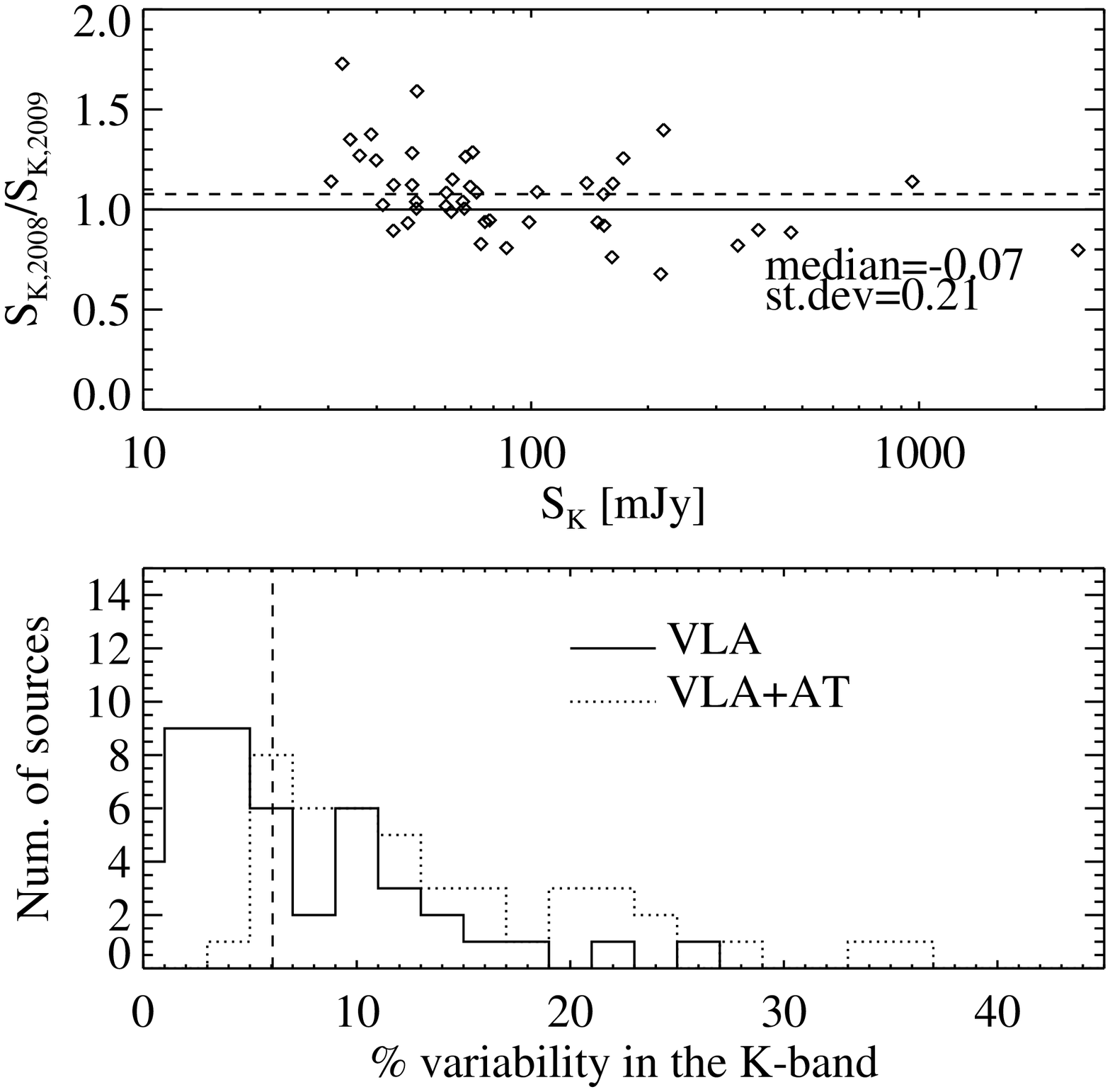} \\
\includegraphics[width=8cm]{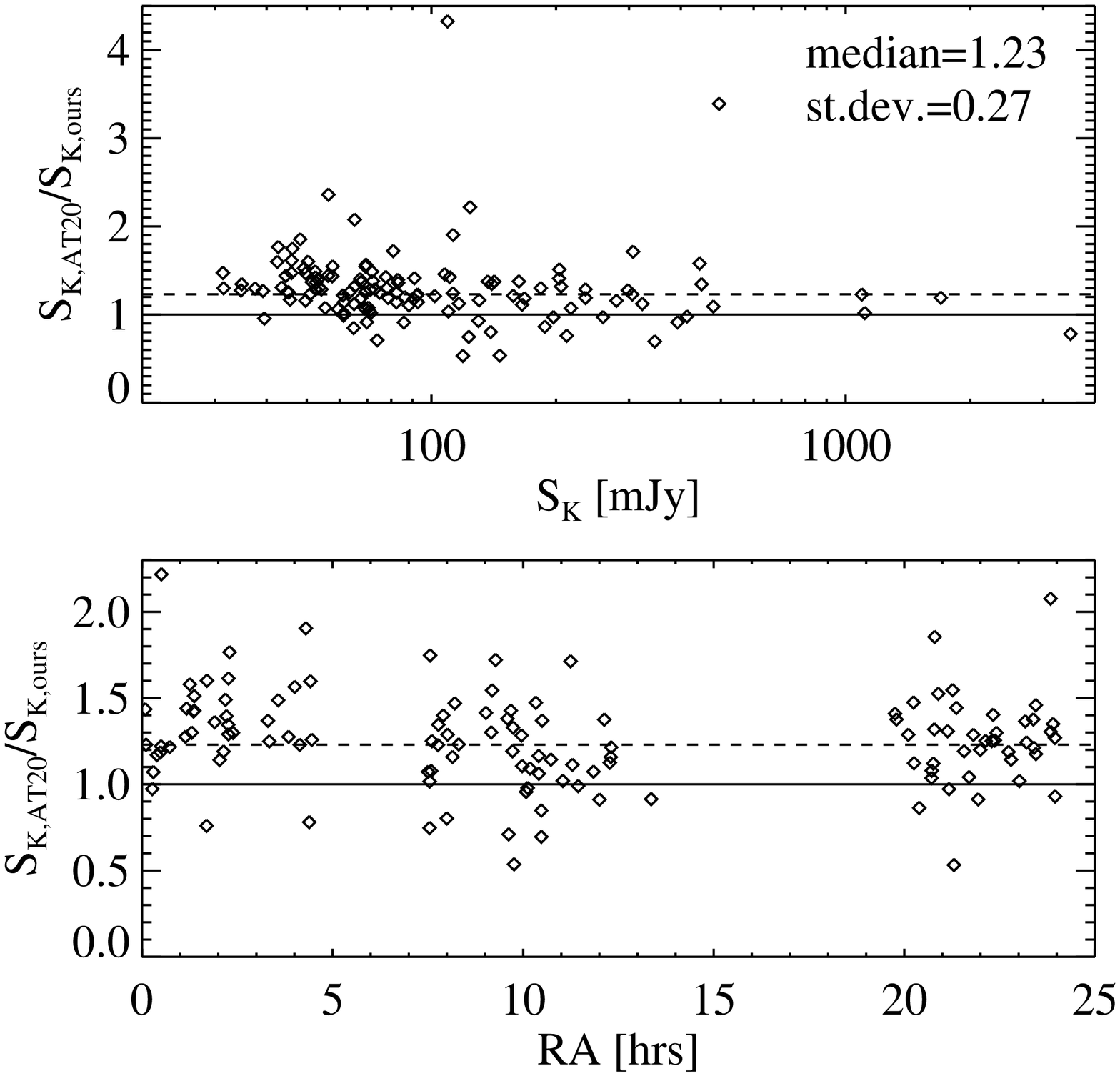}
\end{center}
\caption{The top panel shows the ratio of the two epochs of VLA K-band flux estimates as a function of K-band flux (from November 2009).  The second panel shows the variability index based on the VLA data alone (solid histogram), and based on the VLA+AT data (dotted histogram). The dashed line shows the median variability based on the VLA data alone.  The third panel shows the ratio of the AT20G fluxes and our K-band fluxes of unresolved sources, after correction for difference in effective frequency. The bottom panel shows what appears to be a dependence on RA, suggesting some systematic effects might also be at play.  For clarity, this figure excludes J220643+003103 which has a ratio of 4.4 (see top panel). The dashed lines in both panels show the median ratio.  \label{fig_var}}
\end{figure}

Figure\,\ref{fig_var}{\it top} shows a comparison of the K-band fluxes for sources that were observed both in 2008 and again in November 2009. The 2009 measurements were made in better weather, so any decoherence caused by atmospheric turbulence should affect the 2008 values more. Instead, we find marginal evidence that the November 2009 flux densities were on average slightly lower once we had flagged and removed some manifestly variable sources. We find a median offset of 7\% with a standard deviation of 21\%.  Of the 45 point-like sources we observed in the K-band in both epochs, 2 have varied by more than 50\%.  

We estimate the variability index of each source following the prescription in \citet{sadler06}.  We include only our own VLA K-band flux density measurements. The second panel of Figure\,\ref{fig_var} shows a histogram of the K-band variability indices for our sources, where available. We get a median value of 6.1\% (dashed line in Figure\,\ref{fig_var}), which is slightly lower than but still compares well with the median value of 6.9\% found in a follow-up in the K-band of 170 of the AT20 sources \citep{sadler06,murphy10}.   Figure\,\ref{fig_var} also shows the variability index histogram obtained when we include the AT20G K-band flux densities as well, where we have corrected for the effective frequency difference and have divided the AT20G fluxes by the median offset we discover in Section\,\ref{sec_compat20}. 

\subsection{Comparison with AT20G \label{sec_compat20}}

The majority of the AT20G flux densities at 19.904\,GHz reference frequency \citep{murphy10} were determined in October 2007, that is roughly 10 months before our K-band observations began. The AT20G flux densities for the point sources in our sample are plotted against our measurements in Figure\,\ref{fig_var}. Some sources are evidently strongly variable. The most salient example is J220643 - 003103, for which we measure a flux $<$1/4 of that found earlier by the Australia Telescope. 

 In addition to evidence for variability, we found a systematic difference in flux densities, in the sense that our flux densities on average tended to be lower than those measured earlier at the Australia Telescope.  Some of the difference is due to the different center frequencies used in the two surveys, 22.46\,GHz for our VLA work and 19.90\,GHz for the AT20G survey. To make a first order correction for this effect, we used the spectral indices we determined between 8.46 and 22.46\,GHz to interpolate our flux densities to the AT20G frequency. It is these interpolated values that we plot in Figure\,\ref{fig_var}. We were also concerned that our measurements at the VLA, made with antenna spacings as large as $\sim$\,1\,km, could be missing some flux as a consequence of resolution effects. We thus excluded from Figure\,\ref{fig_var} and the remaining discussion the 26 sources that were evidently resolved or showed structure in any of our observations. Whenever available, the Nov, 2009 K-band fluxes are used, as generally those data were of better quality. 
 
 After all these steps and after we exclude the strongest outliers, the median ratio is $S_{AT,19.9GHz}$/$S_{VLA,19.9GHz}$ =  1.23 with a standard deviation of 0.27.  The mean ratio is 1.19\,$\pm$\,0.02. We note that such an offset was first reported for a small number of VLA calibrators in \citet{murphy10}  where they find a median AT20G/VLA ratio of 1.08\,$\pm$\,0.10. This ratio is smaller than we find here; however, their comparison is done for calibrator (hence higher flux density) sources. Therefore the first test we perform is to look at the median ratio of bright sources in our sample.  In order to have reasonable numbers we select $>$\,200\,mJy as the criterion for a `bright' source.  With this flux density cutoff, we find a median ratio of 1.16 and an rms of 0.28.  Figure\,\ref{fig_var} also shows the ratio with respect to RA. The dashed line shows the median ratio for the full sample (i.e. 1.23).  We note that sources in the RA range 0-5h show particularly high ratios that are nearly always above the median for the sample (this may extend up to RA\,$\sim$\,9h). If we look at RA$>$5h sources only, we find a median ratio that drops significantly to 1.14\,$\pm$\,0.04 where the error is rms/$\sqrt{N}$.  We can make the further restriction of looking at bright sources at RA$>$5h. There are only 9 such sources, but the median ratio  has now dropped to 1.07\,$\pm$\,0.11.  We note that our K-band data for RA\,$\sim$\,20\,--\,5h sources were all taken on the same day (July 25th, 2008, see Table\,\ref{table_obs}), and were subjected to the same setup, atmospheric conditions, and subsequent data reduction. Therefore, it is unlikely that there is a systematic difference between the flux scale we measure for the RA\,=\,0\,--\,5 and RA\,=\,20\,--\,24h sources.  We note that the AT20G survey is at its noisiest in the equatorial strip where we overlap, and therefore it is not clear that this flux density offset necessarily would translate to an overall offset between the flux density scales of ATCA and the VLA.  Our K-band fluxes are fully consistent with the absolute flux scale of the VLA at that frequency as can be seen by the flux densities for both the primary and secondary calibrators we observe (see Section\,\ref{sec_datared}). 
 
 \citet{murphy10} quote a comparison with the WMAP fluxes yielding a mean value of $\langle(S_{AT20}/S_{WMAP})\rangle$\,=\,1.01\,$\pm$\,0.03. This is based on very bright sources ($S_{AT20G}$\,$\gs$\,500\,mJy). We do not have enough sources in this flux range for good statistics.  We can, however, compare some individual bright sources in our sample with the 7 year, five band WMAP point source catalog \citep{gold10}. The calibrator source 0725-009 has $S_K$\,=\,1.99Jy which is sufficiently different from the  $S_{K,WMAP}$\,=\,1.0Jy to suggest variability is at play.  On the other hand, 1150-003 has $S_{K}$\,=\,0.803Jy which is very close to the  $S_{K,WMAP}$\,=\,0.8Jy, while  2134-018 has $S_{K}$\,=\,1.73Jy which is about 11\% lower (after accounting for the difference in central frequency and the spectral index) than the $S_{K,WMAP}$\,=\,2.0Jy .  Other bright sources such as J042315-012033 and J074554-004418 are not useful here as they show too strong a variability, while J230107-0158804 was not found in the WMAP catalog. Finally, we can look at the two primary calibrators we use, 0137+331 ($S_{K,WMAP}$\,=\,0.9Jy)  and 1331+305 ($S_{K,WMAP}$\,=\,2.3Jy). In both cases, our K-band flux densities are actually larger, in the case of 0137+331 by $\sim$\,20\%.  From so few sources it is difficult to draw conclusions, but these values suggest that, for bright sources, there is no evidence  that our VLA K-band flux densities are systematically underestimated with respect to WMAP. 
 
 Since variability plays a large role here, the only secure way to address this is through simultaneous ATCA-VLA observations of the same set of sources, including a range in flux densities. Work is underway to resolve this issue. For the purposes of this paper, we will use our K-band flux densities as measured.  However, we keep in mind the possibility that these are underestimated. Therefore, for example, when we discuss spectral indices involving the K-band, we add/subtract 0.09 or $\log(1.23)$, where 1.23 is the median AT/VLA ratio, in order to test the effect of this discrepancy on our conclusoins.

\subsection{Strongly variable sources \label{sec_var}}

The comparison of the VLA K-band observations from 2008 and 2009 revealed two sources that vary by $>$50\%. These are: J042315-012033 and J080512-011114. The comparison between the AT20G and VLA K-band fluxes revealed 3 sources where the AT20G/VLA ratio is $>$2 (after accounting for the $\sim$\,20\% systematic offset discussed above). These are:  J080512-011114, J220643-003103, and J235013-020614.  The first of these (J08512-011114) overlaps with one of the strongly variable sources revealed in the K-band data alone. For J042315-012033, the AT20G K-band flux density is in between the two values in our data.  We address in more detail the spectral and polarization properties of these four sources in Section\,\ref{sec_strongvar}.

\section{Results}

\subsection{Morphology}

A total of 26 of our sources show extended morphologies at one or more frequencies (16\% of the sample). Figure\,\ref{fig_extended} shows the images of all extended sources, where the greyscale shows the higher-resolution image and the contours the lower-resolution image, as indicated in the figure caption.  Of these extended sources, only $\sim$\,12 appear to be classic double-lobe radio galaxies, 1 shows a single strong jet, and 1 is a one-sided double-lobed galaxy; the rest are extended but without clearly defined jets, i.e. `blob'-like.  Overall, roughly 2/3 of these resolved sources show clear cores.  Combined with the low incidence of extended sources, this suggests that on the whole our sources are heavily core-dominated. In Figure\,\ref{fig_extended}, we can also see that the sources tend to get more extended with decreasing frequency (the contours are typically C or X-band, while the greyscale shows the X or K-band). We also note that for J094123-014251, the nominal AT20G position, on which we centered the VLA beam, actually corresponds to the northern lobe of a complex source. Our K-band image suggests a classic radio galaxy with two roughly symmetric prominent lobes and a faint core in between.  In Table\,\ref{table_fluxes}, we present the flux densities for both cores (when such can be distinguished) and for the total sources as measured using the AIPS routine IMSTAT with a rectangular aperture enclosing all visible extended emission. 

Anecdotal evidence would suggest that 16\% of the sample showing extended structures is significantly smaller than what one normally expects for lower frequency radio surveys. To test this assumption, we looked at the 1.4\,GHz image of the {\sl Spitzer} First Look Survey (FLS) field by \citet{condon2003}. We chose these data because they were obtained in the B configuration of the VLA resulting in a 5\arcsec\ FWHM beam, which is comparable to our observations, unlike the much worse resolution of all-sky surveys.  Next, we look at the 16 sources in the FLS sample that have $S_{1.4GHz}$\,$>$20\,mJy to be roughly comparable with our sources. We examined the image of each of these sources for extended or multi component morphology, but ignoring fainter extended structures. We find that 9/16 (or 56\%) show extended/complex morphology. This of course is not directly comparable to our results due to the differences in sensitivity and resolution. However, it does confirm that 20\,GHz-selected sources are typically more compact than 1.4\,GHz-selected sources, as expected.

\begin{figure}[h!]
\begin{center}
\includegraphics[width=10cm]{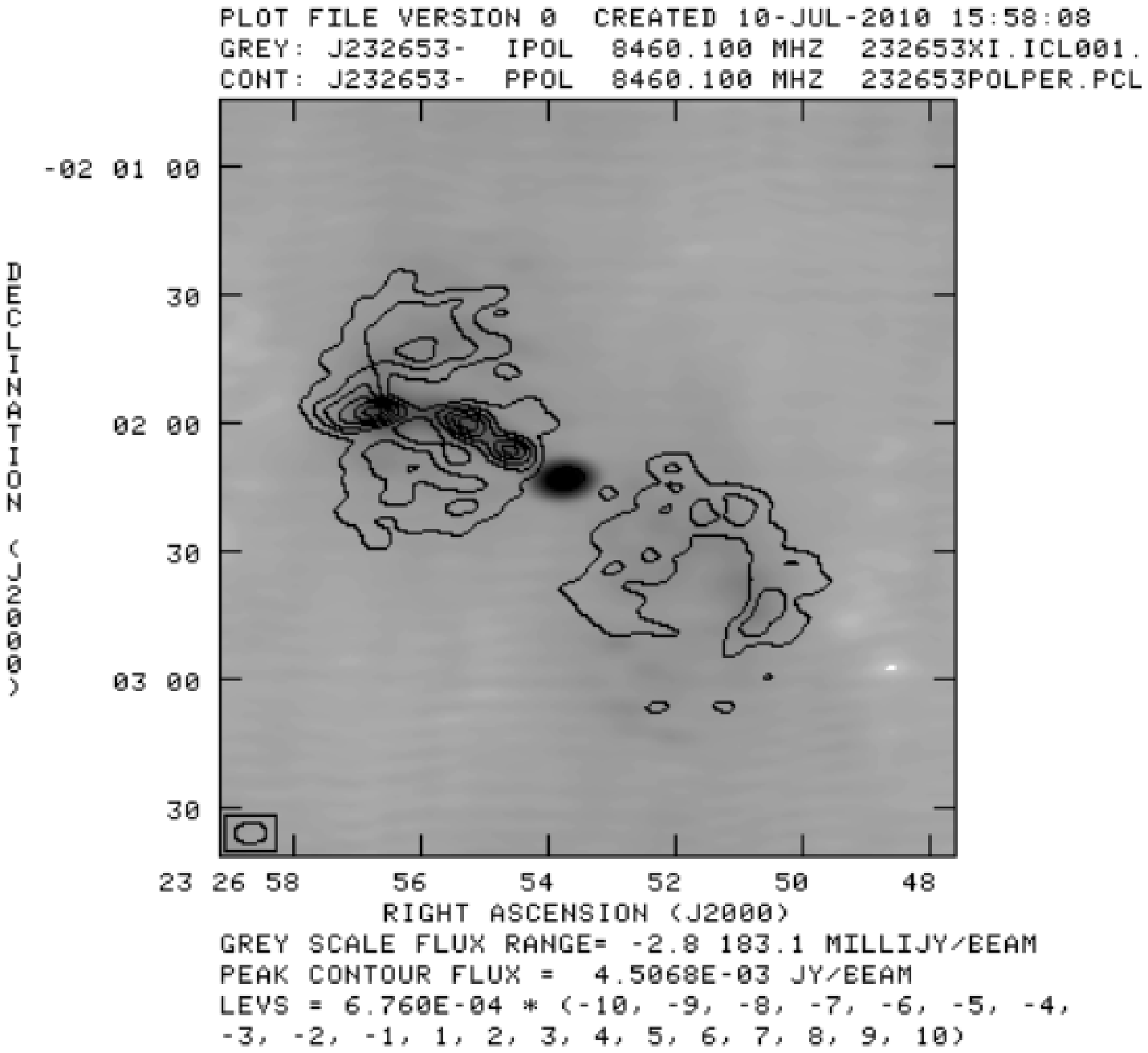} \\
\includegraphics[width=10cm]{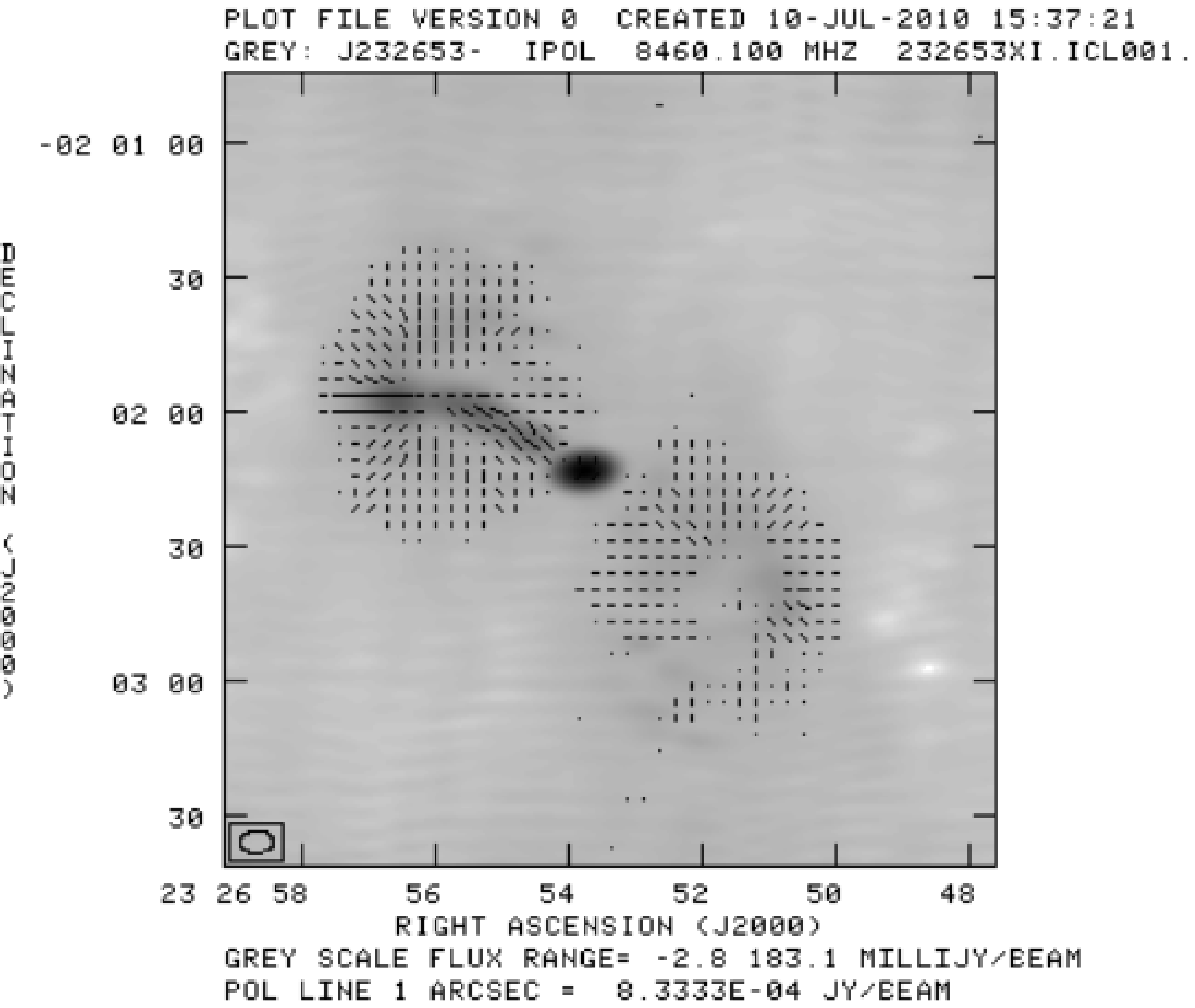}
\end{center}
\caption{The X-band polarization percentage map of J232653-020213, showing a prominent highly polarized (up to $\sim$\,15\%) jet that also has a kink in it.  The top panel shows the total linear polarization overlaid on the Stokes I image in greyscale. The bottom panel shows the polarization angle map. Note that the core in this source is very weakly polarized. In both panels, the X-band beam is shown in the bottom-left of the image. \label{fig_polmap}}
\end{figure}

Lastly, we briefly address polarization in extended sources. The core is typically the dominant source of polarized emission in the extended sources; hence the polarization percentages given in Table\,\ref{table_fluxes} are either those of unresolved sources or the cores in the case of extended sources. However, in one case, we find that the jets and lobes show rather spectacularly in polarized emission (see Figure\,\ref{fig_polmap}). A more detailed analysis of this extended polarized emission, however, is beyond the scope of this paper. These highly polarized lobes in general have steep spectral indices and hence make negligible contributions at the frequencies used in most CMB and SZ experiments. 

\begin{figure}[!ht]
\begin{center}
\plottwo{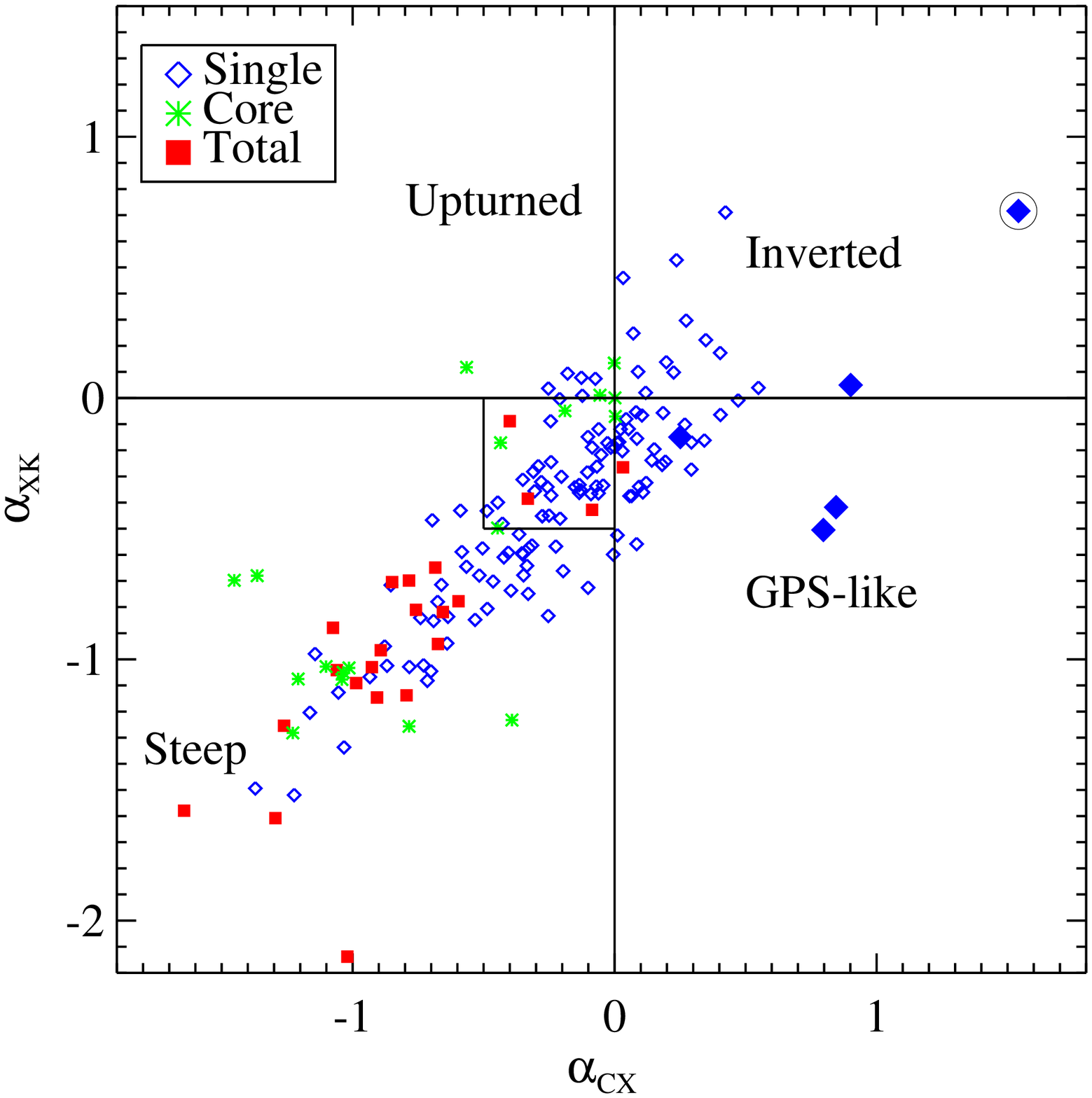}{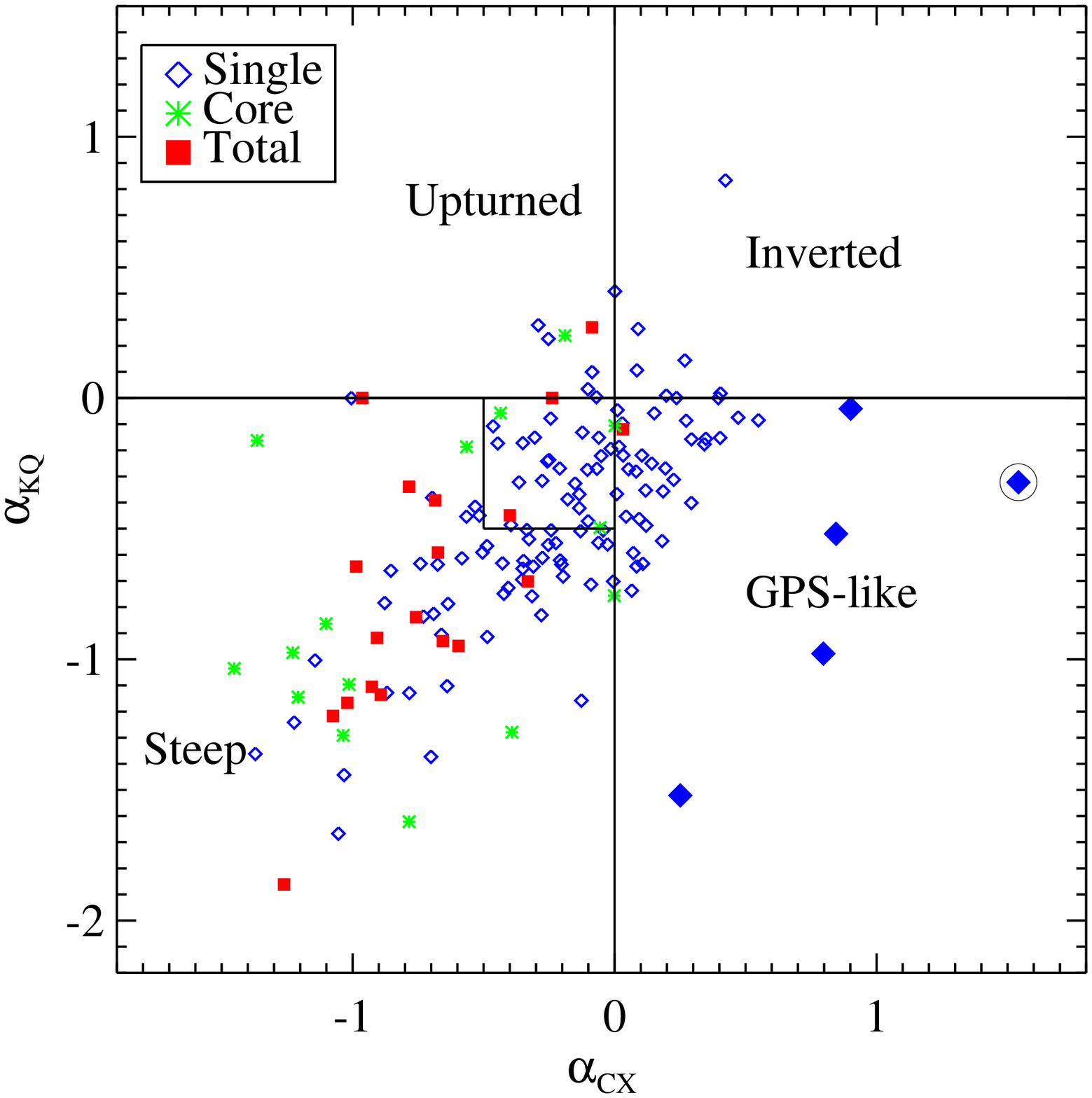}
\end{center}
\caption{Radio color-color diagram for our sources based on the X-K spectral index ({\it left}) and the K-Q spectral index ({\it right}). The upper panel has artificially lower scatter due to the use of the X-band flux in both colors. We have labeled the locations of different classes of objects with the central square representing flat-spectrum objects as usually defined ($\alpha$\,$>$\,-0.5). The boundaries are largely historic and serve to facilitate classification, rather than having a rigorous physical meaning. Most interesting are the GPS-like or "peaked spectrum" sources, which represent 23-27\% of the sample depending on the colors used.  The circled source is our only Ultra Inverted Spectrum source \citep[following the definition in][]{murphy10}. \label{fig_slopes}}
\end{figure}

\subsection{Spectral indices and color-color diagnostics \label{sec_alpha}}

We determine spectral indices (using the convention $S_{\nu}$\,$\propto$\,$\nu^{\alpha}$) in various frequency intervals. In Figure\,\ref{fig_slopes} we plot two `color-color' diagnostic diagrams\footnote{We compute the spectral indices based on pairs of observations as close in time as available (for example $\alpha_{\rm{KX}}$ is based entirely on 2008 K-band data, whereas $\alpha_{\rm{KQ}}$ uses the November 2009 K-band data for the sources whose Q-band observations are from November 2009.}. Such plots allow us to distinguish different classes of sources based on the shape of their SEDs, including in particular the Gigahertz Peak Spectrum sources \citep[GPS; for a review see][]{odea98}. Our $\alpha_{\rm{CX}}$ vs. $\alpha_{\rm{XK}}$ plot (Figure\,\ref{fig_slopes}{\it top}) looks very similar to the equivalent plot in \citet{murphy10}, when we consider our much smaller sample ($\sim$160 vs. $\sim$\,6000).  Comparing the distribution of spectral properties in our sources with the whole AT20G survey we find:  62\% of our sources have steep or flat spectra\footnote{Flat spectrum sources have both spectral indices in the range -0.5 to 0.0.}, with both spectral indices $<$\,0 (vs. 57 percent in AT20G); 22.5 percent fall in the lower right-hand (peaked-spectrum sources) quadrant (vs. 21 percent in AT20G); 11\% have inverted spectra as compared to 14\% in AT20G; and 4 percent (vs. 8\%) show a spectral upturn above $\sim$8\,GHz. The slightly smaller numbers of upturned and inverted spectrum and higher numbers of steep/flat and peaked spectrum sources can be accounted for in part by our slightly higher K-band frequency (22.46 vs. 19.9\,GHz), and in part by the $\sim$\,20\% discrepancy between our flux densities and those tabulated in the AT20G catalog. \citet{murphy10} define 45 sources (1.2\% of the AT20G sample with 5 and 8GHz data) as Ultra Inverted Spectrum (UIS) based on $\alpha_{\rm{XK}}$\,$>$\,0.70. Therefore, we should have 2\,$\pm$\,1 UIS sources in our sample. We find 1 such object (J033427-015358), consistent with expectations.  
 
 The availability of the Q-band (43GHz) data; however, allows us to go beyond what was done for the AT20G survey. Figure\,\ref{fig_slopes}{\it bottom} shows the $\alpha_{\rm{CX}}$ vs. $\alpha_{\rm{KQ}}$ (our highest frequency spectral index) plot. We can immediately draw some conclusions: 1) our one UIS source now appears to be a peaked spectrum source,  2) the number of inverted spectrum sources is now lower (5\%), and 3) the fraction of peaked spectrum sources is now slightly larger (rising from 23\% to 26\%). 
 
 Both panels show some correlation between the low frequency spectral index and higher frequency spectral indices, especially for the steep spectrum sources. However, there is also considerable scatter, consistent with earlier AT20G results \citep{murphy10}. This scatter and changing distributions in spectral indices show why extrapolations of source counts from low frequency surveys can produce misleading predictions of the numbers of sources at higher frequency, and hence of the contamination produced by such sources at the frequencies ($>$\,30\,GHz) typically used in searches for CMB anisotropies. We return to this point in the Discussion. 

Finally, as expected, we have a connection between morphology and spectral indices. In particular, the total flux densities of resolved sources generally have steep spectra. The same is true for most cores of resolved sources. In contrast, the single (i.e. unresolved) sources have somewhat flatter spectra and comprise almost exclusively the inverted spectrum and peaked spectrum sources.  
	
\subsection{Peaked Spectrum Sources \label{GPS}}

Figure\,\ref{fig_slopes} shows that a significant fraction of our sample have GPS-like (i.e. "peaked") spectra.  In total we have 36 such sources, based on the $\alpha_{\rm{KQ}}$ vs $\alpha_{\rm{CX}}$ diagnostic plot (26\%).  If we account for the fact that our $K$-band fluxes may be $\sim$\,20\% too low, this would still leave us with 30 GPS-like sources (22.5\%). This is comparable to expectations given the results in \citet{sadler06}. Figure\,\ref{fig_slopes} however suggests that the bulk of these are in the main "cloud" of sources, rather near the $\alpha$\,=\,0 lines. Due both to flux density errors (for the sources near $\alpha$\,$\sim$\,0), and variability, it is unclear whether some of the sources displaying GPS-like spectra are bona fide GPS \cite[as discussed in][]{torniainen07}.  Therefore, we generally use terms such as "GPS-like" or "peaked spectrum" throughout this paper. 
	
However, there are a handful of sources that stand apart by being much more likely GPS sources. These are: J021121-014515, J033427-015358, J101956-002411, J211022-012658, and  J230107-015804 (marked by the filled symbols in Figure\,\ref{fig_slopes}). Of these, only J101956-002411 has a known redshift ($z$\,=\,1.13), therefore it is difficult to say where is the spectral peak in these sources in the rest-frame. However, they all have an observed  spectral peak somewhere in the frequency interval 8-22\,GHz, and are thus extreme examples of GPS sources, or High Frequency Peakers in the nomenclature of \citet{dallacasa00}.  

\begin{figure*}[h!]
\begin{center}
\plotone{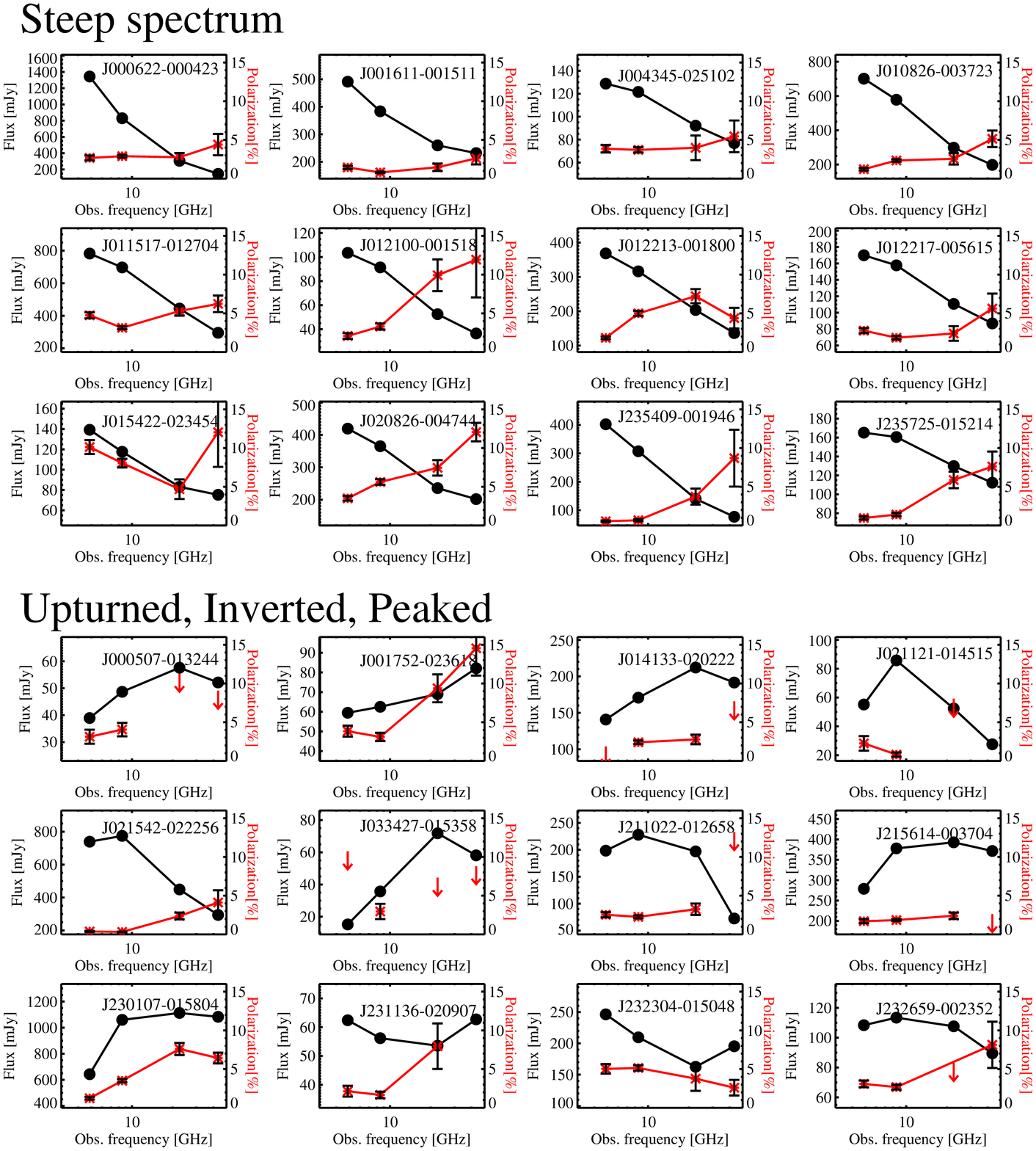}
\end{center}
\caption{SEDs and polarized fractions for sources observed at a single epoch. The large filled dots and black lines indicate the flux densities. The red lines and errorbars are for the polarization fractions. Dashed lines mean significantly different epoch of observations, whereas solid lines connect the different frequencies for which data were obtained over a couple of weeks or less. \label{fig_polseds}}
\end{figure*}

\begin{figure*}[h!]
\begin{center}
\plotone{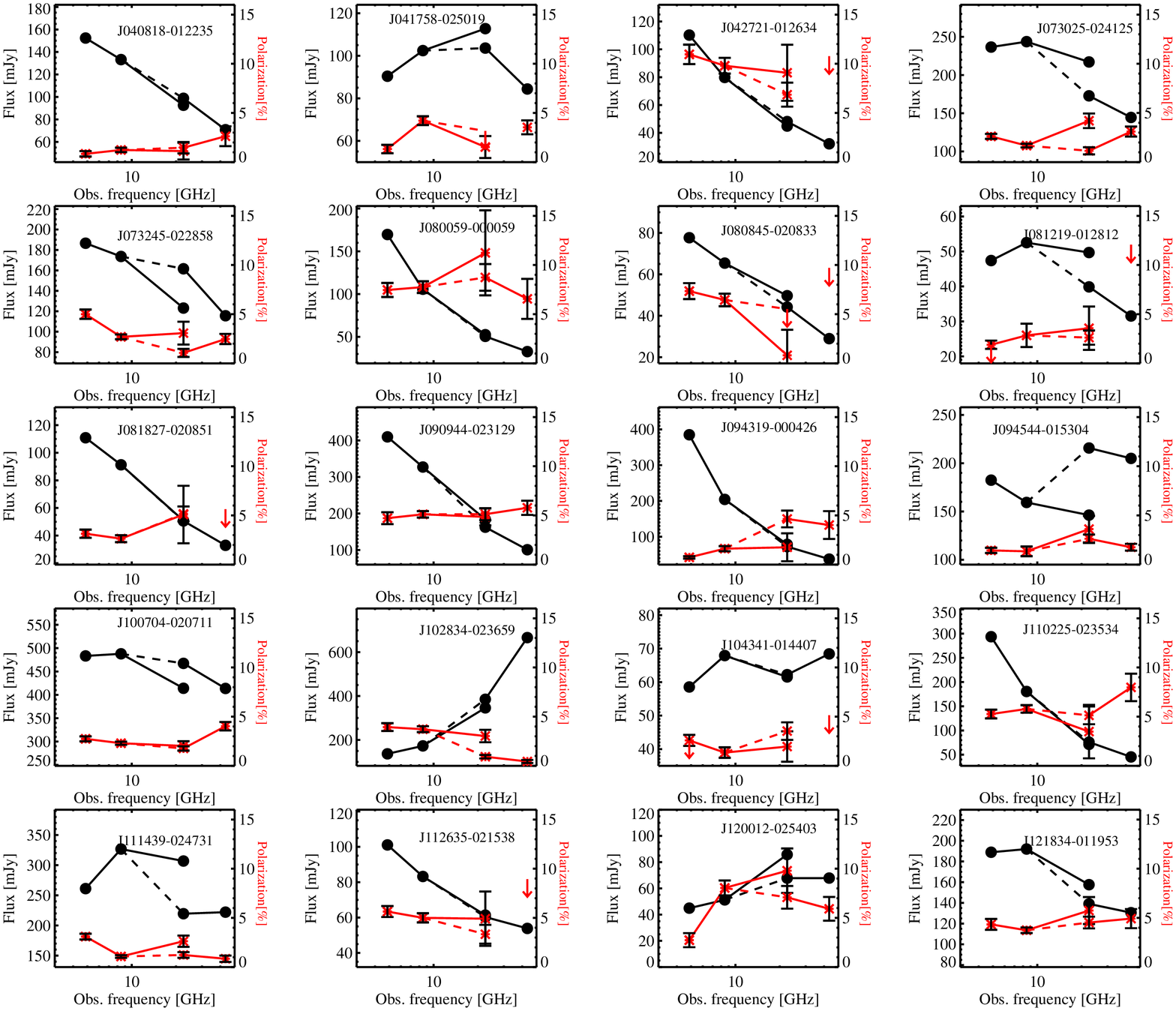}
\end{center}
\caption{SEDs and polarized fractions for sources observed in two epochs (summer 2008 and November 2009). The large filled dots and black lines indicate the flux densities. The red lines and errorbars are for the polarization fractions.  \label{fig_polseds2}}
\end{figure*}

\subsection{SEDs and polarization fractions}

Figure\,\ref{fig_polseds} shows examples of the SEDs of our sources (separated into 2 classes --  steep spectrum and inverted/GPS-spectrum sources), where we show both the total intensity SED and the polarized fraction of that SED. Figure\,\ref{fig_polseds2} shows examples of the SEDs and polarization fractions of sources observed in two different epochs (2008 and 2009 see Section\,\ref{sec_obs}). These SEDs illustrate the spread in spectral classes in our sample as well as the typical variation of spectral index with frequency which complicates extrapolations from lower frequency data to the higher frequency regime (as in modeling the radio source population for CMB or Sunyaev-Zeldovich experiments). 

Typically, we find low polarization fractions of only a few percent, although there are exceptions. When K or Q-band polarized emission is detected, it tends to imply higher polarization fractions than seen in the the C or X bands. This implied trend in rising polarization fraction with frequency may simply reflect the fact that the K and Q-bands have lower sensitivities with regards to polarization percentages and hence only the more strongly polarized sources with have detections in those bands. In Section\,\ref{Pol}, we examine trends with polarization fraction for the whole sample including trends with frequency, spectral class, or flux density levels. 

\subsection{The properties of strongly variable sources in our sample \label{sec_strongvar}}

In Section\,\ref{sec_var}, we listed four sources with strongly variable (i.e. $>$\,50\%) K-band flux densities based either on our own two epochs of K-band observations or on the comparison between our K-band measurements and the AT20G values. These are: J042315-012033, J080512-011114, J220643-003103, and J235013-020614. None of these sources are resolved in any of the frequency bands we study here. All except J080512-011114 are peaked spectrum sources. The first two show overall low polarization (typically a few percent) with the second two showing higher high frequency polarization (though for J220643-003103 the significance is questionable due to the large errors).   For J042315-012033 and J080512-011114 although there is significant variability in the K-band Stokes I values from the two observing epochs, there is no detectable variability in the fraction of polarization which are in both cases low. 

\begin{figure}
\begin{center}
\plottwo{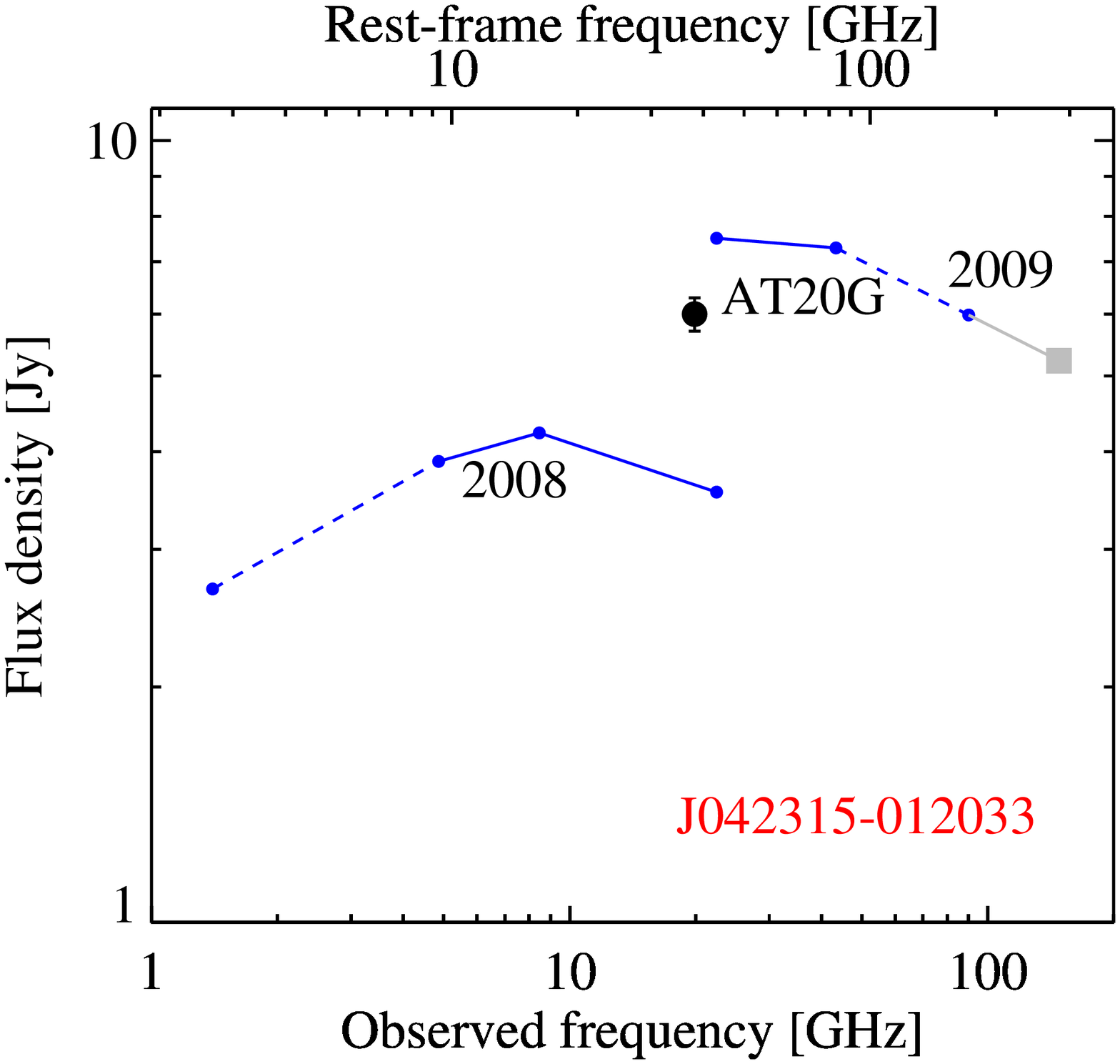}{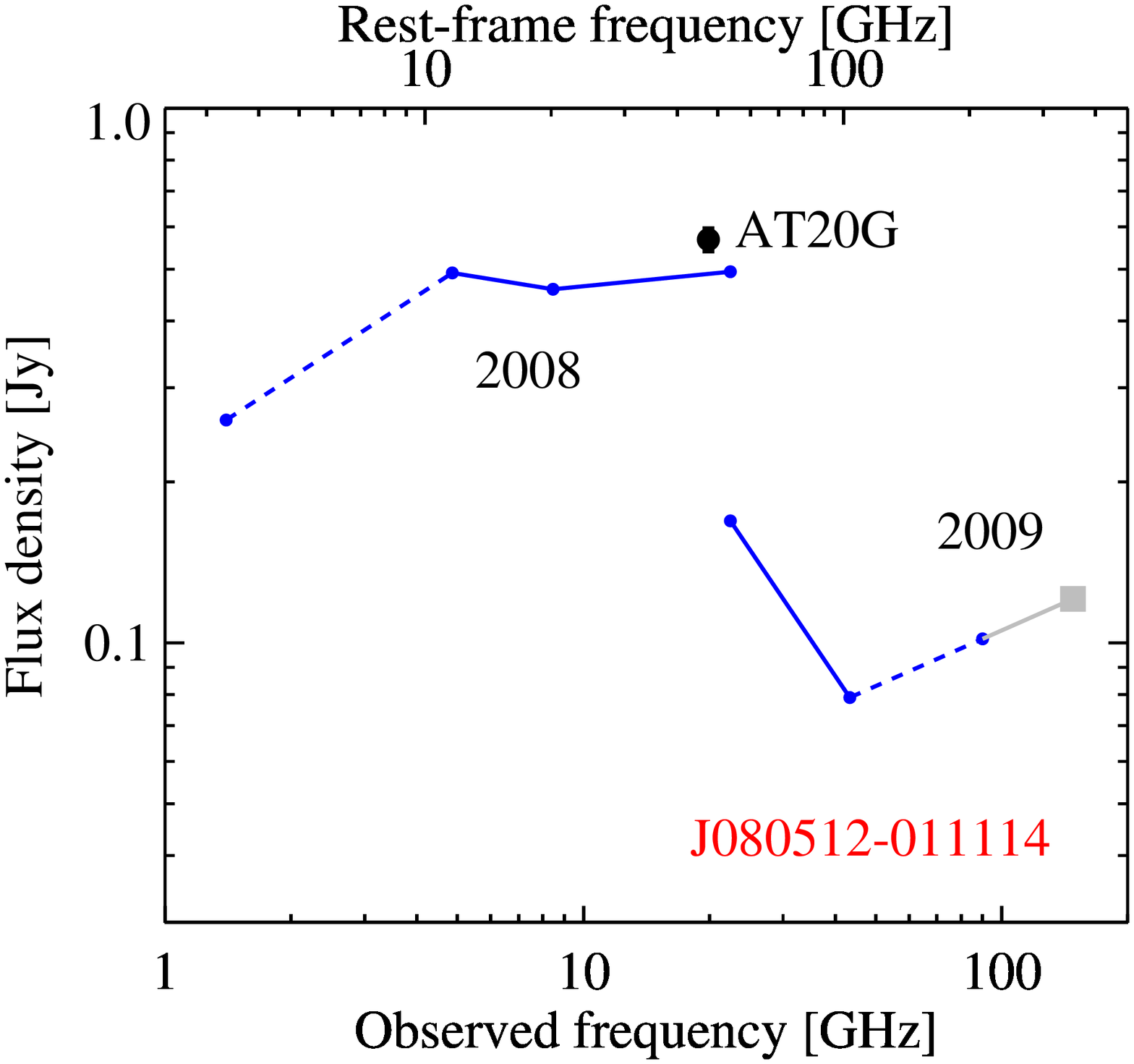}
\end{center}
\caption{SEDs of highly variable sources. {\it Left:} The SED of J042315-012033. This is a known BLLAC and RL QSO with a redshift of $z$\,=\,0.914.  {\it Right:} The SED of J080512-011114 which has a redshift of $z$\,=\,1.388. Note that as in Figure\,\ref{fig_polseds2}, solid lines indicate near-simultaneous observations, while dashed lines indicate non-simultaneous observations.  In both cases we also include the GBT 90\,GHz data which was obtained about a month after the 2009 VLA data as well as archival NVSS 1.4GHz flux densities. Finally we also show the extrapolation to the ACT 148\,GHz frequency (see Section\,\ref{sec_gbt} for more details). \label{fig_j0423}}
\end{figure}

In Figure\,\ref{fig_j0423} we plot the available data for J042315-012033. This source flared dramatically between our 2008 and 2009 observations. The AT20G K-band flux is in between our two measurements.  This source was also among the first we observed at the GBT in December 2009 and hence only a month after the November K and Q-band observations. Figure\,\ref{fig_j0423} shows a summary of the SED and its changes from 2008 to 2009 (see caption of Figure\,\ref{fig_seds90} for the meaning of the lines and symbols). Other monitoring programs (Anne L\"ahtenm\"aki, private communication) confirm the strong variability of this source during our observations.  The polarized flux density seems to have increased roughly consistently with the total intensity flare, as the polarized fractions in the K-band are comparable in the two epochs.  This is a well studied flaring source, whose 2009 flare is not even at the maximum observed in the past. \citet{hovatta08} report results of long term monitoring of this and other blazar sources and find a maximum of $S_{37GHz}$\,=\,15.7Jy for this source.

In Figure\,\ref{fig_j0423} we also show the available data for J080512-011114 which by contrast faded between 2008 and 2009. The AT20G data, from 2007, is consistent with our 2008 observations. Both of the sources highlighted in Figure\,\ref{fig_j0423} show the hazards of spectral classification in the presence of significant variability. 

\begin{figure}[h!]
\begin{center}
\plottwo{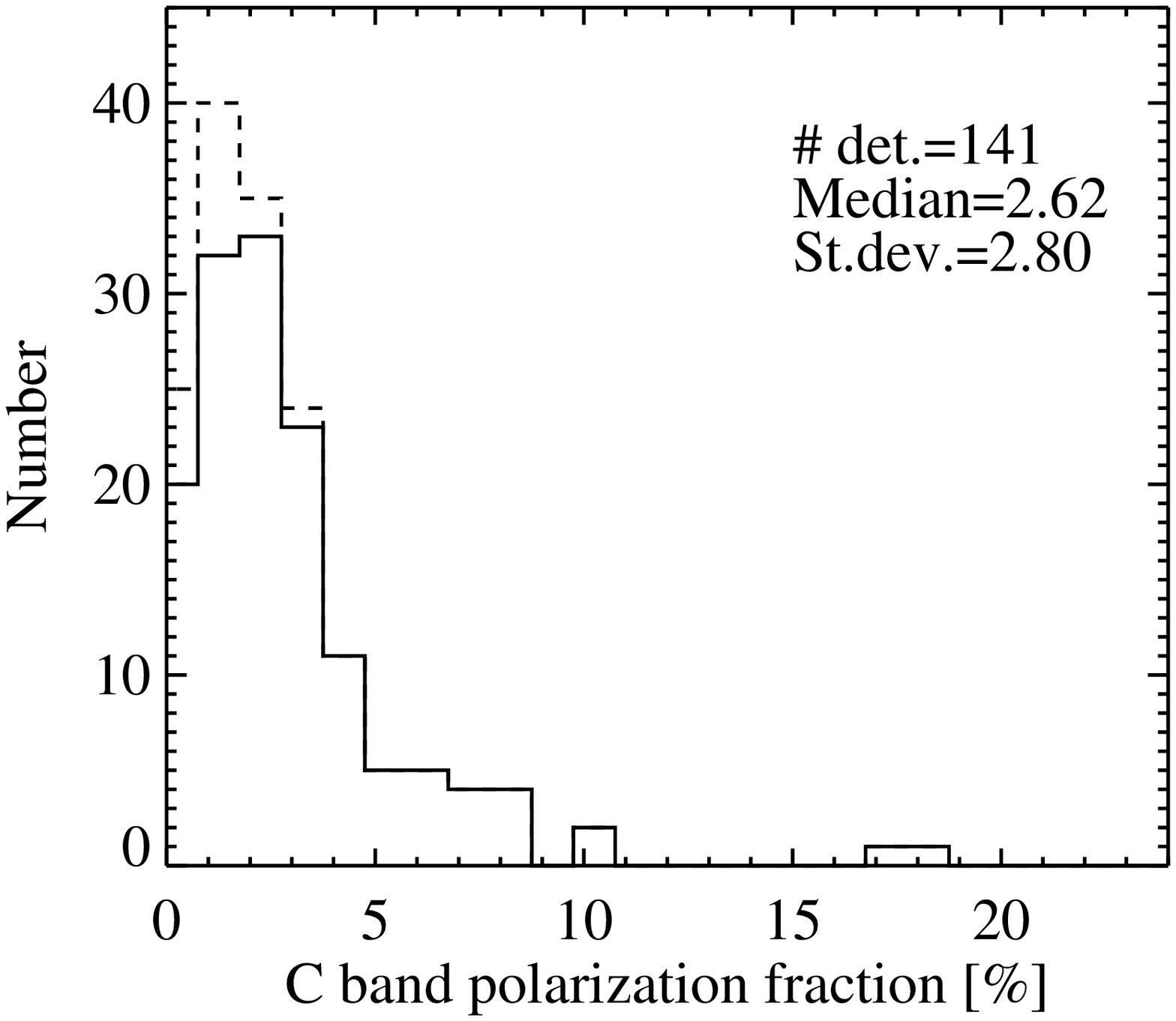}{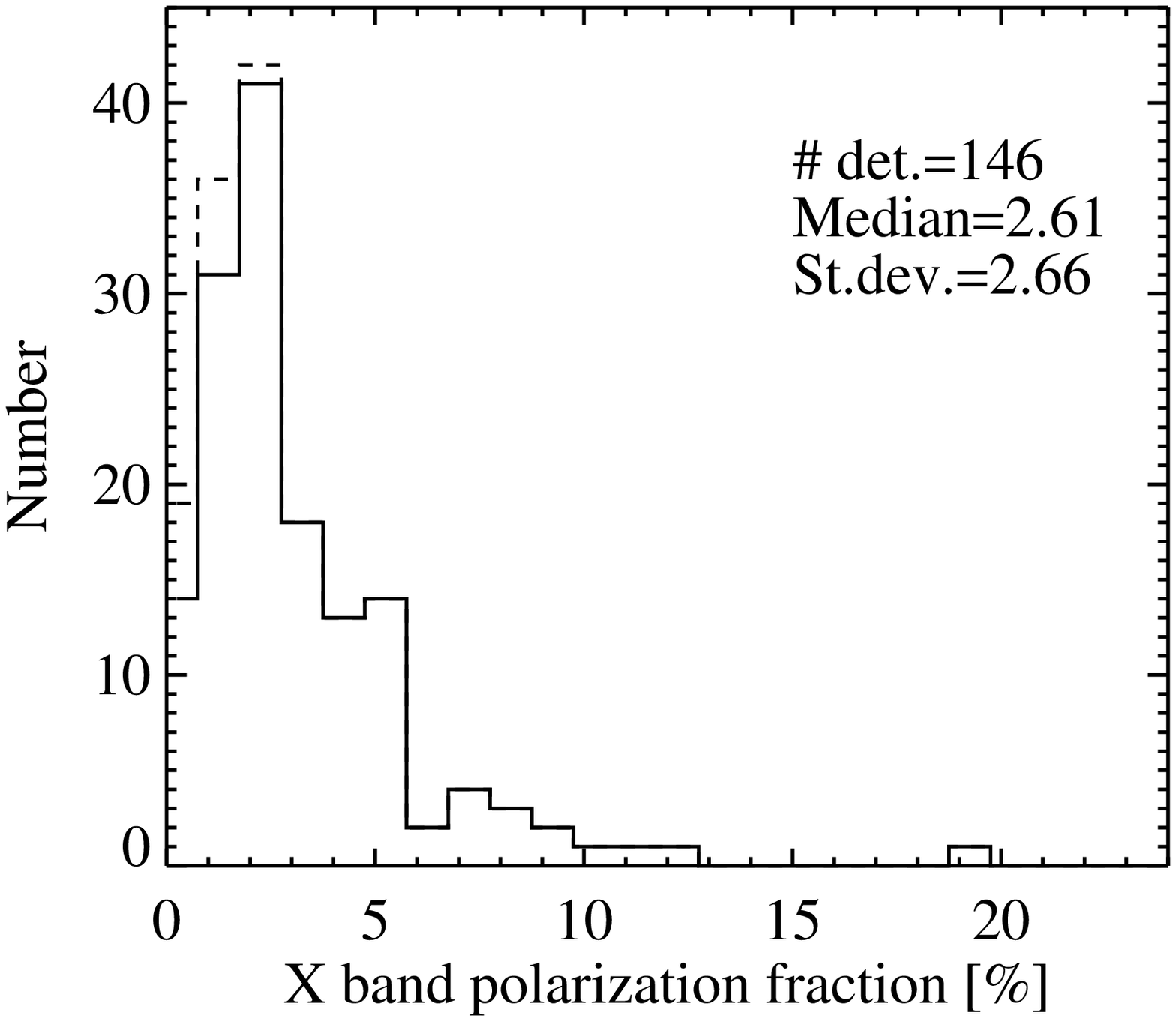}
\plottwo{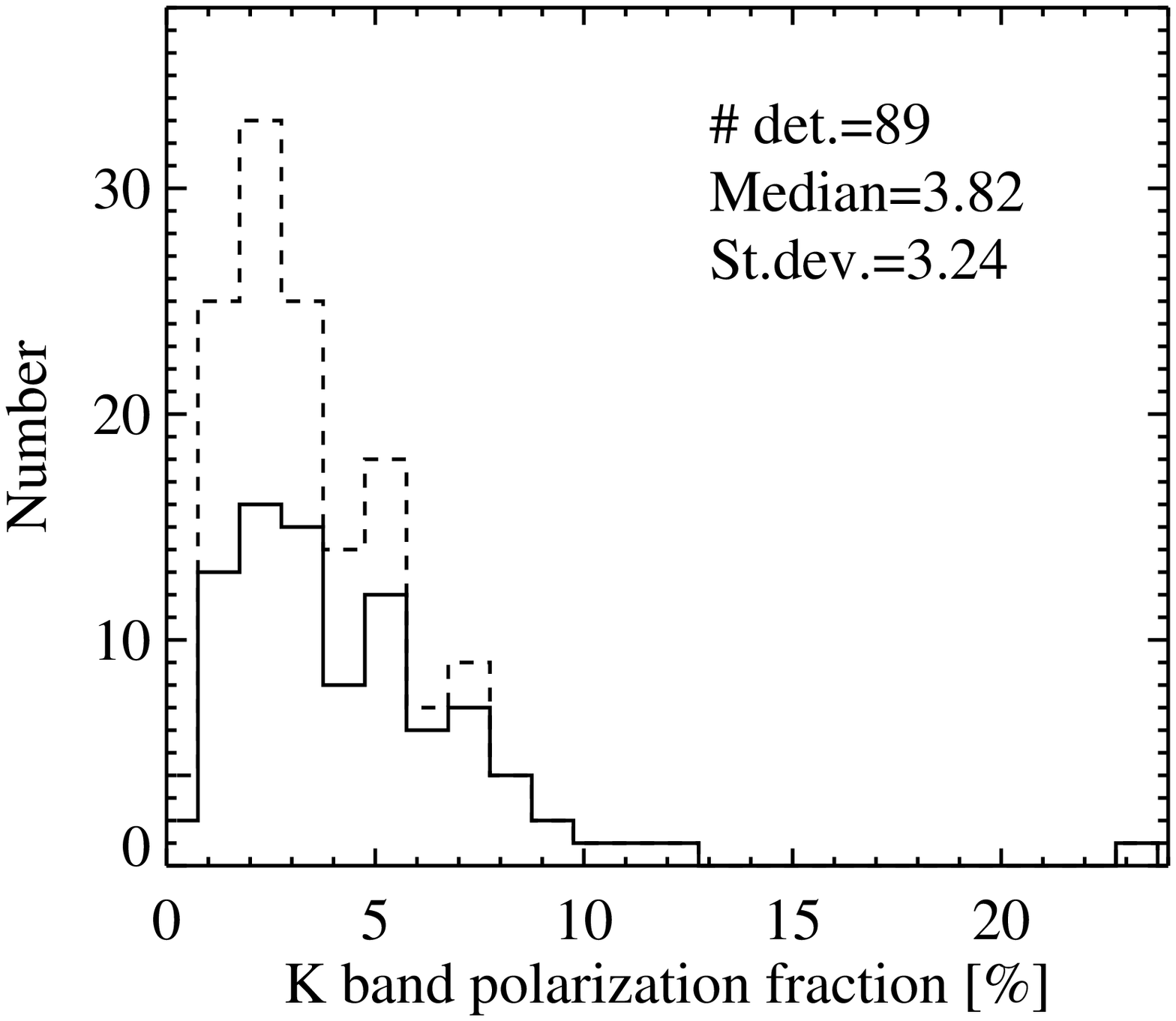}{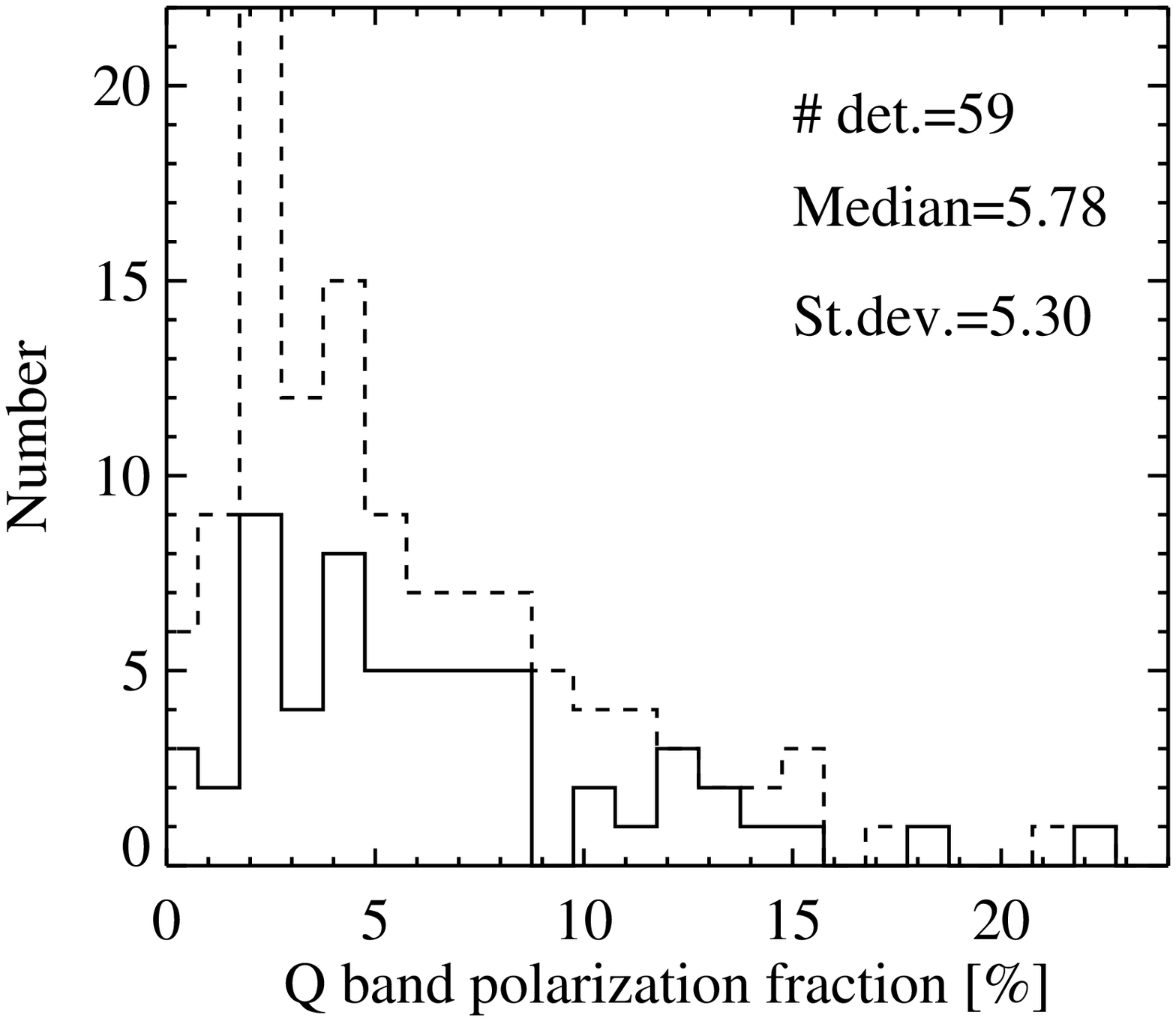}
\end{center}
\caption{The polarization fraction histograms for the different bands.  The solid histograms are for sources with detected polarization, while the dashed histograms include all polarizations as measured regardless of whether or not they are formally detected. Note the 95\% upper limits for these cases are significantly larger and would push the histograms to the right (especially for the K and Q-bands). \label{fig_polhist}}
\end{figure}

\subsection{Polarization fraction trends \label{Pol}}

\begin{figure}[h!]
\begin{center}
\plottwo{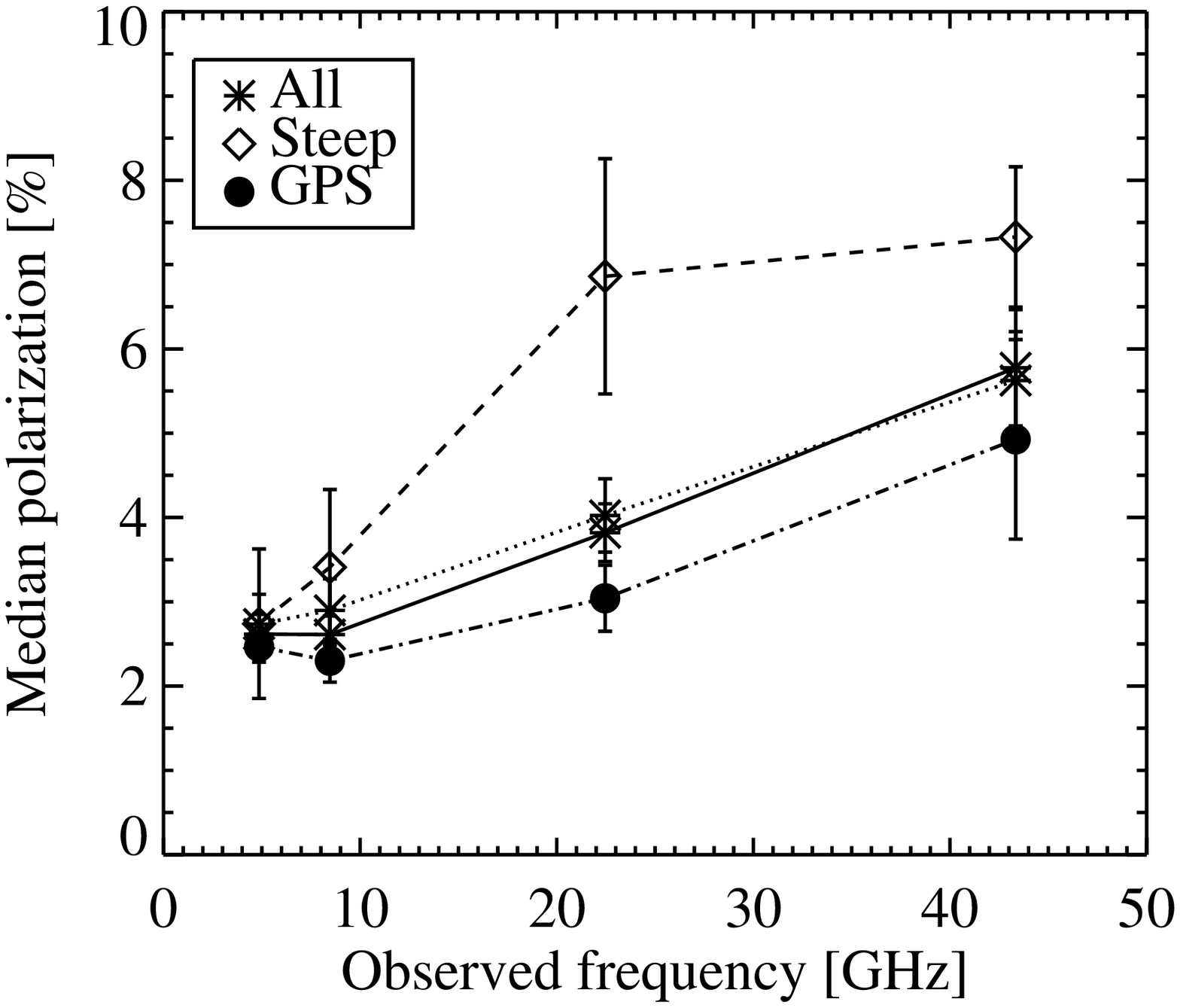}{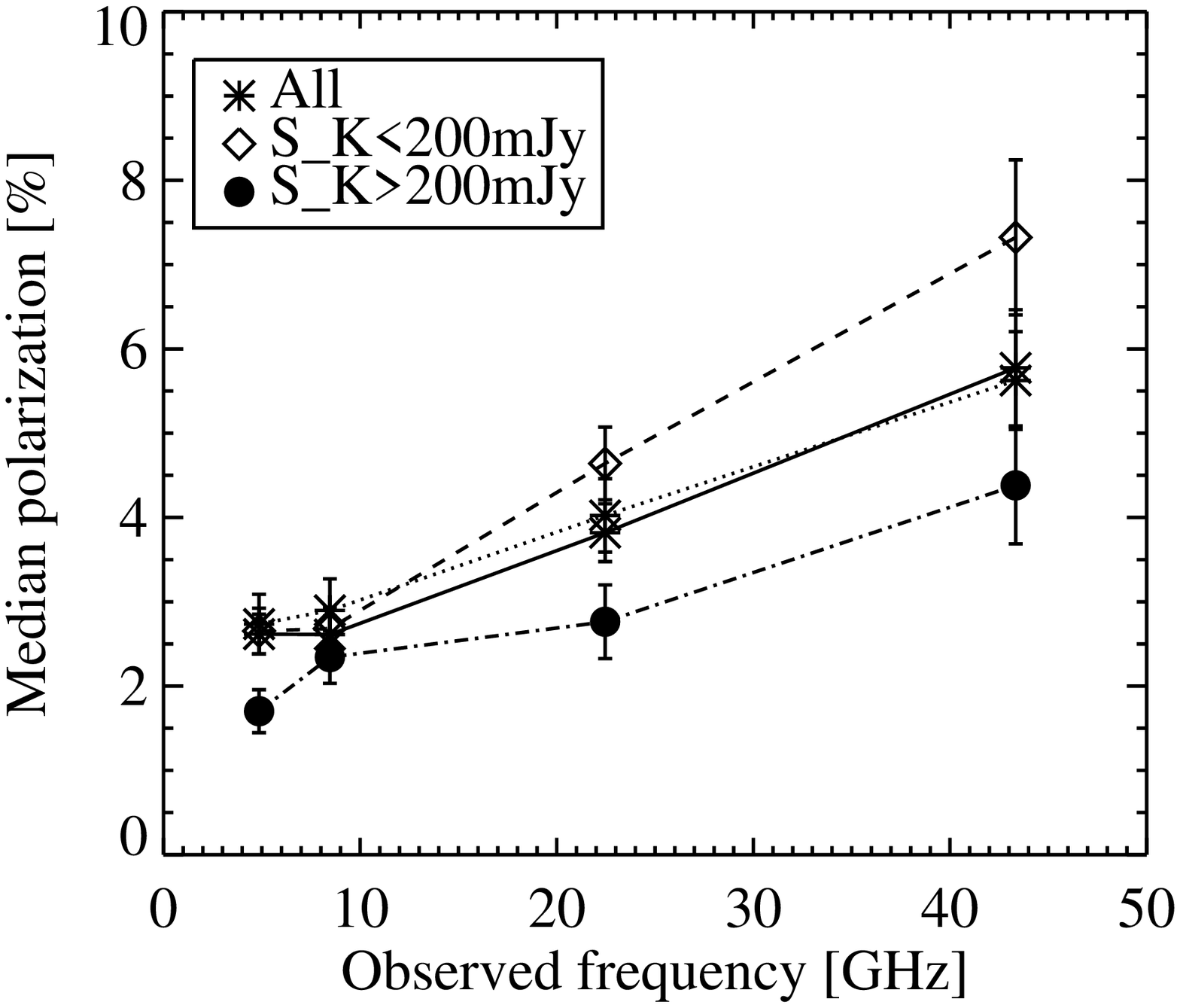}
\end{center}
\caption{Median polarization fractions for the different bands as a function of spectral type ({\it left}) and flux density ({\it right}).   \label{fig_polmeds}}
\end{figure}

While our sample is smaller than found in some other work on the polarization of microwave sources such as \citet[][]{tucci04,murphy10,battye10}, it has some advantages.

1. We used a high frequency selected sample, drawn from the AT20G survey \citep[see][for a discussion]{sadler06,murphy10}. We are thus selecting an extreme population, dominated by cores or unresolved sources. These sources, with their typically flat spectra, are those most likely to have high enough high frequency flux to interfere with SZ and CMB measurements, typically made in the 30-150\,GHz region. High frequency selected sources are also likely to be less affected by Faraday depolarization.

2. Our sensitivity in both polarization and total power is significantly higher than the much larger AT20G survey, so we could push to lower values of polarization percentage, $p$. Because our sample is unbiased with respect to its parent sample (the AT20G survey), as shown for instance by the very similar distribution in spectral properties, our conclusions here are likely to apply for the whole AT20G sample. 

3. Because of our nearly simultaneous observations, at least in sets of C+X-band and K+Q-band, any trends in polarization fraction with frequency we find would have a higher significance than a collection of observations at different epochs. 

We have computed the linear polarization fractions or upper limits for each band following the prescription in Section\,\ref{sec_polest} (see Table\,\ref{table_fluxes}).  For extended sources, we compute only the polarization of the core. Some of the missing values in Table\,\ref{table_fluxes} correspond to extended sources where the core is not clearly distinguishable in a given band  (others are genuinely missing data, such as the calibrator 0725-009 which was observed only in the K and Q-bands).  Finally, for some of the weakest Q band sources, which we could not self-calibrate, we could not obtain useful values of $S_Q$ or $S_U$. 

Figure\,\ref{fig_polhist} shows the polarization fraction histograms for each of the bands (C,X, K and Q). Here we include all the $p_{corr}$ values estimated following Eq-n\,\ref{eq_pcorr}. A number of these sources, especially in the Q-band are not actually detected in polarized flux (using the upper limits for those would shift these histograms to the right).  The polarization fractions we find are typically a few percent, although a tail of more strongly polarized (up to $\sim$\,10\%) sources exists in all frequencies. The most striking aspect of Figure\,\ref{fig_polhist} however is that the polarization fraction seems to increase with frequency. 

Figure\,\ref{fig_polmeds} shows the median polarization fractions of our sources separated by spectral type and flux density. The trend of increasing polarization with increasing frequency (at least for the sources with detected polarizations in the K and Q-bands) is quite evident. In this figure the dotted line shows the trend for all sources that have polarization detections in all four bands. This suggests that at least for these $\sim$\,60 sources, this trend is real. However, where data are available, we have 62 sources with measured Q-band polarization and 65 with upper limits on $p_{Q}$. Would the observed trends hold if we had detections in these 65 sources as well? We looked at the spectral classes of these 65 Q-band upper limits and find only 9 "steep" spectrum sources (defined as $\alpha$\,$<$\,-0.5), vs. 18 "gps" sources. This suggests that, even if all these sources have very low Q-band polarizations, the trend seen in Figure\,\ref{fig_polmeds}{\it top} for steep spectrum sources will be maintained. Looking at their fluxes virtually all of these 65 sources have $S_K$\,$<$200\,mJy (their median flux density is 67\,mJy). This is certainly not surprising as polarization is harder to detect in fainter sources; however, it does suggest that the trend in Figure\,\ref{fig_polmeds}{\it bottom} is less robust to the effects of these upper limits.  %We return to this figure in relation to polarization studies of other radio sources in Section\,\ref{sec_polfrac}.
%Because of the many upper limits and the uneven coverage at the different bands, understanding this finding of increasing polarization fraction with frequency is aided by looking directly at sources with detected polarization fractions for  all four bands. This is addressed in the following section.

\subsection{Extension to 90\,GHz \label{sec_gbt}}
\begin{figure}[!ht]
\begin{center}
\plottwo{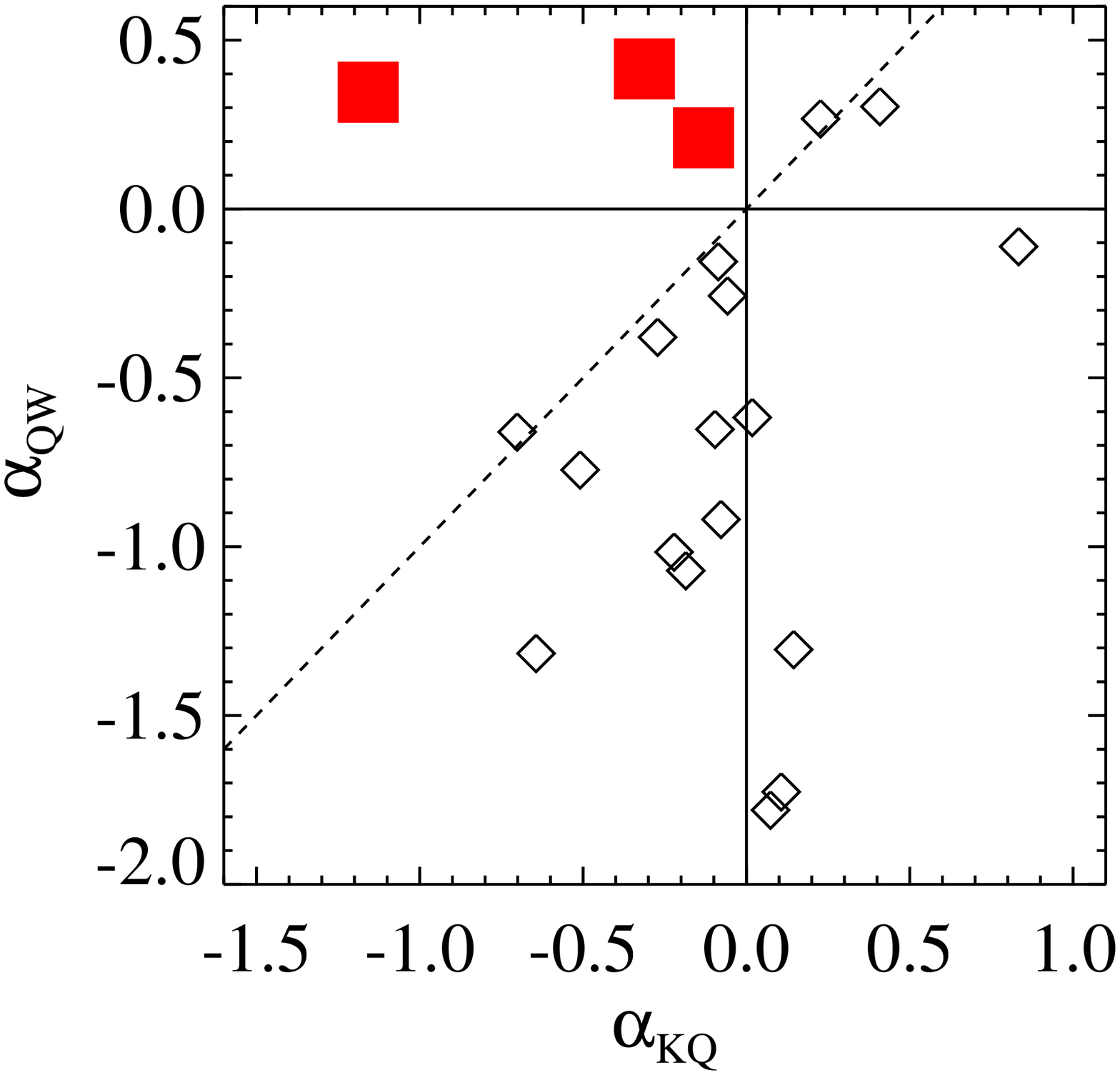}{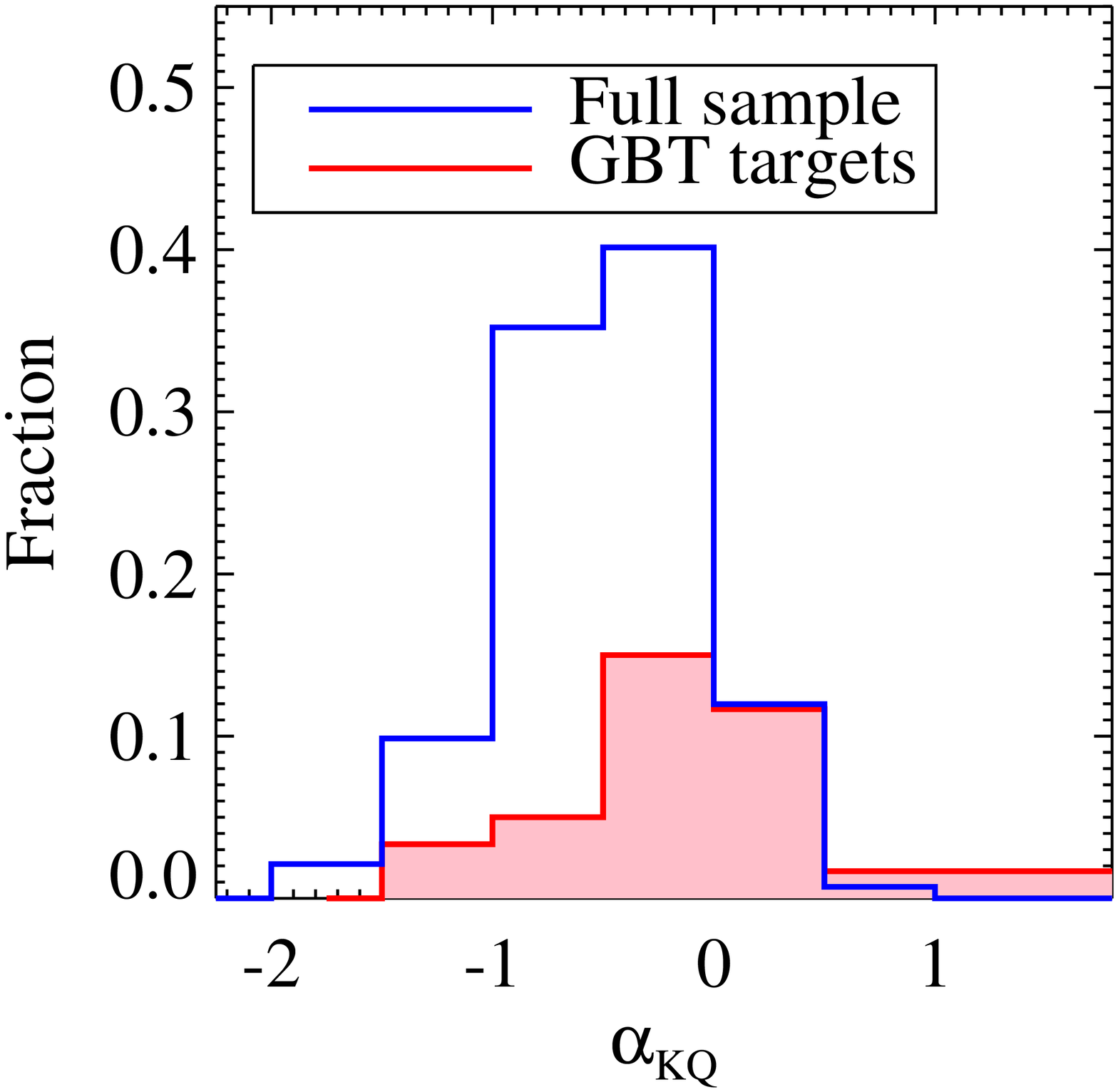}
\end{center}
\caption{{\it Left: }The spectral index diagnostics from the K-band(22\,GHz) to the W-band (90\,GHz).  The dashed line shows the $y=x$ line where extrapolations from the K and Q bands predict perfectly the 90\,GHz fluxes.  Due to the likelihood of resolved out flux, we omit extended sources from this plot. {\it Right:} The K-Q spectral index for the whole sample compared with the GBT targets.  The GBT sub-sample is biased toward flatter spectrum sources. \label{fig_slopes_q90}}
\end{figure}

\begin{figure}[!ht]
\begin{center}
\plotone{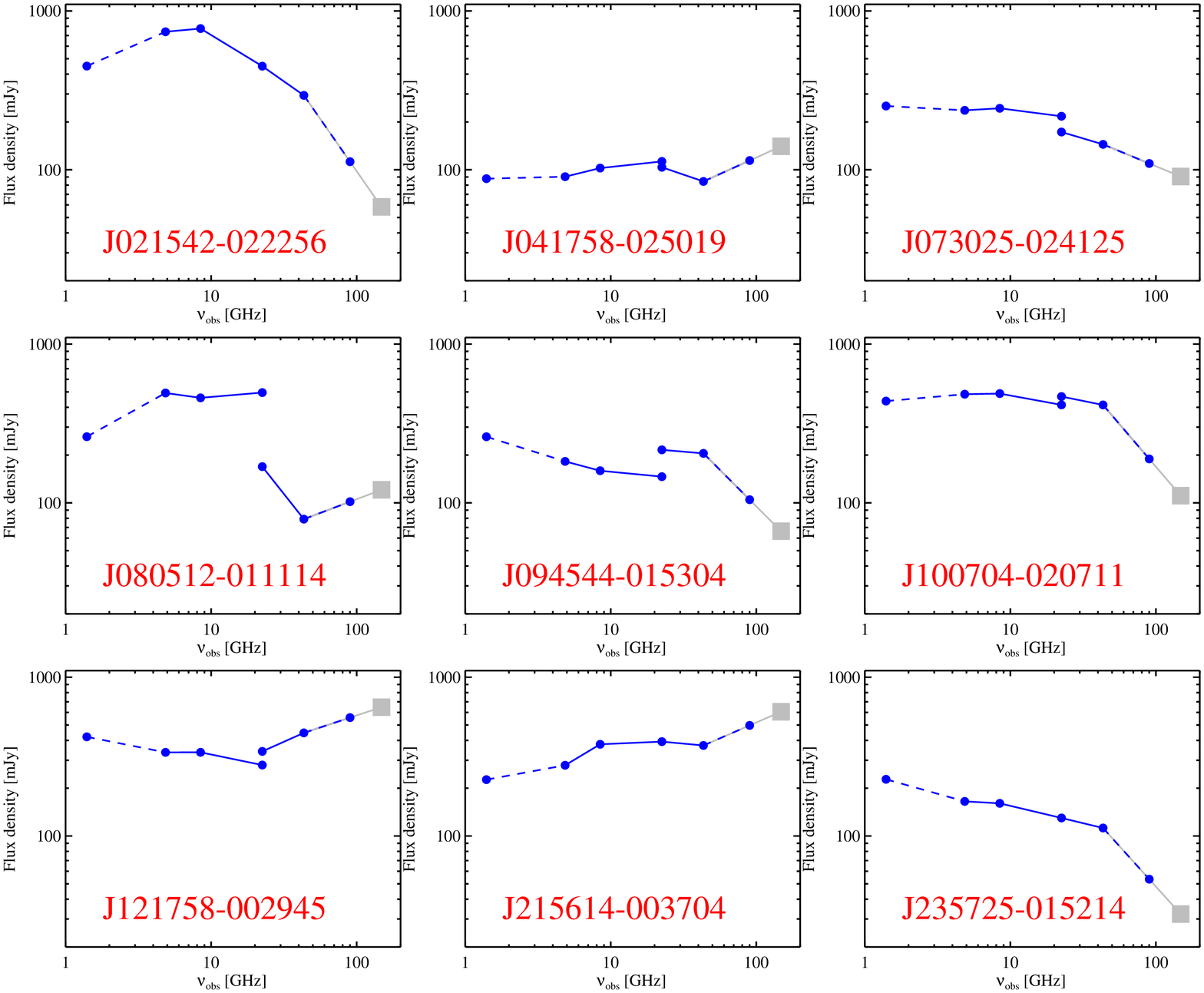}
\end{center}
\caption{The SEDs of some of our sources where we also have 90\,GHz observations that were taken within about a month of the VLA observations. The blue lines show the SEDs based on the data reported here (we also include earlier NVSS 1.4\,GHz flux densities based on NED data). Dashed lines indicate non-simultaneous observations. The thin grey lines are constant slope extrapolations of our data to the ACT frequency of 148\,GHz (large grey square). \label{fig_seds90}}
\end{figure}

Table\,\ref{table_gbt} shows the measured 90\,GHz fluxes compared with extrapolations based on our $\alpha_{KQ}$ spectral indices.  The last column shows the ratio of measured-to-expected flux densities. Some of the deviation from unity may be due to variability. However, a much stronger factor is likely changes in spectral index in the $\sim$\,20\,--\,90\,GHz range. In particular, it is noticeable that for the bulk of the sample the measured flux densities are below the extrapolated ones.  Figure\,\ref{fig_slopes_q90} shows a color-color diagnostic diagram extending to 90\,GHz. Note that 19/24 of the sources are below the $y=x$ line, suggesting simple extrapolation from 43\,GHz will overestimate the true 90\,GHz flux densities. The three outliers which actually show a significant upturn from 43 to 90\,GHz are intriguing.  This may partially be due to variability as the 90\,GHz observations were not simultaneous with the K and Q-band observations. Figure\,\ref{fig_seds90} shows a few examples of  the SEDs of compact/point-like sources with 90\,GHz data. Most of these were observed at the GBT in December 2009 and the last three were in our November 2009 K+Q sample. Therefore the 22-90\,GHz observations were all done within about a month of each other.  One of the these sources, J080512-011114, is one of the sources showing an upturn between 43 and 90\,GHz. However, as discussed in Section\,\ref{sec_strongvar} this source is also among the most strongly variable sources in our sample. Potentially variability could explain the other two outliers here.     

The extrapolated values shown in Table\,\ref{table_gbt} of course depend on the $\alpha_{KQ}$ values which in turn depend on the K-band flux densities. However, in Section\,\ref{sec_compat20}, we showed that our K-band flux densities may be underestimated compared with the AT20G survey flux densities. We test the effect of this by multiplying our K-band values by the median offset from the AT20G values, 1.22, and re-extrapolate to 90\,GHz. Table\,\ref{table_gbt} shows the resulting values. The ratios of these modified extrapolated values to the observed 90\,GHz data are clearly closer to unity than the the ratios before the above modification. 

\citet{sadler08} followed up 70 AT20G sources at 95GHz (with no overlap with our GBT sample). They found a median $\alpha_{KW}$ value of -0.39.  For our smaller sample, we find a median of -0.32 and a standard deviation of 0.40.  Without adjusting for the difference in central frequencies (for either the K or W bands), the two median values are in reasonably good agreement.  The spread in our $\alpha_{KW}$ is comparable to that in \citet{sadler08} as seen in their Figure 5.  Although of marginal significance, the slightly steeper slope found by \citet{sadler08} might be an indication of further downturn in the spectra (as their W-band is of higher central frequency than ours), or again might reflect the difference between our VLA K-band flux densities and the AT20G ones. 

\section{Discussion}

\subsection{The degree of polarization of 20\,GHz-selected sources \label{sec_polfrac}}

Figure\,\ref{fig_polmeds} shows the median polarization fractions in our four observed bands as a function of spectral type and flux density.  These values exclude upper limits. Nevertheless, it is clear that the median polarization fraction rises with frequency. This trend is reminiscent of earlier results such as the multi-frequency polarization estimates of a 400MHz-selected sample presented in \citet{klein03} and a small sample of 15 AGN whose 43-300\,GHz polarization fractions were studied by \citet{jorstad2007}. Similar to \citet{klein03}, we find that this effect is stronger for the steeper spectrum sources, while the polarization fraction is nearly  constant with frequency  for the peaked-spectrum sources. However, \citet{klein03} find a median polarization of $\sim$\,4.5\% at 15\,GHz for their low-frequency selected sample (and nearly 6\% for their steep spectrum sources).  These values are higher than our data interpolated to 15\,GHz. It is important to recognize, however, that we report polarization figures only for unresolved sources or the cores of resolved sources, thus ignoring the potentially more polarized lobes found in many radio galaxies (see for instance Figure\,\ref{fig_polmap}).  Nevertheless, this comparison suggests that the degree of polarization of our 20\,GHz-selected sample is somewhat lower than that of low frequency-selected samples.  The general trend of decreasing polarization with increasing wavelength is also expected from Faraday depolarization \citep{burn66}. 

\citet{battye10} analyze the 8\,--\,43\,GHz polarization properties of WMAP sources, in particular as a means to determine the level of point source contamination in CMB polarization measurements. They find polarization fractions of typically $\sim$\,2\% and little change with frequency. This is at first glance contradictory to our finding. However, WMAP point sources have significantly higher flux densities and tend to have even flatter spectra than our sample. The increase in polarization fraction with frequency is most significant in steep spectrum sources as observed by \citet{klein03} and confirmed for our high frequency-selected sources here (Figure\,\ref{fig_polmeds}). In addition,  the spectra of our sources flatten (as estimated by $\alpha_{KQ}$) as the Q-band flux density increases (Figure\,\ref{fig_slopes_q90}).  The apparent contradiction between our results and \citet{battye10} can be easily resolved by some combination of these two effects. 

\subsection{High-frequency GPS}

Our sample includes 36 peaked or GPS-like sources, many of which are by selection high frequency peakers (at $\sim$\,10\,--\,20\,GHz in observed frequencies).  This high turnover frequency can be the result of younger ages \citep{dallacasa00} or lower redshifts than low frequency selected GPS. We used NED\footnote{The NASA Extragalactic Database} to determine likely optical associations and redshifts for our sources. We found that 7/12 of our GPS-like sources with available redshifts have $z$\,$>$\,1 (the median value is 1.2\,$\pm$\,0.6 with the lowest redshift being 0.112 and the highest 2.160). These moderate redshifts suggest that an intrinsic spectral difference (perhaps resulting from their youth) is a more likely explanation.  In addition to the Stokes I flux densities, we also report here the polarized fractions for our 4 observational bands.  All of our GPS sources have measurable polarizations  of typically $\sim$\,1\,--\,2\% but in some cases higher (e.g. J121834-011953 has polarization fractions of $\sim$\,4\% across all bands). By contrast, the earliest polarization studies of GPS sources \citep{rj82} conducted in the cm regime suggested very low polarization fractions of under a percent for these sources.  

\subsection{Comparison with cluster radio galaxies}

\begin{figure}[h!]
\begin{center}
\plotone{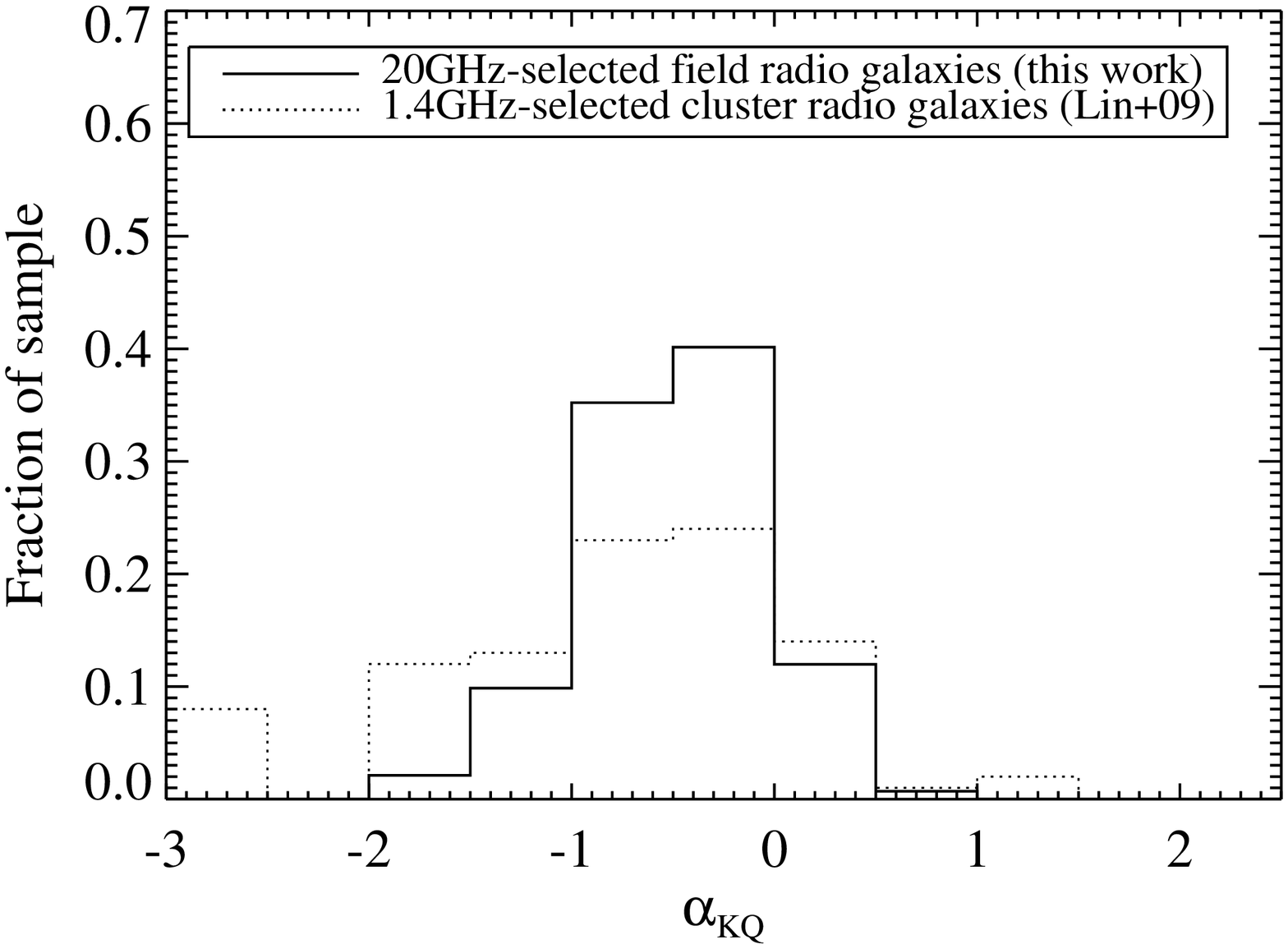}
\end{center}
\caption{The spectral index distribution for our sources compared with the known cluster radio galaxies in \citet{lin09}. \label{fig_indices}}
\end{figure}

One of the principal motivations for this work was to better understand the radio galaxy contamination in SZ surveys. Such an analysis was performed by \citet{lin09} for the case of 1.4\,GHz-selected, low redshift radio galaxies that are known cluster members. That work concludes that a significant fraction of these sources have an upturn between 22\,GHz and 43\,GHz ($\alpha_{KQ}$\,$>$\,-0.5) contrary to simple extrapolation from 1.4\,GHz as has been done in the past. In Figure\,\ref{fig_indices}, we compare the $\alpha_{KQ}$ distribution for our sample (including unresolved sources plus total fluxes for resolved sources) with the sample distribution for the Lin et al. sample. We find that the range covered by the two samples is very similar, with the exception that we are missing the most extreme steep spectrum sources.  However, our distribution is more `peaked'  than the 1.4\,GHz-selected sample.  In particular, in our sample 56\% of the sources have $\alpha_{KQ}$\,$>$\,-0.5, while only 41\% of the 1.4\,GHz-selected cluster radio galaxies meet the same criterion.  Moreover the significant fraction of GPS-like sources in our sample supports the well known fact that single power-law extrapolation from 1.4\,GHz-based luminosity functions in modeling the radio point sources at higher frequencies is unlikely to be adequate. Note that the \citet{lin09} paper assumes the same VLA flux scale as used here and hence this direct comparison is valid even though there is a possibility of our K-band fluxes being underestimated (Section\,\ref{sec_compat20}). 

\subsection{Implications for mm-wave cosmology surveys  \label{sec_implications}}
	
 The differences we find in this work (usually reinforcing earlier results) regarding the nature of high-frequency selected sources, as compared with low-frequency selected sources, in terms of morphology, spectral index and polarization fraction all suggest that an accurate modeling of the radio galaxy contamination  of mm-wave cosmological surveys requires starting with high frequency-selected samples and simultaneous multifrequency observations (to model the spectral index and control for variability).  In this fairly speculative section, we use the $\alpha_{KQ}$ spectral index to extrapolate our measured flux densities to 148\,GHz to estimate the expected flux density distribution of these sources in the lowest frequency band employed by ACT. The resulting extrapolated fluxes for our sources tend to cluster in the 10\,--\,100\,mJy  range, and hence are observable by ACT. However,  in Section\,\ref{sec_gbt}, we found that measured 90\,GHz flux densities were somewhat below extrapolations based on the $\alpha_{KQ}$ indices. Some of this can be attributed to variability;  however, the one-sidedness of the effect suggests that the spectra do indeed steepen from $\sim$\,40GHz to $\sim$\,90GHz. This by necessity would also affect the extrapolated 148\,GHz fluxes. Because of the effects of variability (in general the 90GHz data were not simultaneous with the 43GHz data) and the fact that GBT observations were only performed for the brightest sources in the Q-band, we cannot safely generalize these conclusions to the whole sample. 

We next ask what is the contribution of radio galaxies (as selected by the AT20G survey) to the ACT survey? This can be done by extrapolating the AT20G number counts to 148GHz.  A significant excess between the measured ACT number counts and this extrapolation should reveal either: a) a population not present in the AT20G survey, or b)  an upturn in the AT20G sources' spectra as might be expected from for example free-free or even dust emission. A significant deficit might reveal that the AT20G spectra actually steepen as our GBT results for example suggest. Of course these effects (if present) might cancel each other on average and not leave a significant imprint on the number count. Therefore, once the equatorial ACT data become available, we plan to perform a source by source comparison between our results and the measured 148GHz fluxes for this particular sample. 

\begin{figure}[!ht]
\begin{center}
\plottwo{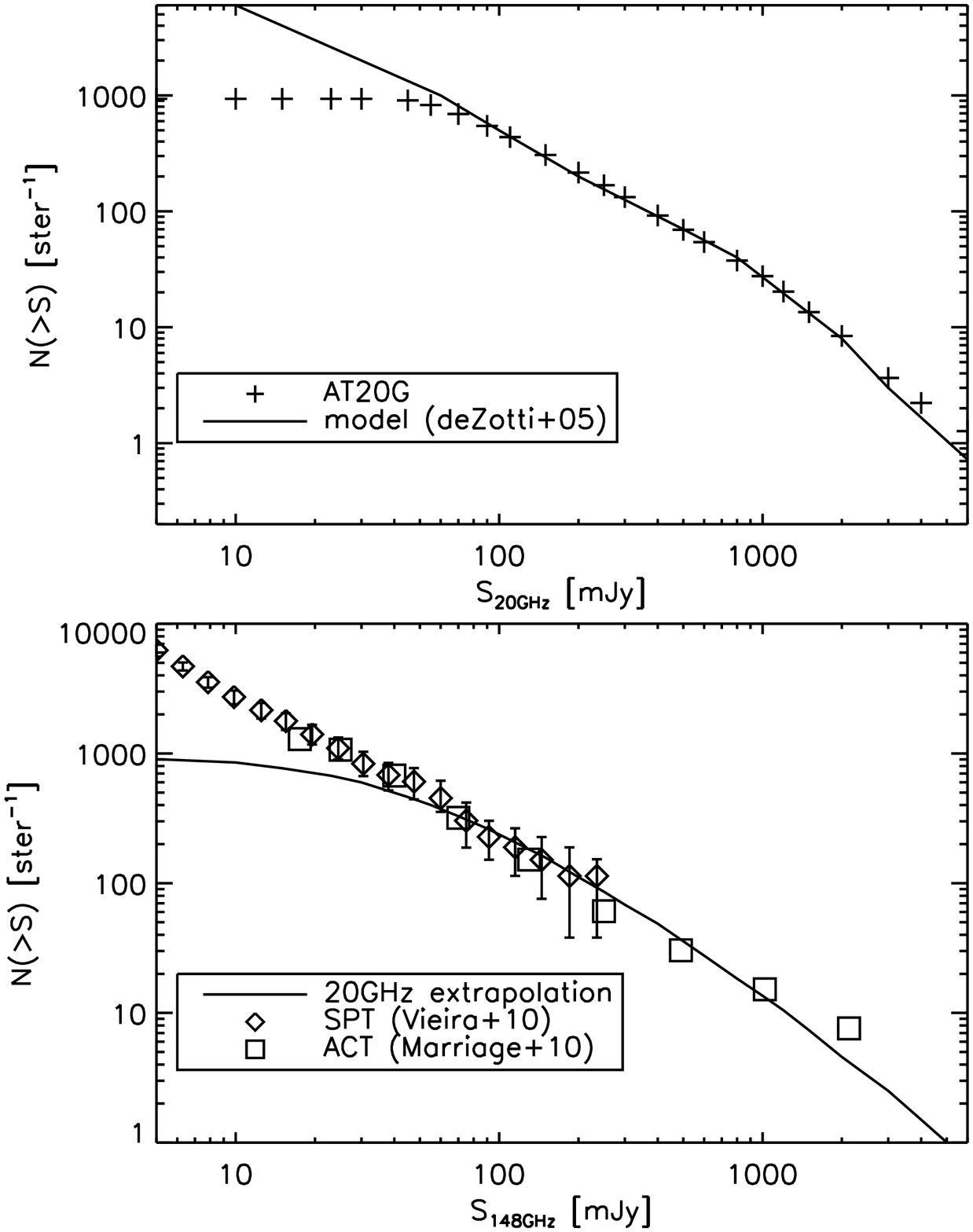}{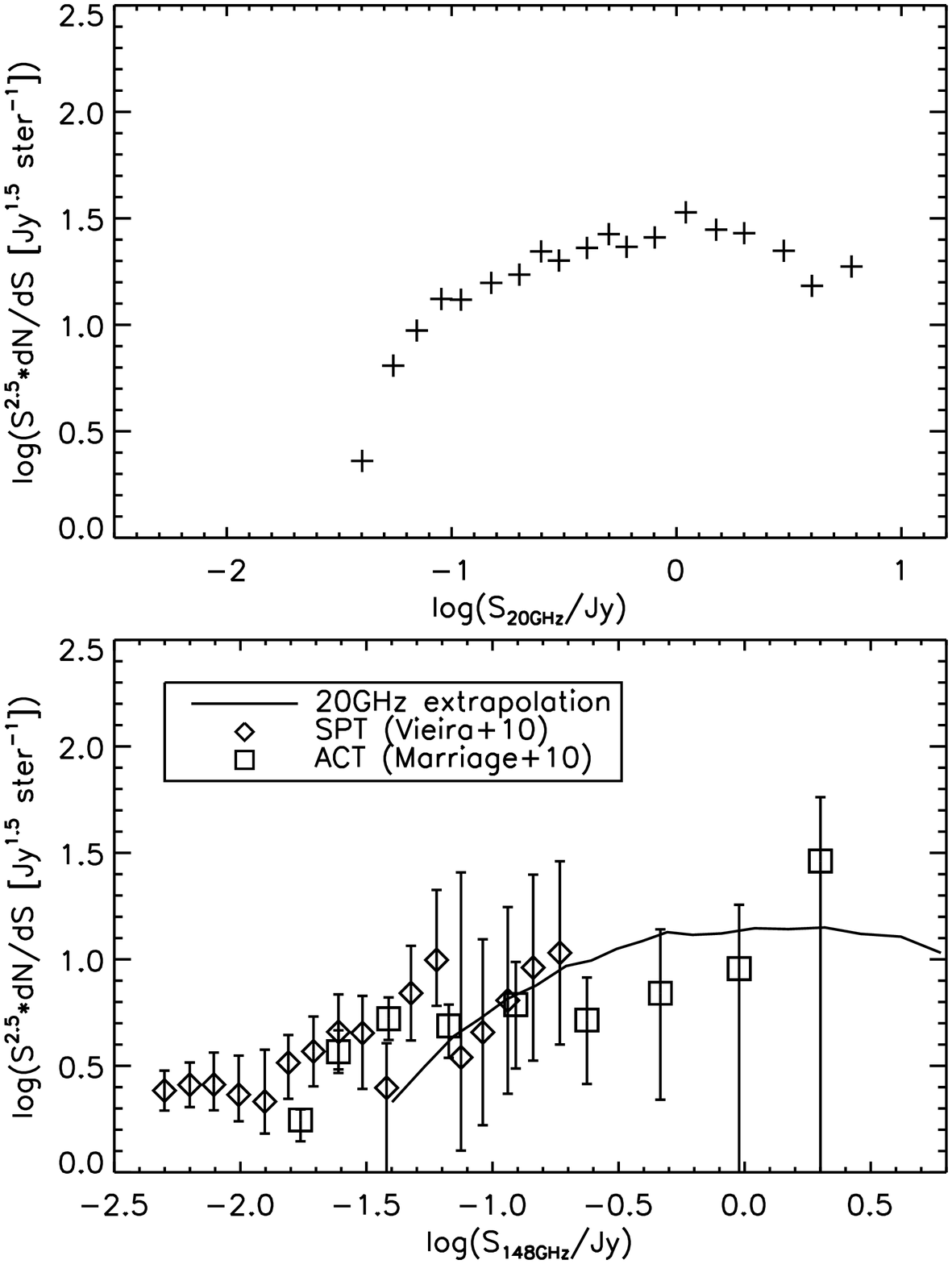}
\end{center}
\caption{Here the top panels are the AT20G number counts where integral counts are shown on the left and normalized differential counts on the right. The extrapolated 148GHz counts are shown on the bottom (solid line) where for comparison we also overlay the SPT \citep{vieira09} and ACT number counts \citep{marriage10}.   \label{fig_nc}}
\end{figure}

Figure\,\ref{fig_nc}{\it top left} shows the AT20G integral number counts, where for comparison we also overlay the model from \citet{dezotti05}. Note the survey limit of $\sim$\,40\,mJy at 20GHz. In order to extrapolate to 148GHz, we assume that our $\alpha_{KQ}$ distribution (Figure\,\ref{fig_indices}) is applicable for the whole AT20G sample (a good assumption) and that it is independent of flux density (likely a poor assumption given the discussion above).  Figure\,\ref{fig_nc}{\it bottom left} shows the resulting extrapolated number counts at 148GHz compared with the SPT results as given by \citet{vieira09}, and preliminary ACT results as given by \citet{marriage10}. We find that while we come reassuringly close to the observed number counts for the higher flux densities ($S_{148}$\,$>$\,60\,mJy), we come under the observed counts at lower flux densities. Our 20GHz limit of $\sim$\,40\,GHz, combined, with the an on average negative $\alpha_{KQ}$ (median=-0.46), suggests that incompleteness cannot fully explain this shortfall (that should come at flux densities $<$\,40\,mJy). We can conclude from this rough analysis that the higher flux density SPT sources are fully accounted for by the radio population as seen by the AT20G, while an additional population or emission component to the radio galaxies might be present  in the SPT counts at a few 10s of mJy.  The right-hand side of Figure\,\ref{fig_nc} shows the same but for the 20\,GHz and extrapolated 148GHz differential number counts.  Here again we find that the faint end of our extrapolated counts at 148\,GHz is slightly lower than the observed SPT counts.  If our K-band flux densities are underestimated by $\sim$\,20\%, that would depress the extrapolated 148GHz counts.  The preliminary ACT number counts \citep{marriage10} are indeed systematically lower than than our extrapolation. While a potential underestimate of our K-band flux densities (and hence overestimate of the KQ spectral index) might account for some of this discrepancy, a more likely explanation is the steepening of the SEDs of radio sources.  Our finding of a median $\alpha_{KQ}$\,=\,-0.14 and a median $\alpha_{QW}$\,=\,-0.47 supports this view. 

\section{Summary and Conclusions}

1) In our sample only 26 out of 159 sources (16\%) are extended. A qualitative comparison with the 1.4\,GHz \citet{condon03} data, which has comparable resolution to our observations, imply that, as expected, at higher frequencies, sources become more core-dominated.  \\

2) Using K-band observations at two epochs, we explore the level of variability in our sources. We find a median variability of $\sim$\,6\%. We highlight J042315-012033, which flared dramatically in 2009.\\

3)  On average, the spectral indices tend to steepen  between 20 and 90\,GHz -- such that some inverted spectrum sources in the AT20G survey appear "peaked spectrum"  sources when the 43\,GHz VLA data are included. The trend continues with the 90\,GHz data, where available, although a small subset of these sources shows an upturn from 43 to 90\,GHz. These data will be very useful to modeling efforts to understand the general radio source population contamination to SZ surveys. \\

4) The observed polarization fractions are typically $\sim$\,2\,--\,5\%, with a tail extending up to $\sim$\,15\%. We observe an increase of the median polarization fraction with frequency.  This effect is strongest for the steep spectrum sources. Flatter spectrum and higher flux density sources tend to have lower high frequency median polarization fractions. This is broadly consistent with recent results on WMAP sources \citep{battye10} and suggests that radio point sources polarization is unlikely to be a dominant foreground for CMB polarization measurements. \\

5)  As expected, our 20\,GHz-selected galaxies show a higher fraction of flat and inverted spectrum sources than seen in recent studies of 1.4\,GHz cluster radio galaxies. Our  observed distribution of $\alpha_{KQ}$ in conjunction with the AT20G number counts provides a better constraint on the number of discrete radio galaxies  that can be expected in current and upcoming SZ surveys.  

6) In this work, we discover an inconsistency between the VLA and ATCA 20\,GHz absolute flux scale, in the sense that the VLA fluxes tend to be $\sim$\,20\% lower. Such a discrepancy was first presented in \citet{murphy10}, but at a lower level, although we argue that this is mostly due to the higher flux densities probed in that work.  While we discuss various possibilities here, the cause of this offset is as yet unknown. Planned, simultaneous, ATCA and VLA observations should be able to resolve the issue. Throughout this paper, we address the effect of this discrepancy where appropriate. \\

{\it Acknowledgements:} 

We are grateful to the anonymous referee for a careful reading of the manuscript and helpful suggestions which improved the content and style of this paper. We thank the AT20G survey team (PI Ron Ekers) for their work in producing such an excellent resource for the study of high-frequency radio sources. We are particularly grateful to Mark Devlin for his involvement with the GBT 90\,GHz proposal. We would also like to thank the staff at the NRAO VLA for help during the preparation of as well as the execution of the observations.  The National Radio Astronomy Observatory is a facility of the National Science Foundation operated under cooperative agreement by Associated Universities, Inc.  We would like to thank the people responsible for writing and maintaining AIPS, which, despite occasionally eliciting some colorful language, was invaluable to our project, as well as many many others. Work on this project at Haverford College was provided by NSF Grant AST-06-06975 and a subcontract to Haverford College from NASA/JPL in support of ESA's {\sl Planck} mission. Two current Haverford students, Quentin Sherman and Ben Walter helped in the data analysis. Additional funds to support student involvement in the project were provided by NSF to the Keck Northeast Astronomy Coalition to support Shannon Stefl and an NRAO Student Observing Grant GSSP09-0013 to support the work of Tyler Evans. PK and SD were partially supported by NSF grant AST-0607654. PK was also funded by the NRAO graduate student support program.

%%%############################################################################
%%% References:

\clearpage

\begin{center}
\begin{deluxetable}{cc rrr rrrrr r c}
\tablecolumns{12}
\tablewidth{7in}
\tabletypesize{\tiny}
\tablecaption{\label{table_fluxes} Flux densities and polarization percentages\tablenotemark{1,2}} 
\tablehead{ \colhead{Source} &  \colhead{Flag\tablenotemark{3}} & \colhead{$z$} & \colhead{$S_{C}$}  & \colhead{$S_{X}$} &\colhead{$S_{K}$} & \colhead{$S_{Q}$} & \colhead{$p_{C}$}  & \colhead{$p_{X}$} &\colhead{$p_{K}$} & \colhead{$p_{Q}$} & \colhead{Class} \\
\colhead{} &\colhead{} & \colhead{} & \colhead{[mJy]} & \colhead{[mJy]} & \colhead{[mJy]} & \colhead{[mJy]}  & \colhead{[\%]} & \colhead{[\%]} & \colhead{[\%]} & \colhead{[\%]} & \colhead{}  }
\startdata
J000507-013244 &   & 1.710 & 38.9$\pm$0.5 & 48.7$\pm$0.4 & 57.6$\pm$1.1 & 52.1$\pm$1.5 & 3.1$\pm$0.9 & 4.1$\pm$0.9 & $<$11.2 & $<$9.08 & GPS \\
J000622-000423 &   & 1.037 & 1343.7$\pm$14. & 830.2$\pm$6.0 & 305.4$\pm$3.3 & 145.5$\pm$2.3 & 2.7$\pm$0.3 & 2.9$\pm$0.2 & 2.8$\pm$0.5 & 4.44$\pm$1.4 &   \\
J001611-001511 &   & 1.575 & 491.0$\pm$5.8 & 383.5$\pm$2.7 & 259.7$\pm$2.8 & 231.7$\pm$2.7 & 1.4$\pm$0.2 & .84$\pm$0.1 & 1.5$\pm$0.5 & 2.6$\pm$0.8 &   \\
J001752-023618 &   &   & 59.5$\pm$0.7 & 62.5$\pm$0.5 & 69.0$\pm$1.2 & 82.1$\pm$1.3 & 3.8$\pm$0.7 & 3.1$\pm$0.5 & 9.43$\pm$1.8 & 14.6$\pm$3.5 & Inverted \\
J001917-010357 & C &   & 65.9$\pm$0.9 & 66.0$\pm$0.5 & 66.0$\pm$1.4 & 61.6$\pm$1.2 & 1.4$\pm$0.4 & $<$1.26 & $<$14.4 & $<$10.7 & GPS \\
  & T &   & 84.0$\pm$2.2 & 85.5$\pm$1.2 & 66.0$\pm$1.1 & 61.0$\pm$1.2 &  &  &  &  &   \\
J002336-025307 &   &   & 156.2$\pm$1.7 & 103.6$\pm$0.8 & 45.6$\pm$1.0 & 30.0$\pm$1.1 & 1.0$\pm$0.2 & 2.1$\pm$0.4 & 8.53$\pm$2.5 & $<$16.5 &   \\
J002901-011341 &   & 0.086 & 249.9$\pm$2.7 & 231.9$\pm$1.6 & 167.7$\pm$1.8 & 127.2$\pm$1.4 & 3.2$\pm$0.3 & 2.4$\pm$0.2 & 2.88$\pm$1.0 & $<$5.43 &   \\
J003007-000007 &   & 0.475 & 82.0$\pm$0.9 & 79.0$\pm$0.6 & 61.2$\pm$1.1 & 51.2$\pm$1.1 & $<$2.03 & 1.1$\pm$0.4 & 6.74$\pm$2.3 & $<$9.00 &   \\
J003031-021156 &   &   & 279.1$\pm$3.0 & 231.8$\pm$1.6 & 123.9$\pm$1.6 & 88.9$\pm$1.4 & 2.5$\pm$0.2 & 2.8$\pm$0.2 & $<$4.45 & $<$8.16 &   \\
J004345-025102 &   &   & 128.8$\pm$1.6 & 121.6$\pm$0.9 & 92.2$\pm$1.3 & 77.0$\pm$1.4 & 3.9$\pm$0.5 & 3.7$\pm$0.3 & 4.02$\pm$1.6 & 5.50$\pm$2.0 &   \\
J005734-012327 & C &   & 290.0$\pm$40. & 96.9$\pm$0.9 & 59.6$\pm$1.5 & 70.8$\pm$1.4 & 18.7$\pm$4.5 & 4.0$\pm$0.3 & $<$8.88 & 10.3$\pm$4.3 & Upturned \\
  & T & 0.045 & 2200.0$\pm$29. & 1250.0$\pm$13. & 155.0$\pm$3.2 & 72.0$\pm$2.1 &  &  &  &  &   \\
J010622-015538 &   & 2.201 & 419.0$\pm$5.8 & 366.6$\pm$2.6 & --- & 93.4$\pm$1.6 & 9.00$\pm$1.3 & 7.6$\pm$0.5 & --- & 3.39$\pm$1.3 &   \\
J010826-003723 &   & 1.374 & 701.0$\pm$9.9 & 578.2$\pm$4.1 & 298.2$\pm$3.2 & 198.0$\pm$2.5 & 1.2$\pm$0.2 & 2.3$\pm$0.2 & 2.6$\pm$0.7 & 5.19$\pm$1.1 &   \\
J011013-021952 &   & 0.961 & 113.7$\pm$1.3 & 91.3$\pm$0.7 & 44.5$\pm$1.1 & 32.3$\pm$1.2 & 2.6$\pm$0.5 & 2.0$\pm$0.5 & 4.70$\pm$2.0 & $<$12.7 &   \\
J011517-012704 &   & 1.365 & 782.1$\pm$7.8 & 696.9$\pm$4.9 & 444.3$\pm$4.6 & 295.4$\pm$3.1 & 4.7$\pm$0.4 & 3.1$\pm$0.2 & 5.3$\pm$0.6 & 6.27$\pm$1.1 &   \\
J011815-012030 &   & 1.162 & 289.5$\pm$3.1 & 135.4$\pm$1.0 & 31.5$\pm$0.8 & 12.9$\pm$1.9 & 1.7$\pm$0.2 & 3.4$\pm$0.3 & $<$14.0 & $<$52.4 &   \\
J012100-001518 &   & 0.864 & 103.5$\pm$1.2 & 91.4$\pm$0.7 & 52.5$\pm$1.0 & 36.5$\pm$1.2 & 2.1$\pm$0.4 & 3.3$\pm$0.4 & 9.97$\pm$2.0 & 11.9$\pm$4.9 &   \\
J012213-001800 &   & 1.325 & 368.4$\pm$4.0 & 316.2$\pm$2.2 & 203.6$\pm$2.2 & 136.2$\pm$1.8 & 1.8$\pm$0.2 & 5.0$\pm$0.3 & 7.2$\pm$0.9 & 4.43$\pm$1.3 &   \\
J012217-005615 &   &   & 170.0$\pm$1.9 & 157.7$\pm$1.1 & 110.7$\pm$1.4 & 86.9$\pm$1.3 & 2.8$\pm$0.3 & 1.9$\pm$0.2 & 2.4$\pm$0.9 & 5.67$\pm$1.9 &   \\
J014133-020222 &   & 1.281 & 140.8$\pm$1.7 & 170.8$\pm$1.2 & 212.2$\pm$2.3 & 191.6$\pm$2.5 & $<$1.87 & 2.4$\pm$0.2 & 2.8$\pm$0.6 & $<$7.70 & GPS \\
J014214-011458 &   & 0.425 & 108.1$\pm$1.2 & 102.3$\pm$0.8 & 50.4$\pm$1.0 & 36.9$\pm$1.4 & 2.1$\pm$0.3 & 3.8$\pm$0.4 & $<$13.1 & $<$24.2 &   \\
J014316-011901 & C &   & 259.9$\pm$2.6 & 122.0$\pm$5.1 & 62.8$\pm$1.2 & 56.5$\pm$1.5 & 2.5$\pm$0.3 & 3.7$\pm$0.5 & $<$6.42 & $<$18.5 &   \\
  & T & 0.520 & 275.0$\pm$4.1 & 178.0$\pm$5.2 & 90.0$\pm$2.1 & 72.0$\pm$2.1 &  &  &  &  &   \\
J015422-023454 &   & 0.082 & 139.3$\pm$1.5 & 117.6$\pm$0.8 & 83.1$\pm$1.3 & 75.3$\pm$1.8 & 10.1$\pm$0.9 & 8.1$\pm$0.6 & 4.74$\pm$1.3 & 12.1$\pm$4.5 &   \\
J020214-001748 &   &   & 63.9$\pm$0.8 & 83.0$\pm$0.6 & 82.3$\pm$1.2 & 78.4$\pm$1.9 & 2.7$\pm$0.7 & 2.0$\pm$0.4 & 2.60$\pm$1.0 & $<$9.01 & GPS \\
J020826-004744 &   & 2.025 & 419.5$\pm$5.0 & 365.2$\pm$2.6 & 235.5$\pm$2.5 & 201.6$\pm$2.2 & 3.5$\pm$0.3 & 5.6$\pm$0.4 & 7.49$\pm$1.0 & 12.1$\pm$1.2 &   \\
J021120-025027 &   &   & 43.3$\pm$0.6 & 41.3$\pm$0.4 & 34.4$\pm$0.9 & 36.7$\pm$1.6 & 6.13$\pm$1.6 & 4.38$\pm$1.1 & $<$15.1 & $<$25.6 & Upturned \\
J021121-014515 &   &   & 55.2$\pm$0.7 & 85.8$\pm$0.6 & 52.4$\pm$1.0 & 27.5$\pm$1.4 & 2.3$\pm$0.9 & .87$\pm$0.3 & $<$8.04 & $<$25.2 & GPS \\
J021301-001815 &   &   & 80.4$\pm$1.1 & 97.2$\pm$0.7 & 83.0$\pm$1.2 & 73.9$\pm$1.4 & .94$\pm$0.3 & 1.4$\pm$0.4 & $<$9.95 & $<$7.80 & GPS \\
J021542-022256 &   & 1.178 & 739.4$\pm$10. & 774.6$\pm$5.4 & 449.0$\pm$4.7 & 294.0$\pm$3.6 & .42$\pm$0.1 & .37$\pm$0.1 & 2.4$\pm$0.4 & 4.17$\pm$1.6 & GPS \\
J021605-011803 &   &   & 155.3$\pm$1.8 & 139.3$\pm$1.0 & 73.0$\pm$1.2 & 46.6$\pm$1.7 & $<$1.75 & .80$\pm$0.2 & $<$4.66 & 13.4$\pm$5.0 &   \\
J021612-010518 &   & 1.493 & 119.3$\pm$1.3 & 103.7$\pm$0.7 & 45.9$\pm$1.0 & 31.8$\pm$1.4 & 1.6$\pm$0.4 & 1.6$\pm$0.3 & $<$9.51 & $<$25.9 &   \\
J021755-012150 &   &   & 109.4$\pm$1.4 & 84.6$\pm$0.6 & 42.7$\pm$1.1 & 39.8$\pm$1.4 & .66$\pm$0.2 & 1.5$\pm$0.4 & $<$16.9 & 22.1$\pm$7.7 &   \\
J022054-015655 & C &   & --- & 262.0$\pm$2.7 & 60.4$\pm$1.5 & 16.5$\pm$4.9 & 7.9$\pm$0.7 & 11.0$\pm$0.7 & 6.87$\pm$2.4 & --- &   \\
  & T &   & 815.0$\pm$8.3 & 405.0$\pm$5.7 & 119.0$\pm$2.2 & 35.0$\pm$5.0 &  &  &  &  &   \\
J022313-020507 &   &   & 141.4$\pm$1.7 & 121.3$\pm$0.9 & 78.0$\pm$1.2 & 63.4$\pm$1.6 & 3.9$\pm$0.6 & 5.7$\pm$0.4 & 4.55$\pm$1.6 & $<$21.0 &   \\
J022508-003530 & C &   & 403.2$\pm$4.8 & 219.0$\pm$2.5 & 80.3$\pm$1.1 & 45.5$\pm$1.5 & 8.58$\pm$1.0 & 9.7$\pm$0.6 & 12.0$\pm$1.8 & 8.61$\pm$3.2 &   \\
  & T & 0.687 & 410.0$\pm$6.5 & 250.0$\pm$5.3 & 97.4$\pm$1.6 & 46.2$\pm$1.5 &  &  &  &  &   \\
J024240-000046 & C &   & --- & --- & 161.3$\pm$2.0 & 78.7$\pm$2.1 & .51$\pm$0.1 & .97$\pm$0.1 & 10.9$\pm$1.4 & $<$9.31 &   \\
  & T & 0.004 & 1620.0$\pm$34. & 980.0$\pm$12. & 320.0$\pm$10. & 175.0$\pm$5.3 &  &  &  &  &   \\
J030313-001459 & C &   & 393.0$\pm$4.9 & 201.2$\pm$1.6 & 70.4$\pm$1.2 & 33.2$\pm$2.0 & 1.1$\pm$0.1 & 3.1$\pm$0.2 & $<$5.52 & $<$30.8 &   \\
  & T & 0.700 & 406.0$\pm$5.0 & 235.0$\pm$5.3 & 81.0$\pm$1.2 & 53.0$\pm$2.1 &  &  &  &  &   \\
J031542-015123 & C &   & 146.0$\pm$1.6 & 141.5$\pm$1.3 & 143.0$\pm$2.3 & 103.2$\pm$1.6 & 3.6$\pm$0.6 & 2.3$\pm$0.3 & 1.3$\pm$0.4 & $<$6.32 &   \\
  & T &   & 181.0$\pm$2.7 & 145.0$\pm$2.2 & 133.0$\pm$2.3 & 99.0$\pm$5.1 &  &  &  &  &   \\
J031814-002949 &   &   & 104.7$\pm$1.2 & 85.6$\pm$0.6 & 51.5$\pm$20. & 41.6$\pm$1.3 & 1.3$\pm$0.4 & 2.7$\pm$0.4 & $<$13.1 & $<$13.7 &   \\
J032028-020839 &   &   & 265.6$\pm$3.2 & 186.7$\pm$1.3 & 82.5$\pm$1.3 & 49.2$\pm$1.2 & 3.9$\pm$0.6 & 5.5$\pm$0.4 & 7.27$\pm$1.8 & $<$22.9 &   \\
J033427-015358 &   &   & 15.2$\pm$0.4 & 35.7$\pm$0.3 & 71.8$\pm$1.0 & 58.1$\pm$1.3 & $<$10.7 & 3.01$\pm$1.0 & $<$7.32 & $<$8.81 & GPS \\
J035045-010847 &   &   & 106.7$\pm$1.4 & 79.4$\pm$0.6 & 34.7$\pm$0.9 & 26.4$\pm$1.8 & .65$\pm$0.2 & 2.3$\pm$0.9 & $<$18.3 & $<$41.9 &   \\
J040028-015207 &   &   & 54.5$\pm$0.7 & 60.8$\pm$0.5 & 69.5$\pm$1.1 & 70.0$\pm$3.9 & $<$2.77 & $<$6.77 & $<$5.76 & --- & Inverted \\
J040818-012235 &   &   & 152.4$\pm$1.9 & 133.2$\pm$1.0 & 92.6$\pm$1.3 & --- & .89$\pm$0.3 & 1.2$\pm$0.2 & $<$3.50 & --- &   \\
  &  &  &  &  & 98.8$\pm$0.5 & 70.9$\pm$0.6 &  &  & 1.5$\pm$0.6 & 2.68$\pm$1.0 &   \\
J041758-025019 &   &   & 90.4$\pm$1.2 & 102.4$\pm$0.7 & 112.8$\pm$1.5 & --- & 1.3$\pm$0.4 & 4.2$\pm$0.4 & $<$4.62 & --- &   \\
  &  &  &  &  & 103.7$\pm$0.5 & 84.4$\pm$1.1 &  &  & $<$3.17 & 3.5$\pm$0.7 & GPS \\
J042315-012033 &   & 0.914 & 3887.4$\pm$39. & 4227.7$\pm$29. & 3492.5$\pm$31. & --- & 1.1$\pm$0.1 & 2.6$\pm$0.2 & 4.0$\pm$0.4 & --- &   \\
  &  &  &  &  & 7500.3$\pm$3.5 & 7221.2$\pm$3.5 &  &  & 3.0$\pm$0.3 & 2.0$\pm$0.2 & GPS \\
J042511-024735 &   &   & 61.3$\pm$0.8 & 59.0$\pm$0.4 & 42.4$\pm$1.4 & --- & 3.65$\pm$1.5 & 1.6$\pm$0.6 & $<$13.6 & --- &   \\
  &  &  &  &  & 41.5$\pm$0.6 & 41.6$\pm$0.4 &  &  & $<$4.22 & 2.96$\pm$1.1 & Upturned \\
J042721-012634 &   &   & 110.2$\pm$1.2 & 79.8$\pm$0.6 & 44.9$\pm$1.3 & --- & 11.0$\pm$1.0 & 9.8$\pm$0.8 & 9.13$\pm$2.9 & --- &   \\
  &  &  &  &  & 48.2$\pm$0.5 & 32.2$\pm$0.6 &  &  & 6.91$\pm$1.2 & $<$10.7 &   \\
0725-009 &   & 0.128 & --- & --- & 2050.3$\pm$16. & --- & --- & --- & 2.7$\pm$0.5 & --- &   \\
  &  &  &  &  & 2570.6$\pm$0.7 & 2364.8$\pm$0.9 &  &  & .57$\pm$0.1 & 2.7$\pm$0.2 &   \\
J073025-024125 &   &   & 236.6$\pm$2.6 & 243.7$\pm$1.7 & 217.1$\pm$1.8 & --- & 2.6$\pm$0.3 & 1.7$\pm$0.2 & 4.2$\pm$0.7 & --- &   \\
  &  &  &  &  & 172.8$\pm$0.5 & 144.6$\pm$0.8 &  &  & 1.1$\pm$0.4 & 3.1$\pm$0.5 & GPS \\
J073245-022858 &   & 2.750 & 186.6$\pm$2.2 & 173.6$\pm$1.2 & 123.1$\pm$1.1 & --- & 5.0$\pm$0.5 & 2.7$\pm$0.2 & 3.12$\pm$1.2 & --- &   \\
  &  &  &  &  & 161.6$\pm$0.5 & 115.7$\pm$0.8 &  &  & 1.1$\pm$0.4 & 2.5$\pm$0.5 &   \\
J073251-012858 &   &   & 72.3$\pm$0.9 & 72.6$\pm$0.6 & 61.5$\pm$0.7 & --- & 1.9$\pm$0.7 & $<$2.86 & $<$9.02 & --- &   \\
  &  &  &  &  & 74.3$\pm$0.5 & 58.4$\pm$0.6 &  &  & 3.35$\pm$1.3 & 2.9$\pm$0.9 & GPS \\
J073333-002636 &   &   & 96.9$\pm$1.2 & 80.8$\pm$0.6 & 46.2$\pm$0.6 & --- & .90$\pm$0.3 & $<$2.82 & $<$14.3 & --- &   \\
  &  &  &  &  & 34.2$\pm$0.6 & 24.0$\pm$0.6 &  &  & $<$9.47 & $<$8.14 &   \\
J073520-002953 &   &   & 110.7$\pm$1.4 & 84.5$\pm$0.6 & 55.4$\pm$0.6 & --- & 2.6$\pm$0.5 & 3.0$\pm$0.6 & $<$12.0 & --- &   \\
  &  &  &  &  & 49.4$\pm$0.4 & 34.1$\pm$0.6 &  &  & 5.48$\pm$1.1 & 7.37$\pm$1.6 &   \\
J073625-015726 &   &   & 95.0$\pm$1.1 & 84.9$\pm$0.6 & 63.3$\pm$0.7 & --- & 2.8$\pm$0.4 & 4.3$\pm$0.5 & $<$10.8 & --- &   \\
  &  &  &  &  & 49.4$\pm$0.4 & 32.5$\pm$0.6 &  &  & $<$5.03 & $<$10.0 &   \\
J073833-020422 & C &   & 361.6$\pm$3.8 & 203.6$\pm$1.5 & 73.0$\pm$0.8 & --- & 6.7$\pm$0.8 & 7.1$\pm$0.5 & 17.0$\pm$4.1 & --- &   \\
  &  &  &  &  & --- & 31.3$\pm$0.7 &  &  & --- & 6.87$\pm$2.1 &   \\
  & T & 1.033 & 361.0$\pm$3.7 & 216.0$\pm$2.5 & 79.0$\pm$2.1 & --- &  &  &  &  &   \\
  &  &  &  &  & --- & 38.2$\pm$0.7 &  &  &   &   &   \\
J074554-004418 &   & 0.994 & 1970.0$\pm$20. & 1964.2$\pm$13. & 1094.7$\pm$8.8 & --- & 1.7$\pm$0.1 & 1.5$\pm$0.1 & 1.7$\pm$0.3 & --- &   \\
  &  &  &  &  & 960.2$\pm$0.8 & 605.4$\pm$0.7 &  &  & 4.0$\pm$0.4 & 4.6$\pm$0.4 &   \\
J074631-001930 &   &   & 124.1$\pm$1.5 & 87.1$\pm$0.6 & 34.8$\pm$0.5 & --- & .52$\pm$0.2 & $<$.866 & $<$13.1 & --- &   \\
  &  &  &  &  & 30.6$\pm$0.4 & 14.8$\pm$0.6 &  &  & $<$9.14 & $<$18.6 &   \\
J075408-001741 &   &   & 154.3$\pm$1.9 & 107.0$\pm$0.8 & 53.2$\pm$0.7 & --- & 3.3$\pm$0.5 & 3.5$\pm$0.4 & $<$15.9 & --- &   \\
  &  &  &  &  & 38.7$\pm$0.5 & 21.4$\pm$0.6 &  &  & 3.86$\pm$1.4 & 7.37$\pm$3.0 &   \\
J075447-024734 & C &   & 59.3$\pm$0.6 & 59.4$\pm$0.4 & 55.5$\pm$0.7 & --- & 2.7$\pm$0.9 & 1.5$\pm$0.6 & $<$12.6 & --- &   \\
  & T &   & 295.0$\pm$4.2 & 173.0$\pm$3.2 & --- & --- &  &  &  &  &   \\
J075953-022803 &   &   & 89.5$\pm$1.2 & 104.1$\pm$0.8 & 139.1$\pm$1.2 & --- & 2.4$\pm$0.8 & .65$\pm$0.2 & $<$4.04 & --- &   \\
  &  &  &  &  & 148.5$\pm$0.5 & 140.4$\pm$0.4 &  &  & $<$1.66 & 2.5$\pm$0.3 & GPS \\
J080059-000059 &   &   & 169.7$\pm$2.2 & 105.7$\pm$0.8 & 52.5$\pm$0.7 & --- & 7.5$\pm$0.7 & 7.8$\pm$0.6 & 11.2$\pm$4.3 & --- &   \\
  &  &  &  &  & 50.6$\pm$0.5 & 32.8$\pm$0.5 &  &  & 8.78$\pm$1.4 & 6.60$\pm$2.0 &   \\
J080512-011114 &   & 1.388 & 492.0$\pm$5.6 & 458.6$\pm$4.1 & 495.0$\pm$4.0 & --- & .39$\pm$0.1 & .65$\pm$0.1 & 3.1$\pm$0.4 & --- &   \\
  &  &  &  &  & 169.2$\pm$1.0 & 79.1$\pm$1.0 &  &  & 2.9$\pm$0.5 & $<$2.87 &   \\
J080537-005819 & C &   & 175.2$\pm$2.4 & 141.0$\pm$1.1 & 42.3$\pm$0.9 & --- & 3.34$\pm$1.1 & 5.3$\pm$0.6 & $<$7.24 & --- &   \\
  &  &  &  &  & --- & 18.3$\pm$0.7 &  &  & --- & $<$22.5 &   \\
  & T & 0.090 & 381.0$\pm$10. & 210.0$\pm$3.3 & 89.0$\pm$2.1 & --- &  &  &  &  &   \\
  &  &  &  &  & --- & 40.0$\pm$3.0 &  &  &   &   &   \\
J080845-020833 &   &   & 77.7$\pm$1.0 & 65.4$\pm$0.5 & 49.7$\pm$0.7 & --- & 7.3$\pm$0.8 & 6.4$\pm$0.6 & $<$9.09 & --- &   \\
  &  &  &  &  & 44.3$\pm$0.4 & 29.0$\pm$0.6 &  &  & $<$5.52 & $<$9.70 &   \\
J081219-012812 &   &   & 47.4$\pm$0.6 & 52.5$\pm$0.4 & 49.7$\pm$0.7 & --- & $<$1.89 & 2.89$\pm$1.2 & $<$9.60 & --- &   \\
  &  &  &  &  & 39.9$\pm$0.3 & 31.6$\pm$0.5 &  &  & 2.6$\pm$0.7 & $<$12.0 & GPS \\
J081827-020851 &   &   & 110.9$\pm$1.3 & 91.3$\pm$0.7 & 50.9$\pm$0.6 & --- & 3.1$\pm$0.4 & 2.6$\pm$0.4 & $<$13.0 & --- &   \\
  &  &  &  &  & 50.7$\pm$0.4 & 33.0$\pm$0.5 &  &  & 5.2$\pm$0.8 & $<$5.64 &   \\
J090125-003702 &   & 1.907 & 187.8$\pm$2.1 & 157.7$\pm$1.1 & 91.0$\pm$0.9 & --- & 1.4$\pm$0.2 & 2.2$\pm$0.3 & $<$5.01 & --- &   \\
  &  &  &  &  & 70.7$\pm$0.4 & 43.0$\pm$0.5 &  &  & $<$3.14 & $<$4.04 &   \\
J090944-023129 &   & 0.957 & 409.8$\pm$4.4 & 327.2$\pm$2.3 & 183.7$\pm$1.5 & --- & 4.7$\pm$0.6 & 5.1$\pm$0.3 & 4.8$\pm$0.9 & --- &   \\
  &  &  &  &  & 162.5$\pm$0.4 & 100.9$\pm$0.8 &  &  & 5.1$\pm$0.6 & 5.8$\pm$0.7 &   \\
J091107-020727 &   &   & 80.9$\pm$1.0 & 87.5$\pm$0.7 & 69.3$\pm$0.8 & --- & 2.5$\pm$0.5 & 2.6$\pm$0.2 & $<$9.62 & --- &   \\
  &  &  &  &  & 66.7$\pm$0.5 & 56.6$\pm$1.0 &  &  & 3.24$\pm$1.1 & $<$10.1 & GPS \\
J091643-025910 &   &   & 98.9$\pm$1.1 & 93.5$\pm$0.7 & 80.9$\pm$0.8 & --- & 1.8$\pm$0.4 & 3.0$\pm$0.4 & $<$15.3 & --- &   \\
  &  &  &  &  & 50.8$\pm$0.5 & 52.0$\pm$0.5 &  &  & $<$4.53 & $<$4.04 & Upturned \\
J093517-024107 &   &   & 156.5$\pm$1.8 & 128.7$\pm$0.9 & 72.2$\pm$0.8 & --- & 2.4$\pm$0.3 & 2.5$\pm$0.2 & $<$9.61 & --- &   \\
  &  &  &  &  & 62.8$\pm$0.5 & 39.8$\pm$0.6 &  &  & 2.3$\pm$0.8 & $<$5.87 &   \\
J093713-010025 &   &   & 99.1$\pm$1.2 & 105.2$\pm$0.8 & 74.0$\pm$0.8 & --- & 2.5$\pm$0.4 & 2.6$\pm$0.3 & $<$8.45 & --- &   \\
  &  &  &  &  & 78.2$\pm$0.4 & 51.6$\pm$0.5 &  &  & $<$5.39 & $<$5.49 & GPS \\
J094040-002801 &   & 2.325 & 177.8$\pm$2.1 & 140.7$\pm$1.0 & 77.6$\pm$0.8 & --- & 2.8$\pm$0.3 & 2.7$\pm$0.7 & $<$5.45 & --- &   \\
  &  &  &  &  & 69.7$\pm$0.4 & 42.6$\pm$0.4 &  &  & $<$5.57 & 4.1$\pm$0.8 &   \\
J094123-014251 & J &   & 167.4$\pm$1.9 & 94.0$\pm$0.8 & 32.8$\pm$0.6 & --- & 4.2$\pm$0.5 & 5.8$\pm$0.5 & 8.52$\pm$2.3 & --- &   \\
  &  &  &  &  & --- & 1.5$\pm$0.7 &  &  & --- & --- &   \\
  & T & 0.382 & 306.0$\pm$3.7 & 170.0$\pm$10. & 61.5$\pm$1.6 & --- &  &  &  &  &   \\
J094319-000426 &   &   & 385.0$\pm$4.3 & 204.3$\pm$1.5 & 78.5$\pm$0.8 & --- & .78$\pm$0.1 & 1.6$\pm$0.3 & $<$5.73 & --- &   \\
  &  &  &  &  & 72.4$\pm$0.5 & 37.4$\pm$0.6 &  &  & 4.6$\pm$0.8 & 4.07$\pm$1.4 &   \\
J094409-015116 &   & 0.112 & 49.5$\pm$0.8 & 51.5$\pm$0.4 & 65.6$\pm$0.6 & --- & $<$5.31 & $<$3.37 & 5.88$\pm$2.3 & --- &   \\
  &  &  &  &  & 60.5$\pm$0.5 & 41.0$\pm$0.6 &  &  & $<$4.23 & $<$4.03 & GPS \\
J094544-015304 &   &   & 182.5$\pm$2.8 & 159.4$\pm$1.1 & 146.2$\pm$1.3 & --- & 1.4$\pm$0.3 & 1.3$\pm$0.5 & 3.65$\pm$1.3 & --- &   \\
  &  &  &  &  & 215.8$\pm$0.5 & 205.1$\pm$1.0 &  &  & 2.6$\pm$0.4 & 1.8$\pm$0.3 &   \\
J095158-000126 & C &   & --- & 203.0$\pm$2.5 & 39.2$\pm$1.0 & --- & 7.4$\pm$0.7 & 4.5$\pm$0.4 & $<$13.5 & --- &   \\
  & T & 1.487 & 660.0$\pm$8.3 & 322.0$\pm$5.5 & 67.0$\pm$1.1 & --- &  &  &  &  &   \\
J095727-015655 &   & 0.860 & 177.6$\pm$2.0 & 147.9$\pm$1.1 & 71.2$\pm$0.8 & --- & .93$\pm$0.2 & .83$\pm$0.3 & $<$9.72 & --- &   \\
J095828-014000 &   &   & 420.0$\pm$4.7 & 250.2$\pm$1.8 & 88.2$\pm$0.9 & --- & 1.0$\pm$0.1 & 3.9$\pm$0.3 & 6.86$\pm$2.1 & --- &   \\
J100445-011917 &   & 1.344 & 43.8$\pm$0.6 & 46.0$\pm$0.4 & 39.5$\pm$0.7 & --- & 4.99$\pm$1.2 & 2.9$\pm$0.7 & $<$18.8 & --- &   \\
  &  &  &  &  & 44.2$\pm$0.9 & 47.4$\pm$0.8 &  &  & $<$3.60 & $<$6.74 & Inverted \\
J100704-020711 &   & 1.215 & 483.3$\pm$5.2 & 487.9$\pm$3.4 & 414.1$\pm$3.5 & --- & 2.7$\pm$0.3 & 2.3$\pm$0.2 & 2.0$\pm$0.5 & --- &   \\
  &  &  &  &  & 467.4$\pm$2.1 & 413.8$\pm$1.8 &  &  & 1.8$\pm$0.2 & 4.0$\pm$0.4 & GPS \\
J101051-020019 &   & 0.890 & 464.5$\pm$4.9 & 446.0$\pm$3.2 & 479.5$\pm$4.0 & --- & 1.2$\pm$0.2 & 2.3$\pm$0.2 & 5.1$\pm$0.7 & --- &   \\
J101956-002411 &   & 1.130 & 43.2$\pm$0.6 & 69.1$\pm$0.5 & 45.9$\pm$0.5 & --- & $<$8.00 & 1.8$\pm$0.7 & 6.38$\pm$2.3 & --- &   \\
  &  &  &  &  & 36.2$\pm$0.5 & 25.7$\pm$0.9 &  &  & $<$7.20 & $<$7.92 & GPS \\
J102429-005258 &   & 2.554 & 557.1$\pm$5.9 & 374.8$\pm$2.6 & 130.4$\pm$1.1 & --- & 1.7$\pm$0.2 & 1.6$\pm$0.1 & 3.6$\pm$0.9 & --- &   \\
J102441-001648 &   &   & 86.4$\pm$1.1 & 75.5$\pm$0.6 & 59.5$\pm$0.7 & --- & 6.8$\pm$0.7 & 5.5$\pm$0.5 & $<$12.4 & --- &   \\
J102834-023659 &   & 0.476 & 136.7$\pm$1.7 & 172.9$\pm$1.2 & 346.1$\pm$3.0 & --- & 3.9$\pm$0.4 & 3.7$\pm$0.3 & 3.0$\pm$0.6 & --- &   \\
  &  &  &  &  & 385.2$\pm$1.2 & 666.2$\pm$3.3 &  &  & .97$\pm$0.2 & .49$\pm$0.2 & Inverted \\
J102837-010027 &   & 1.531 & 136.9$\pm$1.6 & 98.8$\pm$0.7 & 64.9$\pm$0.7 & --- & 1.2$\pm$0.3 & 6.9$\pm$0.5 & 8.07$\pm$2.6 & --- &   \\
J102953-014318 &   &   & 108.7$\pm$1.2 & 94.3$\pm$0.7 & 67.6$\pm$0.8 & --- & 3.4$\pm$0.4 & 2.5$\pm$0.3 & $<$18.4 & --- &   \\
  &  &  &  &  & 67.3$\pm$0.6 & 57.4$\pm$1.0 &  &  & 2.71$\pm$1.0 & $<$5.01 &   \\
J104341-014407 &   &   & 58.6$\pm$0.7 & 67.9$\pm$0.5 & 61.6$\pm$0.7 & --- & $<$2.59 & 1.3$\pm$0.5 & $<$6.28 & --- &   \\
  &  &  &  &  & 62.3$\pm$0.5 & 68.5$\pm$0.9 &  &  & 3.5$\pm$0.9 & $<$5.14 & Inverted \\
J104713-002805 & C &   & 68.9$\pm$1.4 & 68.9$\pm$0.6 & 78.5$\pm$0.8 & --- & 4.7$\pm$0.7 & 2.5$\pm$0.6 & 4.28$\pm$1.9 & --- &   \\
  &  &  &  &  & --- & 47.8$\pm$1.5 &  &  & --- & $<$6.03 &   \\
  & T &   & 105.0$\pm$2.3 & 92.0$\pm$2.1 & --- & --- &  &  &  &  &   \\
J105034-025151 & C &   & 115.2$\pm$1.2 & 51.5$\pm$0.4 & 26.1$\pm$0.5 & --- & 4.0$\pm$0.6 & 4.9$\pm$0.7 & $<$26.2 & --- &   \\
  &  &  &  &  & --- & 13.2$\pm$1.8 &  &  & --- & $<$133. &   \\
  & T &   & 128.0$\pm$2.4 & 89.0$\pm$1.2 & 40.0$\pm$1.0 & --- &  &  &  &  &   \\
  &  &  &  &  & --- & 21.8$\pm$0.8 &  &  &   &   &   \\
J110204-011626 & C &   & 419.2$\pm$4.4 & 212.2$\pm$1.7 & 60.7$\pm$0.7 & --- & 8.33$\pm$1.2 & 11.8$\pm$0.7 & 15.3$\pm$2.2 & --- &   \\
  &  &  &  &  & --- & 32.0$\pm$1.1 &  &  & --- & 18.9$\pm$5.0 &   \\
  & T & 1.554 & 672.0$\pm$7.4 & 202.7$\pm$1.5 & 91.0$\pm$3.1 & --- &  &  &  &  &   \\
  &  &  &  &  & --- & 41.0$\pm$2.0 &  &  &   &   &   \\
J110225-023534 &   &   & 293.6$\pm$3.1 & 180.5$\pm$1.3 & 71.4$\pm$0.7 & --- & 5.3$\pm$0.4 & 5.8$\pm$0.4 & $<$11.0 & --- &   \\
  &  &  &  &  & 76.1$\pm$0.6 & 45.5$\pm$0.6 &  &  & 5.1$\pm$0.9 & 8.02$\pm$1.4 &   \\
J110631-005252 & C &   & --- & 203.2$\pm$1.5 & 86.0$\pm$0.9 & --- & 2.5$\pm$0.2 & 4.5$\pm$0.4 & $<$16.1 & --- &   \\
  &  &  &  &  & --- & 53.2$\pm$1.5 &  &  & --- & 3.99$\pm$1.6 &   \\
  & T & 0.423 & 455.0$\pm$6.8 & 327.0$\pm$3.0 & 153.0$\pm$2.3 & --- &  &  &  &  &   \\
  &  &  &  &  & --- & 82.1$\pm$5.1 &  &  &   &   &   \\
J111331-021259 &   & 0.125 & 348.0$\pm$4.0 & 182.7$\pm$1.4 & 56.4$\pm$0.7 & --- & 6.9$\pm$0.7 & 8.3$\pm$0.5 & 11.0$\pm$3.6 & --- &   \\
J111439-024731 &   & 1.040 & 261.1$\pm$3.0 & 326.7$\pm$2.3 & 306.9$\pm$2.5 & --- & 3.1$\pm$0.3 & 1.1$\pm$0.1 & 2.6$\pm$0.6 & --- &   \\
  &  &  &  &  & 219.6$\pm$1.2 & 222.2$\pm$1.0 &  &  & 1.2$\pm$0.3 & .89$\pm$0.3 & Inverted \\
J111733-023600 &   & 0.463 & 384.1$\pm$4.1 & 261.2$\pm$1.8 & 165.5$\pm$1.5 & --- & 1.1$\pm$0.1 & 1.8$\pm$0.2 & 2.78$\pm$1.1 & --- &   \\
  &  &  &  &  & 153.7$\pm$1.0 & 119.7$\pm$0.8 &  &  & 1.4$\pm$0.5 & $<$2.77 &   \\
J112119-001316 & C &   & 163.3$\pm$1.7 & 127.5$\pm$0.9 & 78.4$\pm$0.8 & --- & 2.8$\pm$0.3 & 3.5$\pm$0.3 & $<$6.54 & --- &   \\
  &  &  &  &  & --- & 8.9$\pm$0.9 &  &  & --- & --- &   \\
  & T & 0.099 & 293.0$\pm$5.8 & 183.0$\pm$2.4 & 92.0$\pm$3.1 & --- &  &  &  &  &   \\
J112635-021538 &   & 2.133 & 101.1$\pm$1.1 & 83.2$\pm$0.6 & 61.4$\pm$0.7 & --- & 5.6$\pm$0.6 & 5.0$\pm$0.5 & $<$12.4 & --- &   \\
  &  &  &  &  & 60.4$\pm$0.6 & 53.9$\pm$1.0 &  &  & 3.38$\pm$1.0 & $<$8.94 &   \\
1150-003 &   & 1.976 & --- & --- & --- & --- & --- & --- & --- & --- &   \\
  &  &  &  &  & 803.3$\pm$0.2 & 642.0$\pm$0.7 &  &  & 1.3$\pm$0.2 & .63$\pm$0.2 &   \\
J120012-025403 &   &   & 44.9$\pm$0.6 & 51.2$\pm$0.4 & 85.8$\pm$0.9 & --- & 2.7$\pm$0.7 & 8.0$\pm$0.8 & 9.84$\pm$2.3 & --- &   \\
  &  &  &  &  & 67.9$\pm$0.7 & 67.9$\pm$0.9 &  &  & 7.13$\pm$1.2 & 5.97$\pm$1.2 & Inverted \\
J120405-002948 & C &   & 73.4$\pm$1.1 & 53.6$\pm$0.5 & 60.2$\pm$0.8 & --- & 8.53$\pm$1.0 & 12.6$\pm$1.4 & $<$9.82 & --- &   \\
  &  &  &  &  & --- & 53.2$\pm$1.1 &  &  & --- & $<$9.29 &   \\
  & T &   & 161.0$\pm$2.6 & 134.0$\pm$5.1 & 92.0$\pm$3.1 & --- &  &  &  &  &   \\
  &  &  &  &  & --- & 58.0$\pm$4.1 &  &  &   &   &   \\
J120741-010630 &   & 1.006 & 157.4$\pm$1.7 & 136.8$\pm$1.0 & 141.8$\pm$1.3 & --- & 5.4$\pm$0.8 & 4.4$\pm$0.3 & $<$2.57 & --- &   \\
  &  &  &  &  & 154.2$\pm$0.7 & 178.9$\pm$1.5 &  &  & 2.0$\pm$0.5 & $<$2.93 & Upturned \\
J121031-013653 & C &   & 171.0$\pm$2.6 & 110.7$\pm$1.5 & 32.5$\pm$0.9 & --- & 1.8$\pm$0.5 & 2.5$\pm$0.3 & $<$15.8 & --- &   \\
  &  &  &  &  & --- & 11.2$\pm$1.1 &  &  & --- & --- &   \\
  & T & 0.273 & 250.0$\pm$6.5 & 161.0$\pm$2.3 & 53.0$\pm$1.1 & --- &  &  &  &  &   \\
J121622-010753 &   & 0.475 & 129.2$\pm$1.5 & 130.9$\pm$1.0 & 116.5$\pm$1.1 & --- & 2.0$\pm$0.4 & 2.7$\pm$0.3 & 5.06$\pm$1.5 & --- &   \\
J121758-002945 &   & 0.418 & 336.3$\pm$3.6 & 336.5$\pm$2.4 & 279.8$\pm$2.3 & --- & 2.8$\pm$0.3 & 2.7$\pm$0.2 & 3.1$\pm$0.7 & --- &   \\
  &  &  &  &  & 341.1$\pm$0.6 & 446.1$\pm$3.5 &  &  & 1.6$\pm$0.2 & 2.8$\pm$0.4 & Inverted \\
J121834-011953 &   & 0.415 & 188.9$\pm$2.0 & 192.0$\pm$1.4 & 157.6$\pm$1.4 & --- & 4.3$\pm$0.6 & 3.7$\pm$0.3 & 5.79$\pm$1.4 & --- &   \\
  &  &  &  &  & 139.1$\pm$0.6 & 130.6$\pm$1.0 &  &  & 4.5$\pm$0.6 & 4.97$\pm$1.0 & GPS \\
J132130-001402 &   &   & 78.7$\pm$0.9 & 70.1$\pm$0.6 & 69.8$\pm$0.8 & --- & 3.4$\pm$0.8 & $<$2.76 & $<$5.58 & --- &   \\
  &  &  &  &  & 86.4$\pm$0.6 & 72.4$\pm$1.1 &  &  & 1.7$\pm$0.6 & $<$6.39 &   \\
J152721-001908 &   &   & 48.0$\pm$0.6 & 59.8$\pm$0.5 & --- & --- & $<$2.51 & 1.3$\pm$0.6 & --- & --- &   \\
J161048-011344 &   &   & 184.7$\pm$2.1 & 105.8$\pm$0.9 & --- & --- & .60$\pm$0.2 & 1.4$\pm$0.4 & --- & --- &   \\
J194522-015322 &   &   & 284.1$\pm$3.0 & 293.2$\pm$2.1 & 203.4$\pm$2.2 & --- & 1.3$\pm$0.2 & 1.4$\pm$0.2 & 1.9$\pm$0.5 & --- &   \\
J194743-010324 &   &   & 190.1$\pm$2.7 & 197.1$\pm$1.4 & 136.8$\pm$1.7 & 84.3$\pm$2.1 & .54$\pm$0.1 & .68$\pm$0.2 & 5.82$\pm$1.4 & 15.6$\pm$4.7 & GPS \\
J200608-022335 &   & 1.457 & 964.0$\pm$13. & 653.7$\pm$4.6 & 235.5$\pm$2.6 & 95.5$\pm$2.4 & 3.2$\pm$0.3 & 2.5$\pm$0.2 & 3.4$\pm$0.7 & $<$22.9 &   \\
J201428-004723 &   &   & 63.6$\pm$1.1 & 50.1$\pm$0.4 & 31.4$\pm$1.2 & 20.7$\pm$1.9 & 3.7$\pm$0.9 & 5.1$\pm$0.9 & $<$14.2 & --- &   \\
J201515-013731 &   &   & 486.0$\pm$8.5 & 462.3$\pm$3.2 & 323.0$\pm$3.9 & 202.1$\pm$3.6 & 1.5$\pm$0.2 & .96$\pm$0.1 & 4.7$\pm$0.5 & 6.81$\pm$1.4 &   \\
J202332-012341 &   &   & 199.2$\pm$2.3 & 186.1$\pm$1.3 & 187.8$\pm$2.5 & 172.3$\pm$3.1 & 3.1$\pm$0.5 & 4.5$\pm$0.3 & 2.7$\pm$0.9 & $<$7.46 &   \\
J204248-014909 &   &   & 116.0$\pm$1.3 & 118.9$\pm$0.9 & 109.9$\pm$1.8 & 81.6$\pm$1.7 & 2.5$\pm$0.4 & 1.8$\pm$0.3 & 3.54$\pm$1.3 & $<$8.70 & GPS \\
J204314-004037 &   &   & 112.5$\pm$1.2 & 96.4$\pm$0.7 & 70.4$\pm$1.2 & 40.8$\pm$1.5 & 5.4$\pm$0.8 & 5.3$\pm$0.4 & $<$16.6 & $<$48.2 &   \\
J204539-012641 &   &   & 65.5$\pm$0.8 & 59.3$\pm$0.5 & 65.0$\pm$1.1 & 50.4$\pm$1.7 & 1.4$\pm$0.4 & 1.7$\pm$0.6 & $<$3.56 & $<$32.7 &   \\
J204710-023622 &   & 0.942 & 866.7$\pm$9.7 & 561.4$\pm$3.9 & 205.6$\pm$2.2 & 97.9$\pm$2.0 & 1.0$\pm$0.1 & .39$\pm$0.1 & 3.4$\pm$0.5 & 10.1$\pm$3.6 &   \\
J204745-024603 &   &   & 61.9$\pm$0.9 & 66.2$\pm$0.5 & 48.2$\pm$1.1 & 35.0$\pm$0.9 & 3.23$\pm$1.2 & 2.1$\pm$0.6 & $<$6.25 & $<$65.9 & GPS \\
J205329-010013 &   &   & 81.7$\pm$1.3 & 82.2$\pm$0.6 & 49.2$\pm$1.0 & 47.8$\pm$1.3 & $<$3.09 & 1.5$\pm$0.3 & $<$15.2 & $<$47.2 & GPS \\
J210808-010009 &   &   & 100.8$\pm$1.5 & 76.2$\pm$0.6 & 43.5$\pm$0.9 & 29.5$\pm$1.4 & 6.2$\pm$0.7 & 6.0$\pm$0.6 & 8.38$\pm$3.0 & --- &   \\
J211022-012658 &   &   & 198.4$\pm$2.8 & 228.0$\pm$1.6 & 197.0$\pm$2.1 & 72.5$\pm$1.9 & 2.5$\pm$0.3 & 2.3$\pm$0.2 & 3.3$\pm$0.7 & $<$13.2 & GPS \\
J211603-010829 &   & 0.305 & 76.3$\pm$1.1 & 80.4$\pm$0.6 & 57.7$\pm$1.1 & 42.6$\pm$1.3 & 2.7$\pm$0.5 & 2.6$\pm$0.8 & $<$5.60 & $<$19.7 & GPS \\
J211807-025847 &   &   & 74.6$\pm$0.9 & 76.0$\pm$0.6 & 119.1$\pm$1.5 & 103.1$\pm$2.0 & 3.4$\pm$0.6 & 1.8$\pm$0.4 & 2.4$\pm$0.9 & $<$11.1 & GPS \\
J212159-002613 &   &   & 51.6$\pm$0.6 & 55.1$\pm$0.4 & 56.3$\pm$0.8 & 44.6$\pm$1.2 & $<$4.17 & $<$2.70 & $<$4.40 & $<$33.7 & GPS \\
2134-018 &   & 1.285 & 2411.5$\pm$24. & 2353.7$\pm$16. & 1698.4$\pm$15. & 1217.8$\pm$17. & 4.2$\pm$0.3 & 5.5$\pm$0.3 & 7.6$\pm$0.8 & 8.0$\pm$0.7 &   \\
J214156-003600 &   &   & 82.1$\pm$0.9 & 90.8$\pm$0.7 & 70.8$\pm$1.1 & 49.4$\pm$1.3 & 1.3$\pm$0.5 & 7.87$\pm$3.0 & $<$6.22 & $<$26.8 & GPS \\
J214843-012237 &   &   & 220.6$\pm$2.3 & 147.2$\pm$1.1 & 54.3$\pm$1.3 & 31.3$\pm$0.8 & 1.6$\pm$0.2 & 2.2$\pm$0.3 & $<$8.99 & $<$35.7 &   \\
J215614-003704 &   & 0.495 & 278.8$\pm$2.9 & 378.0$\pm$2.7 & 392.9$\pm$7.2 & 371.6$\pm$5.6 & 1.7$\pm$0.3 & 1.8$\pm$0.1 & 2.4$\pm$0.4 & $<$2.63 & GPS \\
J215934-010555 &   &   & 127.3$\pm$1.4 & 123.0$\pm$0.9 & 86.2$\pm$1.2 & 59.9$\pm$1.7 & 1.0$\pm$0.3 & 1.2$\pm$0.3 & $<$5.77 & 7.87$\pm$3.3 &   \\
J220643-003103 &   &   & 110.1$\pm$1.2 & 116.7$\pm$0.8 & 109.4$\pm$1.4 & 94.7$\pm$2.1 & 1.5$\pm$0.3 & 2.3$\pm$0.3 & $<$4.40 & 8.88$\pm$3.6 & GPS \\
J220755-000215 &   &   & 68.6$\pm$0.8 & 63.1$\pm$0.5 & 45.3$\pm$1.0 & 36.5$\pm$1.4 & 2.4$\pm$0.5 & 3.0$\pm$0.5 & $<$11.4 & 28.1$\pm$7.8 &   \\
J221745-023238 &   &   & 233.3$\pm$2.5 & 160.5$\pm$1.1 & 74.9$\pm$1.0 & 49.3$\pm$1.3 & 2.7$\pm$0.3 & 2.0$\pm$0.2 & $<$10.0 & $<$26.4 &   \\
J221947-005132 &   &   & 172.6$\pm$1.8 & 126.2$\pm$0.9 & 67.2$\pm$1.0 & 49.9$\pm$1.3 & 1.3$\pm$0.3 & 2.4$\pm$0.3 & 5.41$\pm$1.7 & $<$18.4 &   \\
J222222-025006 &   &   & 452.2$\pm$4.7 & 255.1$\pm$1.8 & 69.2$\pm$1.1 & 26.8$\pm$2.3 & 1.9$\pm$0.3 & 4.9$\pm$0.3 & 5.62$\pm$1.9 & --- &   \\
J222352-021043 & C &   & 230.3$\pm$2.5 & 131.3$\pm$1.0 & 47.9$\pm$1.5 & 23.3$\pm$2.9 & 17.7$\pm$2.5 & 19.0$\pm$1.2 & 23.1$\pm$4.5 & --- &   \\
  & T &   & 930.0$\pm$13. & 374.0$\pm$4.8 & 80.0$\pm$5.0 & --- &  &  &  &  &   \\
J222501-022430 &   &   & 126.5$\pm$1.4 & 86.3$\pm$0.6 & 37.5$\pm$1.0 & 21.8$\pm$2.0 & 1.6$\pm$0.3 & 1.5$\pm$0.3 & $<$8.28 & --- &   \\
J224405-024039 &   &   & 67.8$\pm$0.8 & 79.8$\pm$0.6 & 67.7$\pm$1.1 & 61.0$\pm$1.6 & 2.8$\pm$0.6 & 2.6$\pm$0.5 & $<$4.04 & 13.2$\pm$4.9 & GPS \\
J224801-015742 &   &   & 239.5$\pm$3.1 & 179.9$\pm$1.3 & 92.7$\pm$1.4 & 69.0$\pm$2.1 & 3.3$\pm$0.4 & 2.3$\pm$0.2 & 4.19$\pm$1.7 & $<$22.5 &   \\
J230107-015804 &   &   & 643.0$\pm$8.1 & 1059.7$\pm$7.4 & 1112.5$\pm$11. & 1083.0$\pm$11. & 1.2$\pm$0.2 & 3.5$\pm$0.2 & 7.6$\pm$0.8 & 6.4$\pm$0.7 & GPS \\
J230545-003608 & C &   & --- & --- & 33.9$\pm$1.2 & 26.9$\pm$2.4 & $<$3.12 & 2.1$\pm$0.2 & 5.74$\pm$2.5 & --- &   \\
  & T & 0.269 & 215.0$\pm$2.9 & 148.0$\pm$1.3 & 59.0$\pm$3.0 & 40.0$\pm$3.0 &  &  &  &  &   \\
J231028-024350 &   &   & 59.9$\pm$0.7 & 66.7$\pm$0.5 & 52.6$\pm$1.1 & 44.1$\pm$1.8 & $<$1.89 & $<$1.48 & $<$5.16 & --- & GPS \\
J231136-020907 & C &   & 62.4$\pm$0.7 & 56.2$\pm$0.4 & 53.6$\pm$1.0 & 62.7$\pm$2.1 & 2.1$\pm$0.7 & 1.7$\pm$0.4 & 8.01$\pm$2.9 & $<$27.3 & Upturned \\
  & T &   & 86.0$\pm$0.9 & 82.0$\pm$1.3 & 54.0$\pm$1.1 & 64.5$\pm$2.9 &  &  &  &  &   \\
J231222-010925 &   & 1.431 & 130.8$\pm$1.5 & 126.5$\pm$0.9 & 112.7$\pm$1.5 & 102.0$\pm$2.4 & 2.1$\pm$0.3 & 2.3$\pm$0.3 & 6.86$\pm$2.6 & $<$11.3 &   \\
J232304-015048 &   & 1.774 & 246.4$\pm$2.7 & 209.7$\pm$1.5 & 162.6$\pm$1.8 & 195.4$\pm$2.3 & 5.0$\pm$0.6 & 5.2$\pm$0.3 & 3.81$\pm$1.6 & 2.64$\pm$1.0 & Upturned \\
J232404-005853 &   &   & 122.3$\pm$1.4 & 120.5$\pm$0.9 & 101.9$\pm$1.7 & 70.5$\pm$1.7 & 2.4$\pm$0.4 & 1.8$\pm$0.3 & $<$6.33 & $<$12.4 &   \\
J232653-020213 & C &   & 237.0$\pm$3.4 & 186.2$\pm$1.8 & 157.6$\pm$1.8 & 151.7$\pm$2.5 & $<$3.97 & .78$\pm$0.1 & 2.4$\pm$0.8 & $<$8.03 &   \\
  & T & 0.188 & 570.0$\pm$30. & 390.0$\pm$20. & 207.0$\pm$10. & 160.0$\pm$10. &  &  &  &  &   \\
J232659-002352 &   & 2.160 & 108.4$\pm$1.4 & 113.4$\pm$0.8 & 107.6$\pm$1.4 & 89.5$\pm$1.7 & 3.1$\pm$0.4 & 2.7$\pm$0.3 & $<$5.81 & 8.19$\pm$3.0 & GPS \\
J232710-004158 &   & 0.099 & 109.9$\pm$1.3 & 109.0$\pm$0.8 & 90.6$\pm$1.2 & 79.8$\pm$1.7 & 1.0$\pm$0.3 & $<$2.38 & 3.36$\pm$1.0 & $<$17.7 &   \\
J235013-020614 &   & 1.675 & 72.4$\pm$0.8 & 85.2$\pm$0.6 & 65.2$\pm$1.1 & 50.1$\pm$1.5 & $<$4.75 & $<$1.69 & $<$7.89 & 12.5$\pm$4.6 & GPS \\
J235025-022441 &   & 1.036 & 465.4$\pm$5.0 & 236.3$\pm$1.7 & 53.6$\pm$0.9 & 23.7$\pm$1.5 & .53$\pm$0.1 & .84$\pm$0.1 & $<$15.0 & --- &   \\
J235156-010909 & C &   & --- & 173.0$\pm$3.2 & 80.9$\pm$1.6 & 49.9$\pm$1.9 & 4.1$\pm$0.4 & 3.5$\pm$0.7 & 6.36$\pm$1.9 & $<$35.8 &   \\
  & T & 0.174 & 618.0$\pm$11. & 406.0$\pm$5.8 & 184.0$\pm$2.5 & 106.0$\pm$2.3 &  &  &  &  &   \\
J235409-001946 &   & 0.462 & 402.3$\pm$4.2 & 307.3$\pm$2.2 & 139.8$\pm$1.6 & 76.7$\pm$1.9 & .60$\pm$0.1 & .72$\pm$0.1 & 3.78$\pm$1.0 & 8.74$\pm$3.7 &   \\
J235705-022634 &   &   & 211.2$\pm$2.2 & 117.8$\pm$0.9 & 39.2$\pm$1.0 & 13.1$\pm$2.3 & 4.3$\pm$0.4 & 4.9$\pm$0.4 & $<$7.62 & --- &   \\
J235725-015214 &   & 0.812 & 165.2$\pm$1.8 & 160.6$\pm$1.1 & 129.9$\pm$1.5 & 112.3$\pm$1.9 & 1.0$\pm$0.2 & 1.4$\pm$0.2 & 5.93$\pm$1.1 & 7.67$\pm$1.9 &   \\

\enddata

\tablenotetext{1}{The quoted uncertainties are the quadrature sum of the AIPS IMFIT fitting uncertainty (which is typically slightly larger than the rms) and our estimated systematic uncertainty. For the flux densities is the flux calibration uncertainty, which is equal to 1\%\ for the C band,  0.7\% for the X-band, 0.8\% for the K-band, and 1\% for the Q-band.  For the polarization percentages we estimate an uncertainty (based on the measured spread in polarization of our calibrator sources) of 8\% in the C-band, 6\% in the X-band, 10\% in the K-band and 8\% in the Q-band. }
\tablenotetext{2}{The bulk of these data were taken in the Summer and early Fall of 2008 (see Table\,1). For the subset of sources with K and Q-band observations in November 2009, these flux densities and polarization percentages are listed just below the 2008 values for the same source. }
\tablenotetext{3}{C=Core, J=Jet, T=Total. The jet(or lobe) photometry is quoted for J094123-014251, since for this source the northern jet corresponds to the AT20G position (see Figure\,\ref{fig_extended}). }
 
\end{deluxetable}
\end{center}

\begin{center}
\begin{deluxetable}{cccccc}
\tablecolumns{6}
\tablewidth{5in}
\tabletypesize{\scriptsize}
\tablecaption{\label{table_gbt} 90GHz measured and extrapolated fluxes}
\tablehead{
\colhead{Name} & \colhead{$S_{90GHz,meas}$} & \colhead{$S_{90GHz,extrap}$} & \colhead{Ratio} & \colhead{$S_{90GHz,extrap}$ } & \colhead{Ratio} \\
\colhead{} & \colhead{[mJy]} & \colhead{[mJy] VLA $S_K$} & \colhead{} & \colhead{[mJy] $S_K$\,$\times$\,1.24} & \colhead{} }
\startdata
J021542-022256 &  112.4$\pm$ 18.8 &  181.9 &   0.6 &  143.2 &   0.8 \\
J041758-025019 &  114.3$\pm$ 19.2 &   67.2 &   1.7 &   52.9 &   2.2 \\
J042315-012033 & 5982.8$\pm$966.8 & 7062.3 &   0.8 & 5560.3 &   1.1 \\
J073025-024125 &  109.5$\pm$ 18.3 &  118.5 &   0.9 &   93.3 &   1.2 \\
J073245-022858 &   65.7$\pm$ 11.0 &   79.7 &   0.8 &   62.7 &   1.0 \\
J074554-004418 &  373.6$\pm$ 60.8 &  362.5 &   1.0 &  285.4 &   1.3 \\
J075953-022803 &  125.2$\pm$ 21.0 &  131.7 &   1.0 &  103.7 &   1.2 \\
J080512-011114 &  101.7$\pm$ 17.1 &   33.9 &   3.0 &   26.7 &   3.8 \\
J094544-015304 &  104.7$\pm$ 17.0 &  193.7 &   0.5 &  152.5 &   0.7 \\
J100445-011917 &   13.4$\pm$  2.4 &   51.2 &   0.3 &   40.3 &   0.3 \\
J100704-020711 &  189.1$\pm$ 27.8 &  361.2 &   0.5 &  284.4 &   0.7 \\
J102834-023659 &  614.1$\pm$ 90.2 & 1224.9 &   0.5 &  964.3 &   0.6 \\
J104341-014407 &   26.4$\pm$  4.2 &   76.1 &   0.3 &   59.9 &   0.4 \\
J111439-024731 &  141.4$\pm$ 20.8 &  225.0 &   0.6 &  177.2 &   0.8 \\
J120741-010630 &  217.5$\pm$ 31.9 &  211.2 &   1.0 &  166.3 &   1.3 \\
J121622-010753 &   67.1$\pm$  9.9 & --- & --- & --- & --- \\
J121758-002945 &  556.7$\pm$ 81.7 &  601.3 &   0.9 &  473.4 &   1.2 \\
J121834-011953 &   81.0$\pm$ 11.9 &  121.7 &   0.7 &   95.8 &   0.8 \\
J202332-012341 &  201.0$\pm$ 15.9 &  168.9 &   1.2 &  133.0 &   1.5 \\
J211807-025847 &   72.1$\pm$  6.1 &   87.5 &   0.8 &   68.9 &   1.0 \\
J215614-003704 &  497.1$\pm$ 39.3 &  347.1 &   1.4 &  273.3 &   1.8 \\
J230107-015804 &  303.0$\pm$ 24.2 &  941.8 &   0.3 &  741.5 &   0.4 \\
J232304-015048 &  139.7$\pm$ 11.2 &  236.6 &   0.6 &  186.3 &   0.8 \\
J232653-020213 &  215.1$\pm$ 17.1 &  120.2 &   1.8 &   94.6 &   2.3 \\
J235725-015214 &   53.4$\pm$  4.5 &   95.6 &   0.6 &   75.3 &   0.7 \\

\enddata
\end{deluxetable}
\end{center}

\begin{center}
\begin{deluxetable}{ccccc}
\tablecolumns{5}
\tablewidth{5in}
%\tabletypesize{\scriptsize}
\tablecaption{\label{table_indices} Mean and median\tablenotemark{*} spectral indices}
\tablehead{
\colhead{Subset} & \colhead{$\alpha_{CX}$} & \colhead{$\alpha_{XK}$} & \colhead{$\alpha_{KQ}$} & \colhead{$\alpha_{QW}$} }
\startdata
Singles & -0.16(-0.13) & -0.37(-0.35) &  -0.45(-0.41) & \\
Singles+Totals & -0.24(-0.21) & -0.43(-0.37) &  -0.49(-0.45) &  \\
90GHz sample & 0.07(0.03) &  -0.06(-0.09) &  -0.14(-0.09) &  -0.47(-0.46)
\enddata
\tablenotetext{*}{The number is brackets is the median.}
\end{deluxetable}
\end{center}

\end{document}